\documentclass[11pt,a4paper]{article}

\usepackage{jheppub}

\usepackage[english]{babel}

\usepackage{float,subfigure}

\allowdisplaybreaks[2]

\newcommand{\half}{{\textstyle\frac{1}{2}}}

\newcommand{\tvec}[1]{\boldsymbol{#1}}

\newcommand{\eq}[1]{\eqref{eq:#1}}

\newcommand{\vect}[1]{\boldsymbol{#1}}
\newcommand{\df}{\mathrm{d}}

\newcommand{\ms}{\mskip 1.5mu}
\newcommand{\bs}{\mskip -1.5mu}

\newcommand{\lsim}{\raisebox{-4pt}{%
    $\,\stackrel{\textstyle <}{\sim}\,$}}
\newcommand{\gsim}{\raisebox{-4pt}{%
    $\,\stackrel{\textstyle >}{\sim}\,$}}

\subheader{\hfill DESY 15-187}

\title{Cancellation of Glauber gluon exchange in the double Drell-Yan process}

\author[a]{Markus Diehl,}
\author[a,1]{Jonathan R.~Gaunt,}
\author[b]{Daniel Ostermeier}
\author[b]{Peter Pl\"o{\ss}l}
\author[b]{and Andreas Sch\"afer}

\affiliation[a]{Deutsches Elektronen-Synchroton DESY, 22603 Hamburg,
  Germany}

\affiliation[b]{Institut f\"ur Theoretische Physik, Universit\"at
  Regensburg, 93040 Regensburg, Germany}

\note[1]{Present address: Nikhef Theory Group and VU University Amsterdam,
  De Boelelaan 1081, 1081 HV Amsterdam, The Netherlands}

\emailAdd{markus.diehl@desy.de}
\emailAdd{j.gaunt@nikhef.nl}
\emailAdd{Daniel.Ostermeier@physik.uni-regensburg.de}
\emailAdd{Peter.Ploessl@physik.uni-regensburg.de}
\emailAdd{Andreas.Schaefer@physik.uni-regensburg.de}

\abstract{%
  An essential part of any factorisation proof is the demonstration that
  the exchange of Glauber gluons cancels for the considered observable.
  We show this cancellation at all orders for double Drell-Yan production
  (the double parton scattering process in which a pair of electroweak
  gauge bosons is produced) both for the integrated cross section and for
  the cross section differential in the transverse boson momenta.  In the
  process of constructing this proof, we also revisit and clarify some
  issues regarding the Glauber cancellation argument and its relation to
  the rest of the factorisation proof for the single Drell-Yan process.}

\begin{document}

\maketitle



\section{Introduction}
\label{sec:introduction}

As the LHC continues to run after its energy upgrade, hopes are high that
signals for physics beyond the Standard Model will show up in the coming
years.  To identify such signals and to understand their nature will be
challenging for experiment and theory.  This provides strong motivation to
strive for a detailed and quantitative understanding of the final state in
high-energy proton-proton collisions.  Steady progress in perturbative
calculations allows for increasing precision in describing the final state
of a single parton-level scattering.  However, much remains to be
understood even at the conceptual level for multiple hard interactions,
where several parton-level scatters occur in one proton-proton collision.
While such multiple interactions are power suppressed in sufficiently
inclusive cross sections, they are unsuppressed in specific regions of
phase space \cite{Diehl:2011yj}.  Moreover, their importance grows with
increasing collision energy.  Since many search channels for new physics
involve high multiplicity, a thorough understanding of multiple hard
scattering is highly desirable.  Recent years have seen considerable
progress, from theory to phenomenology to experimental studies, as is for
instance documented in the proceedings
\cite{Bartalini:2011jp,Platzer:2012gla,Bansal:2014paa,Astalos:2015ivw}.

The theory for single hard scattering in proton-proton collisions rests on
factorisation formulae, which describe the cross section in terms of
parton distributions, hard-scattering cross sections at parton level, and
possibly other quantities like soft factors.  Substantial effort has gone
into proving these formulae to all orders of perturbation theory in QCD,
see e.g.\ \cite{Collins:1989gx,Sterman:1995fz,Collins:2011zzd} for
comprehensive accounts.  A crucial part of such proofs is to cast the
exchange of soft gluons between the two colliding protons into a form
consistent with the factorisation formula.  In particular kinematics,
which is referred to as the Glauber region and which essentially describes
small-angle parton-parton scattering, the approximations required to
achieve such a form break down.  Factorisation therefore only holds if the
net effect of all Glauber gluon exchanges cancels in the observable
considered.  A prominent example where such a cancellation fails to occur
(and factorisation is strongly violated) is hard diffraction
\cite{Alvero:1998ta,Affolder:2000vb}.  Observables in Drell-Yan production
for which Glauber gluon exchange breaks factorisation are discussed in
\cite{Gaunt:2014ska,Zeng:2015iba}.

Factorisation formulae can also be written down for multiple hard
scattering, where multi-parton distributions appear instead of the
familiar single-parton distributions.  Together with model assumptions
connecting these multi-parton distributions with their single-parton
counterparts, this forms the basis of most phenomenological
investigations, see for instance \cite{Humpert:1984ay, Ametller:1985tp,
  Mangano:1988sq, Godbole:1989ti, Drees:1996rw, Eboli:1997sv,
  DelFabbro:1999tf, Kulesza:1999zh, DelFabbro:2002pw, Cattaruzza:2005nu,
  Maina:2009vx, Domdey:2009bg, Maina:2009sj, Berger:2009cm, Maina:2010vh,
  Gaunt:2010pi, Kom:2011bd, Baranov:2011ch, Berger:2011ep, Kom:2011nu,
  Luszczak:2011zp, Berezhnoy:2012xq, Maciula:2013kd, Blok:2013bpa,
  Cazaroto:2013fua, vanHameren:2014ava, Maciula:2014pla,
  Golec-Biernat:2014nsa, Gaunt:2014rua, Maciula:2015vza, Tao:2015nra,
  Blok:2015rka, Blok:2015afa}.  To put this framework on a more solid
footing, it is necessary to understand whether factorisation actually
holds for multiple hard scattering.  Several pieces of evidence pointing
towards a positive answer have been given in
\cite{Diehl:2011yj,Blok:2011bu,Manohar:2012jr}, but the issue of Glauber
gluons in this context has not been analysed yet.  It is the purpose of
the present paper to fill this gap.  We limit ourselves to double hard
scattering and more specifically to the double Drell-Yan process, i.e.\ to
the production of two electroweak gauge bosons.  To the best of our
knowledge, the cancellation of Glauber gluon exchange in single hard
scattering has also been established only for Drell-Yan production
\cite{Bodwin:1984hc,Collins:1985ue,Collins:1988ig,Collins:2011zzd}.  We
will show that the proof given in \cite{Collins:2011zzd} can indeed be
adapted to the double Drell-Yan process.  In doing so, we will revisit and
clarify some subtle issues in the single Drell-Yan case.

The most commonly used form of factorisation involves parton distributions
integrated over transverse parton momenta, often called ``collinear''
distributions.  Correspondingly, the net transverse momentum of the
particles produced in a hard-scattering subprocess is then integrated over
in the cross section formula.  A different factorisation formalism uses
transverse-momentum dependent parton distributions (TMDs).  It allows one
for instance to compute the transverse-momentum spectrum of the gauge
boson produced in the Drell-Yan process in the region where that
transverse momentum is small compared to the boson invariant mass.  TMD
factorisation for hadron-hadron collisions has only been established for
cases where the particles produced in the hard scattering are colourless,
i.e.\ for the production of electroweak gauge bosons and Higgs particles.
For final states involving hadrons, problems related to gluon exchange
in the Glauber region have so far prevented the formulation of TMD
factorisation \cite{Rogers:2010dm}.
Under the restriction to colourless final states, our proof of Glauber
gluon cancellation for double hard scattering holds in both the collinear
and TMD cases. As already noted for single DY production in
\cite{Collins:2011zzd}, such a proof is no more complicated (if not
simpler) in the TMD framework than for collinear factorisation, and we
treat both cases together in this paper.  A possible extension to final
states involving hadrons, which is of obvious phenomenological relevance,
must await further work.

This paper is organised as follows.  In the next section we give an
account of the overall logic and status of a factorisation proof for the
double Drell-Yan process, in close analogy to the single Drell-Yan case.
Section~\ref{sec:one-gluon} analyses one-gluon exchange in a toy
model, with explicit calculations illustrating the role of complex contour
deformation in avoiding contributions from the Glauber region.  Within
this toy model we show how the exchange of a single Glauber gluon cancels
in the cross section.  In section~\ref{sec:allorder} we present an all-order
proof, following the work in \cite{Collins:1988ig,Collins:2011zzd} on
single Drell-Yan production, where an essential tool is light-cone time
ordered perturbation theory.  We summarise our main results in
section~\ref{sec:sum}.

\section{Overview of the factorisation proof}
\label{sec:fact-overview}

In this section we give a broad overview over the different steps of the
factorisation proof for double hard scattering with colourless final state
particles.  For definiteness, we consider double Drell-Yan production,
$p+p \to V_1 + V_2 + X$, where $V_1$ and $V_2$ are electroweak gauge
bosons and $X$ is the unobserved hadronic final state.  Although many of
the steps do not directly concern the main topic of this work, namely the
cancellation of Glauber gluon exchange, we find it important to exhibit
the interplay of this part of the proof with the other steps, which need
to be taken in a consistent order.

Several parts of the factorisation proof have been elaborated in
\cite{Diehl:2011yj} or \cite{Manohar:2012jr}, but others still require
further work in our opinion.  Many elements of the proof are the same for
TMD and collinear factorisation, so that the discussion can conveniently
treat both cases in parallel.
We largely follow the factorisation proof for single Drell-Yan production
as given in chapter 14 of \cite{Collins:2011zzd}, with some modifications
that we will point out.


\subsection{Overall strategy}
\label{sec:overall-strategy}

\begin{enumerate}
\item \textbf{Power counting.}  Both collinear and TMD factorisation are
  based on a power expansion in a small parameter $\Lambda/Q$.  Here $Q$
  is the hard scale of the process, whereas $\Lambda$ represents small
  kinematic quantities and the scale of nonperturbative QCD interactions.
  The terms ``leading'' and ``dominant'' refer to the limit of small
  $\Lambda/Q$.  

  In double Drell-Yan production, $Q$ is given by the invariant masses of
  the two produced bosons (which we treat as being of the same order).
  For collinear factorisation there is no kinematic scale of size
  $\Lambda$, whereas for TMD factorisation the transverse momenta of the
  bosons are counted to be of order $\Lambda$.  This does \emph{not} mean
  that TMD factorisation is limited to transverse momenta in the
  nonperturbative region; what counts is that they are small compared with
  the hard scale of the process.  Since $\Lambda/Q$ is the parameter used
  to quantify the accuracy of the final factorisation formula, $\Lambda$
  is to be taken as the largest of all scales counted as ``small'' and $Q$
  as the smallest of all scales counted as ``large''.

\item \textbf{Dominant graphs and regions.}\label{enum:leading-regions}
  Following the analysis of Libby and Sterman \cite{Sterman:1978bi,
    Libby:1978bx}, one identifies the dominant regions of loop integration
  for each individual graph that contributes to the cross section.  These
  regions are characterised by the momenta of internal lines being either
  hard, soft (which includes the Glauber region), or collinear to an
  external direction.  They correspond to pinch singularities of the graph
  when all quantities of order $\Lambda$ (masses and transverse momenta)
  are set to zero.  For double Drell-Yan production there are just two
  collinear directions, given by the incoming protons.  Within collinear
  factorisation, unobserved jets in the final state provide additional
  directions for collinear lines, but they disappear in the final result.
  (In TMD factorisation, production of additional jets is power suppressed
  due to the requirement of small transverse momentum for the produced
  gauge bosons.)

  For each leading momentum region, a graph can be organised into
  subgraphs containing only hard, soft or collinear lines.  Dimensional
  analysis shows how these subgraphs can be connected in order to give a
  leading contribution.  In this context, different roles are played by
  gluons with transverse or longitudinal polarisation (respectively having
  a polarisation vector that is approximately transverse or approximately
  proportional to the gluon momentum).  A collinear and a hard subgraph
  must be connected by the minimum number of quarks\footnote{For ease of
    language, we use the term ``quarks'' to denote both quarks and
    antiquarks.}
  or transverse gluons necessary for the process, whereas there can be any
  number of longitudinal gluons.  Between a collinear and a soft graph
  there can be any number of soft gluons.  We only consider soft subgraphs
  that couple to at least two different collinear graphs; soft subgraphs
  connecting to a single collinear graph are treated as a part of that
  graph.  Soft gluons coupling to a hard subgraph are power suppressed.

  For our further discussion we introduce light-cone coordinates,
  $a^\pm = (a^0 \pm a^3) /\sqrt{2}$ for each four-vector and denote its
  transverse components as $\tvec{a} = (a_1, a_2)$.  The components of the
  full vector are given as $(a^+, a^-, \tvec{a})$.  We choose a coordinate
  system in the overall c.m.\ frame where both incoming protons have zero
  transverse momentum, with one proton moving fast to the right and the
  other fast to the left.  The typical size of momenta is then
  \begin{align}
    \label{hard-coll-scaling}
    l & \sim (Q,\ms Q,\ms Q) && \text{for hard lines,}
    \nonumber \\
    l & \sim (Q,\ms \Lambda^2/Q,\ms \Lambda) &&
      \text{for right-moving collinear lines,}
    \nonumber \\
    l & \sim (\Lambda^2/Q,\ms Q,\ms \Lambda) &&
      \text{for left-moving collinear lines.}
  \end{align}
  Virtualities are thus of order $Q^2$ for hard lines and of order
  $\Lambda^2$ for collinear lines.  We note that in low-order graphs, hard
  momenta may have small or zero transverse components, which does however
  not affect the general analysis.  Soft momenta have all components of
  order $\Lambda$ or smaller.  As we will discuss in
  section~\ref{sec:soft-scaling}, there are important differences whether
  soft momentum components are of order $\Lambda$ or $\Lambda^2/Q$.
  
  The analysis described here is based on Feynman graphs in perturbation
  theory.  It is understood that in the final factorisation formula,
  quantities that involve lines with virtuality $\Lambda^2$ or less (i.e.\
  collinear or soft lines) will be represented by hadronic matrix elements
  of quark and gluon operators, which are meaningful beyond perturbation
  theory.

  After these preliminaries, we can now specify the dominant graphs for
  double Drell-Yan production.
\begin{enumerate}
\item For TMD factorisation, the dominant graphs are as shown in figure
  \ref{fig:strategy:leadinggraphs}.  They have the same structure as for
  single Drell-Yan production, except that there are two distinct
  hard-scattering subgraphs on either side of the final-state cut.  For
  ease of notation, we group the hard subgraphs such that $H_i$ ($i=1,2$)
  comprises the two subgraphs (on either side of the final-state cut) that
  produce the gauge boson $V_i$.  There are exactly four quark lines
  entering each subgraph $H_i$.

  We denote the collinear subgraph with right-moving lines by $A$ and the
  one with left-moving lines by $B$.  Each of them has four external quark
  lines and is related to a double-parton distribution that appears in
  the final factorisation formula.  We note that the soft factor $S$ may
  consist of several connected components.  As mentioned above, each of
  these components must couple to both $A$ and $B$.  The simplest
  connected subgraphs of $S$ consist of one gluon directly connecting the
  two collinear graphs.

\begin{figure}
\begin{center}
 \includegraphics[width=0.6\textwidth]{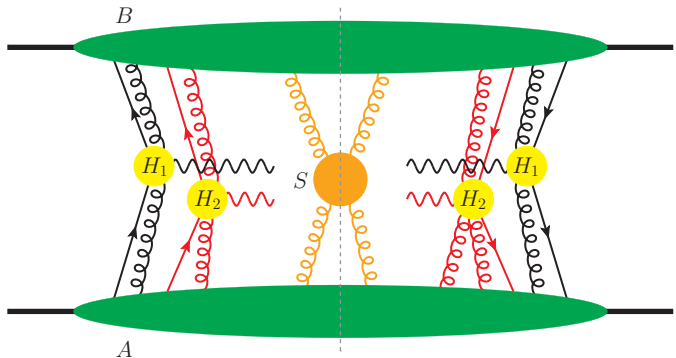}
 \caption{\label{fig:strategy:leadinggraphs} Dominant graphs for double
   Drell-Yan production, $p+p \to V_1 + V_2 + X$, in TMD factorisation.
   $A$ and $B$ denote collinear subgraphs, $S$ the soft subgraph, and
   $H_1$, $H_2$ the hard subgraphs.}
\end{center}
\end{figure}

\item The dominant graphs in collinear factorisation are more complicated
  since each of the two hard scatters can produce additional jets, which
  are part of the unobserved hadronic final state $X$ in the cross
  section.  These additional collinear factors in the final state can
  further couple to the soft subgraph, as shown in figure
  \ref{fig:dDY_inclusive_regions}.  A series of arguments is needed to
  convert this to two hard-scattering graphs with cut parton lines in the
  final state, as in figure~\ref{fig:dDY_inclusive_modified}.  The
  arguments are the same as for the single Drell-Yan process.  We will
  review them in a separate publication.

  \begin{figure}
    \begin{center}
      \includegraphics[width=0.6\textwidth]{%
        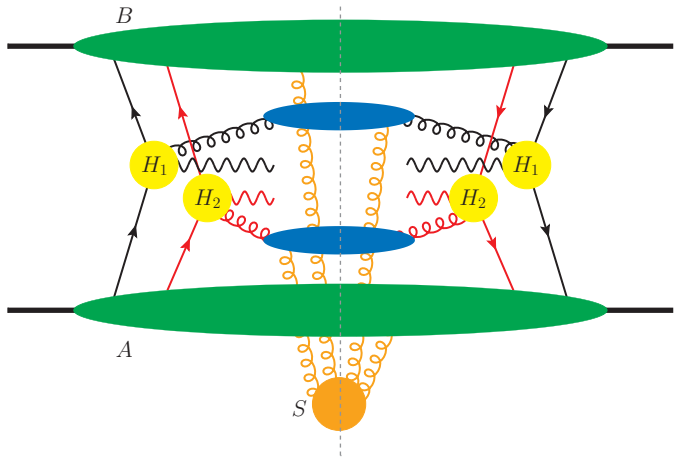}
      \caption{\label{fig:dDY_inclusive_regions} Dominant graph for double
        Drell-Yan production, $p+p \to V_1 + V_2 + X$, in collinear
        factorisation, with one unobserved jet produced by each hard
        scatter.  Not shown for the sake of clarity are additional
        longitudinal gluons that can be exchanged between each pair of a
        hard and a collinear subgraph.}
    \end{center}
  \end{figure}

  \begin{figure}
    \begin{center}
      \includegraphics[width=0.6\textwidth]{%
        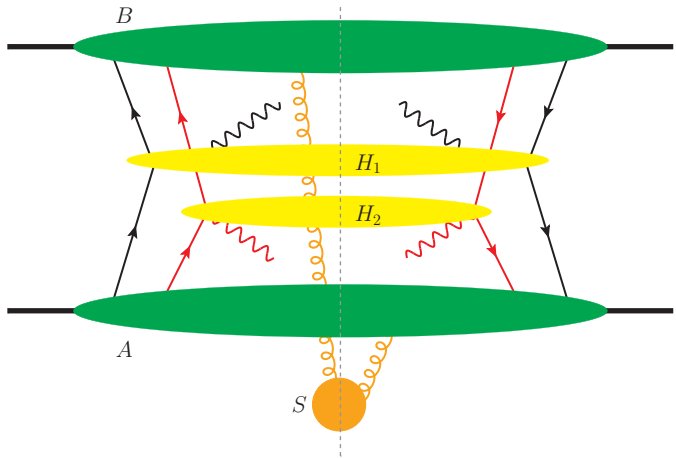} 
      \caption{\label{fig:dDY_inclusive_modified} Simplification of
        figure~\protect\ref{fig:dDY_inclusive_regions} after a series of
        approximations.  Again, an arbitrary number of longitudinal gluons
        can be exchanged between each hard and collinear subgraph.}
    \end{center}
  \end{figure}

  At higher orders in $\alpha_s$, incoming quarks in a hard graph can be
  replaced by gluons with transverse polarisation.  The four partons with
  physical polarisation in each collinear factor thus can be different
  combinations of quarks and gluons.  For brevity we will call these
  ``physical partons''.

\item \label{enum:SPSvsDPS} While the reaction $p + p \to V_1 + V_2 + X$
  can proceed by double hard scattering (double parton scattering, DPS),
  two gauge bosons can also be produced in a single hard scattering
  (single parton scattering, SPS).  In TMD factorisation, SPS and DPS have
  the same power behaviour in $\Lambda/Q$ whereas in collinear
  factorisation SPS is enhanced over DPS by a factor $Q^2/\Lambda^2$.  (On
  the other hand, due to a larger flux of incoming partons, DPS is
  generically enhanced over SPS at small $Q^2/s$, where $s$ is the squared
  $pp$ c.m.\ energy.)  There are also interference terms between SPS and
  DPS, which have the same power behaviour as DPS for both TMD and
  collinear factorisation.  For details we refer to section 2.4 of
  \cite{Diehl:2011yj}.

  As discussed in sections 5.2.3 and 5.3 of \cite{Diehl:2011yj}, certain
  graphs can contribute to both SPS and DPS in different momentum regions.
  (An example is graph b in figure~\ref{fig:spect-model} below.)  The
  existing factorisation formulae for DPS extend into the SPS region and
  vice versa, so that there is a double counting problem.  This problem
  has been discussed in \cite{Cacciari:2009dp, Gaunt:2011xd, Blok:2011bu,
    Ryskin:2011kk, Manohar:2012pe}, but in our opinion it has not been
  solved in a satisfactory way.
\end{enumerate}

\item \textbf{Approximations.}\label{enum:approximations} In each dominant
  momentum region of a graph, one makes appropriate approximations.  As
  discussed in section 10.4.2 of \cite{Collins:2011zzd}, these can be
  grouped into different types:
\begin{enumerate}
\item \textbf{Kinematics.}\label{enum:kin} For collinear lines entering a
  hard subgraph one keeps only large light-cone components, neglecting
  small light-cone components and transverse components.  A right-moving
  momentum $\ell$ entering $H_i$ is thus replaced by
  \begin{align}
    \hat{\ell} &= (\ell^+, 0, \tvec{0}) \,.
  \end{align}
  A small rescaling of the large components can be made so as to have
  exact momentum conservation in the hard subgraph.  In hard subgraphs one
  neglects all quark masses that are small compared with $Q$ (heavy-quark
  contributions are not considered in this work).

  One can also approximate soft momenta $\ell$ that enter a collinear
  subgraph.  As explained in section \ref{sec:soft-approx}, we take here a
  different approximation than the one in \cite{Collins:2011zzd}, which is
  not suitable for the case of double hard scattering.  In section
  \ref{sec:one-gluon} we do not make any kinematic approximation for soft
  gluons, whereas in section \ref{sec:allorder} we neglect the light-cone
  component of $\ell$ that is much smaller than the large component of
  collinear momenta.  For a soft gluon entering graph $A$ we thus replace
  $\ell$ with
  \begin{equation}
    \label{soft-mom-approx}
    \tilde{\ell} = (0, \ell^-, \tvec{\ell}) \,.
  \end{equation}

  After these kinematic approximations, certain loop integrals only
  involve some subgraphs rather than the product of all of them.  This
  induces ultraviolet divergences that are not present in the original
  graph for the cross section and hence not cancelled by the counterterms
  in the QCD Lagrangian.  The individual factors hence require additional
  ultraviolet renormalisation.  At this stage of the argument, all
  computations are to be done in a regularised theory (in practice in $d =
  4-2\epsilon$ dimensions).  The required UV subtractions are done at the
  very end (see point~\ref{enum:UV-subtr} below).

  A comment is in order about the routing of loop momenta.  We choose
  independent loop integration variables such that momentum conservation
  is already satisfied for the collinear and soft factors (if such a
  factor has $n$ external parton lines, there are hence $n-1$ independent
  loop momenta).  A particular choice of independent loop variables may be
  needed in parts of the proof and will be specified when necessary.  For
  the hard subgraphs we write explicit $\delta$ functions to ensure
  momentum conservation.  These $\delta$ functions constrain momentum
  components of the incoming partons, unless they disappear because
  corresponding momentum components of final state particles are
  integrated over.  As part of the kinematic approximations, we neglect in
  the $\delta$ functions small plus or minus components compared to
  large ones (i.e.\ the minus momenta of right-moving and the plus momenta
  of left-moving collinear lines).  $\delta$ functions for transverse
  parton momenta are to be kept unapproximated, even if these momenta are
  neglected inside the hard subgraph.  In the final factorisation formula
  for double parton scattering these $\delta$ functions ``tie together''
  certain transverse momenta in the double parton distributions of the two
  protons, as shown e.g.\ in section~2.1.2 of \cite{Diehl:2011yj}.

\item \textbf{Fermion lines.}\label{enum:Fierz} One performs Fierz
  transformations so that collinear and hard factors have no open Dirac
  indices.  One then selects the large Lorentz components of Dirac
  matrices.  This is shown for double parton scattering in section 2.2.1
  of \cite{Diehl:2011yj}.  Note that even for unpolarised protons, double
  parton distributions have a nontrivial polarisation dependence, since
  the polarisations of two partons can be correlated among themselves.
  Such spin correlations have observable effects in the cross section, as
  shown for instance in \cite{Kasemets:2012pr,Echevarria:2015ufa}.

\item \textbf{Gluons connecting different subgraphs.}\label{enum:GY} For
  the numerator factors of soft or of longitudinally polarised collinear
  gluons, one makes a so-called Grammer-Yennie (also called eikonal)
  approximation, which generalises the treatment by Grammer and Yennie of
  soft photons in QED \cite{Grammer:1973db}.
  For a collinear gluon with momentum $\ell$ flowing from $A$ into a hard
  subgraph $H$, we approximate
  \begin{align}
    \label{GY-collin}
    A^{\mu}(\ell)\, H_{\mu}(\hat{\ell}) & \approx A^+(\ell)\, H^-(\hat{\ell})
    \nonumber \\
     & = A^+(\ell)\, \frac{\ell^+_{\phantom{A}}
              v_A^-}{\ell^+_{\phantom{A}} v_A^- + i \varepsilon}\,
             H^-(\hat{\ell})
     \approx A_{\mu}(\ell)\,
             \frac{v_A^{\mu}}{\ell v_A^{\phantom{+}} + i \varepsilon}\,
             \hat{\ell}^{\nu} H_{\nu}(\hat{\ell}) \,,
  \end{align}
  where $v_A^{} = (v_A^+,v_A^-,\tvec{0})$ is an auxiliary vector that is
  widely separated in rapidity from the right-moving
  gluon.\footnote{\label{rap-def} We define the rapidity of a vector as
    $y = \half \log \frac{|v^+|}{|v^-|}$ with absolute values of the
    components, so that the definition also applies to spacelike $v$.}
  The rapidity of $v_A$ can hence be large and negative
  ($|v_A^-| \gg |v_A^+|$) or central ($|v_A^-| \sim |v_A^+|$).  In the
  first and last steps of \eqref{GY-collin} we used that $A^\mu$ scales
  like a right-moving collinear momentum, so that $A^+ \sim Q$ is its
  largest component.  In the first step we also used that the components
  of $H^\mu$ are all of size $Q$ (the transverse components may be zero
  for symmetry reasons, which does not affect the argument).  The
  $i\varepsilon$ in the denominator provides a regularisation of the pole
  at $\ell v_A = 0$ and will be commented on in
  point~\ref{enum:deform-back}.
  An analogous approximation is made for collinear gluons flowing from $B$
  into $H$, with an auxiliary vector $v_B$ that has either central or large
  positive rapidity ($|v_B^+| \gg |v_B^-|$), and with $\hat{\ell}$ being
  obtained from $\ell$ by keeping only the large component $\ell^-$.

  Similar approximations are made for soft gluons connecting to a
  collinear factor.  For a gluon with momentum flowing from $S$ into $A$,
  the Grammer-Yennie approximation reads
  \begin{align}
    \label{GY-soft}
    S^{\mu}(\ell)\, A_{\mu}(\tilde{\ell})
     & \approx S^-(\ell)\, A^+(\tilde{\ell})
  \nonumber \\
     &= S^-(\ell)\, \frac{\ell^-_{\phantom{R}}
         v_R^+}{\ell^-_{\phantom{R}} v_R^+ + i\varepsilon}\,
         A^+(\tilde{\ell})
  \approx S_{\mu}(\ell)\, 
         \frac{v_R^{\mu}}{\ell v_R^{\phantom{+}} + i\varepsilon}\;
         \tilde{\ell}^{\nu}\bs A_{\nu}(\tilde{\ell}) \,,
  \end{align}
  where $v_R$ has large positive rapidity.  In the first step we have
  again used the scaling properties of $A^\mu$, and in the first and last
  step we have additionally used that the minus component of $S^\mu$ is
  not smaller than its other components (see below).  An analogous
  approximation holds for a gluon flowing from $S$ into $B$, with an
  auxiliary vector $v_L$ that has large negative rapidity.
  As discussed in section \ref{sec:soft-approx}, our approximation
  \eqref{GY-soft} differs from the one in \cite{Collins:2011zzd} because
  we use a different kinematic approximation for soft momenta entering
  collinear subgraphs.  Equation~\eqref{GY-soft} is a slight modification
  of the form used in \cite{Collins:2007ph}, where $\tilde{\ell}$ was
  replaced with the full soft momentum $\ell$ (as we will do in
  section~\ref{sec:one-gluon}).

  In collinear factorisation for single hard scattering, one can take
  lightlike auxiliary vectors, setting either $v^+$ or $v^-$ to zero as
  appropriate.  However, for TMD factorisation this choice would lead to
  divergences in the rapidity of gluon momenta $\ell$ for the factors $A$,
  $B$ and $S$.  For double hard scattering this happens even in collinear
  factorisation.  Different regulators for these divergences have been
  proposed, see e.g.\ \cite{Collins:2011zzd, Ji:2004wu, Becher:2010tm,
    Chiu:2012ir, Echevarria:2015usa}.  We follow \cite{Collins:2011zzd}
  and use non-lightlike vectors $v$.

\item \textbf{Glauber region.}\label{enum:Glauber} A serious complication
  of the proof is that the soft Grammer-Yennie approximation
  \eqref{GY-soft} does \emph{not} work for gluon momenta in the Glauber
  region, which we define as
  \begin{align}
    \label{Glauber-scaling}
    \ell \sim (\Lambda^2 /Q,\ms  \Lambda^2 /Q,\ms \Lambda) \,.
  \end{align}
  More precisely, if $\ell^- \sim \Lambda^2 /Q$ and
  $|\tvec{\ell}| \sim \Lambda$, then the approximation
  $\ell^- A^+ \approx \tilde{\ell}^\nu\bs A_\nu$ in the last step of
  \eqref{GY-soft} fails, because with $A^+ \sim Q$ and
  $|\tvec{A}| \sim \Lambda$ the contribution of transverse components to
  the scalar product $\tilde{\ell}^\nu\bs A_\nu$ cannot be neglected.  A
  similar statement holds in the analogue of \eqref{GY-soft} for a gluon
  connecting $S$ with $B$ if the gluon momentum satisfies
  $\ell^+ \sim \Lambda^2 /Q$ and $|\tvec{\ell}| \sim \Lambda$.  Moreover,
  to derive \eqref{GY-soft} we used that the minus component of $S^\mu$ is
  not smaller than its other components, which can be shown if the gluons
  attached to the soft factor have momenta with
  $\ell_j^+ \sim \ell_j^- \sim |\tvec{\ell}_j^{}|$, but which is not
  obvious if gluon momenta are in the Glauber region.

  One can overcome these problems by deforming the integration contour
  for each soft gluon momentum $\ell$ into a region of the complex plane
  where the Grammer-Yennie approximation works.  This deformation must be
  consistent with the analyticity properties in $\ell$ of the relevant
  subgraphs.  It can be performed if the subgraphs do not have any
  singularities obstructing the contour deformation, or if the additional
  contributions obtained when crossing such singularities cancel after
  summing over all final-state cuts of a given graph.  To show that one of
  these conditions is always met for double Drell-Yan production is the
  main objective of this paper.
\end{enumerate}

\item \textbf{Subtraction formalism.}\label{enum:subtractions} For each
  graph that gives a dominant contribution to the cross section, there are
  one or more terms with approximations made as just discussed, each
  corresponding to a distinct region of the loop momenta (and hence to a
  distinct way of organising the graph into subgraphs).  In each of these
  terms, all loop momenta are integrated over their full range, not only
  over the momentum region for which the approximations have been
  designed.  Using cutoffs to restrict the loop momenta would seriously
  complicate a systematic analysis, for instance because cutoffs break
  Lorentz invariance and lead to complicated nonlocal operators in the
  definition of parton distributions and soft factors.

  The subtraction procedure discussed in sections 10.1 and 10.7 of
  \cite{Collins:2011zzd} provides an alternative to momentum cutoffs and
  ensures that the sum over the terms approximated for different momentum
  regions correctly reproduces the graph, up to power corrections in
  $\Lambda/Q$.  Since we rely on this method in our arguments, we briefly
  sketch its essentials here.  Let $\Gamma$ be a particular Feynman graph
  (integrated over all loop momenta) and $R$ be a particular momentum
  region of that graph, defined by specifying for each line in the graph
  whether its momentum is hard, collinear in a certain direction, or soft.
  We call a region $R'$ smaller than $R$ (denoted as $R'<R$) if hard
  momenta in $R$ are collinear or soft in $R'$, or if collinear momenta in
  $R$ are soft in $R'$.  Examples for the latter case are given in
  figure~\ref{fig:subt-terms}.

  By $T_R\ms\Gamma$ we denote the application of the approximations
  appropriate for the region $R$, as specified in point
  \ref{enum:approximations}.  By construction we have
  $T_R\ms\Gamma_{\text{restr}} \approx \Gamma_{\text{restr}}$ up to power
  corrections, where ``restr'' denotes the restriction of loop integrals
  to the design region $R$ of $T_R$.  However, we need a representation
  without this restriction.  The full loop integral $T_R\ms\Gamma$ can
  contain leading contributions from momentum regions smaller or larger
  than $R$.  Moreover, the application of $T_R$ does in general not give a
  correct approximation in those regions (it was not designed to do so).
  The combined problem of double counting and degraded quality of the
  approximations is solved by the subtraction formalism.

  To cope with momentum regions smaller than $R$, an approximant
  $C_R\ms \Gamma$ for $R$ is defined by subtracting the contributions from
  smaller regions as
  \begin{align}
    \label{subtr-def}
    C_R\ms\Gamma &= T_R\ms\Gamma - \sum_{R'<R} T_R\ms C_{R'}\ms \Gamma \,.
  \end{align}
  Note that the approximant $T_R$ is applied also to the subtraction terms
  (e.g.\ if a line is hard in $R$, one will neglect its mass in $T_R\ms
  C_{R'}\ms \Gamma$ even if it is a collinear or soft line in $R'$.)  In
  the sum one may or may not include regions $R'$ that only provide power
  suppressed contributions.  The recursive definition \eqref{subtr-def} is
  possible since for each momentum the soft region is smallest.  For a
  leading region $R_0$ with no smaller leading region one simply has
  $C_{R_0}\ms\Gamma = T_{R_0}\ms\Gamma$ without subtractions.  Up to power
  corrections one then finds
  \begin{align}
    \label{subtr-sum}
    \Gamma \approx \sum_{R} C_{R}\ms\Gamma
  \end{align}
  with the sum running over all leading regions $R$.  Notice that a term
  $C_R\ms\Gamma$ has no leading contributions from smaller regions because
  of the subtractions in \eqref{subtr-def}, but it may still have leading
  contributions from regions $R'' > R$ (e.g.\ when a collinear momentum in
  $R$ is hard in $R''$).  This unwanted contribution is removed by the
  subtraction term $- T_{R''} C_R\ms\Gamma$ in the approximant
  $C_{R''} \Gamma$ for the larger region.  As shown in
  \cite{Collins:2011zzd}, the subtractions also correctly treat the case
  where two regions intersect each other (collinear momenta in different
  directions have the soft region as a common smaller region and thus
  intersect).

\item \textbf{Back to real momenta.}\label{enum:deform-back} After
  Grammer-Yennie approximations have been made, one can deform the soft
  momenta back to the real axis.  This requires a suitable choice for the
  auxiliary vectors and the $i\varepsilon$ prescription in the
  Grammer-Yennie approximants for soft gluons.  Because of the soft
  subtraction terms discussed in point~\ref{enum:soft-coll-subtr}, the
  same holds for the Grammer-Yennie approximants for collinear gluons.
  The $i\varepsilon$ prescriptions in \eqref{GY-collin} and
  \eqref{GY-soft}, together with the choice of auxiliary vectors specified
  in section~\ref{sec:GY-rearrange} are such that the poles in the
  Grammer-Yennie approximants do not obstruct the contour deformation for
  soft momenta.  If one uses a different prescription, one must show that
  the contributions from poles crossed during the contour deformation are
  power suppressed or cancel in the final factorisation formula.

  Deforming contours back at this stage enables one to initially choose
  momentum routings for individual graphs as one finds suitable for
  establishing the cancellation of Glauber gluon contributions (cf.\ our
  comment in point~\ref{enum:kin}).  In point~\ref{enum:ward}, the
  conditions on loop momentum routings are more stringent since we
  consider sums over different graphs, whose momentum routings must match.
  Having deformed soft gluon momenta back to the real axis, one can
  readily make the required changes of integration variables in the
  graphs.

\item \textbf{Ward identities and Wilson lines.}\label{enum:ward} At this
  point Ward identities are applied to the contractions
  $\hat{\ell}^\nu H(\hat{\ell})_\nu$ obtained by the Grammer-Yennie
  approximation.  This removes collinear gluons entering $H_1$ and $H_2$
  (except for transversely polarised ones) and introduces Wilson lines
  along $v_A$ in $A$ and along $v_B$ in $B$, with one Wilson line
  attaching to each of the four physical partons.
  Likewise, Ward identities are applied to $\tilde{\ell}^{\nu}\bs
  A_{\nu}(\tilde{\ell})$ and $\tilde{\ell}^{\nu}\bs
  B_{\nu}(\tilde{\ell})$.  If necessary, small light-cone components of
  loop momenta are shifted such that the corresponding arguments of
  collinear and soft factors are separate integration variables.  This
  removes all soft gluons entering $A$ or $B$ and provides Wilson lines to
  $S$.  The soft factor is then given as an expectation value of eight
  Wilson lines, one for each physical parton in $A$ and $B$.  With the
  $i\varepsilon$ prescriptions and auxiliary vectors mentioned above, the
  Wilson lines in collinear and soft factors are past pointing, where
  ``past'' refers to the space-time variable over which the gluon
  potential is integrated.

  The result of this procedure is shown in
  figure~\ref{fig:strategy:factorizedform}.  The collinear factors $A$ and
  $B$ are a first version of double parton distributions, to be modified
  by subtractions as discussed below.  They are given by operator matrix
  elements, which for two-quark distributions schematically read
  \begin{align}
    \label{dpd-matr-el}
    \langle\ms p |\,
     (\bar{q}_2^{} W^\dagger)_{k'}^{} \ms\Gamma_2^{}\ms
           (W\bs q_2^{})_{k}^{}\,
     (\bar{q}_1^{} W^\dagger)_{j'}^{} \ms\Gamma_1^{}\ms
           (W\bs q_1^{})_{j}^{}
    \,| p \ms\rangle \,,
  \end{align}
  where $k'$, $k$, $j'$ and $j$ are colour indices in the fundamental
  representation and $W$ is a Wilson line in the appropriate direction.
  $\Gamma_{1,2}$ are Dirac matrices, and the indices $1,2$ on the quark
  fields indicate that (after a Fourier transform) they create or
  annihilate a parton with momentum fraction $x_{1,2}$ in the proton.
  More detail is given in section 2.2 of \cite{Diehl:2011yj}.

  \begin{figure}
    \begin{center}
      \includegraphics[width=0.5\textwidth]{%
        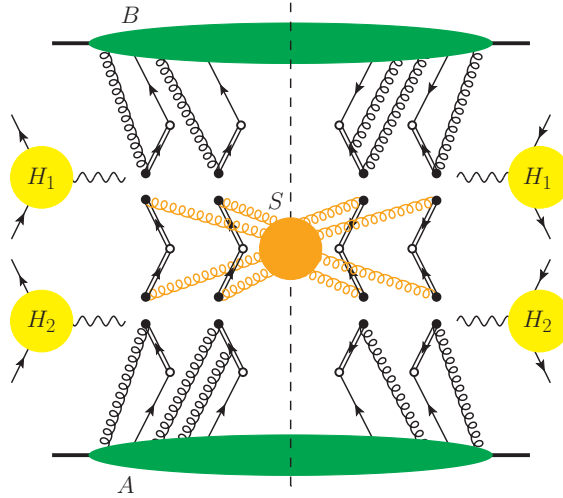}
      \caption{ \label{fig:strategy:factorizedform} Factorised form of the
        double Drell-Yan process in the TMD formalism.  Double lines
        denote Wilson lines.  The colour indices of adjacent black blobs
        of the Wilson lines in the soft and collinear factors are equal
        and summed over.  A corresponding form is obtained in collinear
        factorisation, with hard subgraphs crossing the final-state cut as
        in figure~\protect\ref{fig:dDY_inclusive_modified}.}
    \end{center}
  \end{figure}

  First-order examples for the application of Ward identities and the
  emergence of Wilson lines have been given in sections 3.2 and 3.3 of
  \cite{Diehl:2011yj}; an all-order derivation of the required identities
  in QCD remains to be worked out.  For the single Drell-Yan process an
  all-order proof of the analogous procedure is given in
  \cite{Collins:1988ig}, and for different single-parton scattering
  processes in an Abelian theory in \cite{Collins:2011zzd}.

  Each connected part of the original soft factor couples to both $A$ and
  $B$, so that after applying Ward identities each connected part couples
  to Wilson lines along both $v_R$ and $v_L$.  A corresponding restriction
  applies to the Wilson lines in the collinear factors, since they are
  generated by a Grammer-Yennie approximation for gluons connecting a
  collinear and hard subgraph rather than a collinear subgraph with
  itself.  So-called Wilson line self-interactions must therefore be
  explicitly excluded from the matrix elements that define soft and
  collinear factors.  In the overall factorisation formula, these
  self-interactions cancel.  For further discussion we refer to
  sections~13.3.4 and 13.7.2 of \cite{Collins:2011zzd}.

  Wilson lines are defined in position space, and the structure of the
  factorisation formula is simplified by a Fourier transform from
  transverse momenta in collinear and soft factors to coordinates in
  transverse configuration space (often referred to as $b$ space).  This
  replaces convolution integrals in transverse momenta by ordinary
  products.

\item \textbf{Colour structure.}  The collinear factors have a nontrivial
  structure in the colour of the four physical parton lines (or more
  precisely, of the Wilson lines attached to them, see
  \eqref{dpd-matr-el}).  This is different from single hard scattering,
  where the two parton lines attached to a collinear factor can only
  couple to an overall colour singlet.  Colour decompositions for double
  scattering are given in section 2.3 of \cite{Diehl:2011yj}, see
  \cite{Manohar:2012jr,Mekhfi:1985dv} for alternative choices and
  \cite{Kasemets:2014yna} for a simplification in the gluon sector.  If
  the colours of the two partons with the same momentum fraction ($x_1$ or
  $x_2)$ are coupled to singlets, we speak of ``colour singlet'' double
  parton distributions, which are an immediate generalisation of the
  familiar distributions for a single parton.  In \eqref{dpd-matr-el} one
  then contracts $j'$ with $j$ and $k'$ with $k$.  Other colour
  combinations, referred to as ``nonsinglet'' channels, can be regarded as
  describing colour interference or colour correlations.  In the basis of
  \cite{Diehl:2011yj}, the second colour structure for two-quark
  distributions is called ``colour octet'' combination and obtained by
  contracting \eqref{dpd-matr-el} with $t^a_{j'j}\, t^a_{k'k}$.

  In TMD factorisation, the hard subgraphs for double Drell-Yan production
  have a trivial colour structure since their final state is colourless.
  As a consequence, Wilson lines in the soft factor have their colour
  indices contracted pairwise, with one Wilson line along $v_R$ and the
  other along $v_L$ in each pair.  Regarding colour indices, the cross
  section thus involves a matrix multiplication of the form
  \begin{align}
    \label{Xsect-color-basic}
    B^T \cdot S \cdot A &= B^{c}\, S^{cd}\, A^d
  \end{align}
  where $c$ and $d$ label the different terms in the colour decompositions
  of $A$, $B$ and $S$ and are to be summed over.  (In the basis just
  described, $c$ and $d$ run over the singlet and octet channels.)

  In collinear factorisation the hard factors can produce unobserved jets,
  so that several colour channels are open.  Taking $q\bar{q}$
  annihilation as an example, one can then decompose
  \begin{align}
    H_{jk,j'k'} = \frac{1}{N_c}
          \biggl[ \ms {}^{1}\!H\, \delta_{jk}\ms\delta_{k'j'}
    + \frac{1}{C_F}\, {}^{8}\!H\, t^a_{jk}\ms t^a_{k'j'} \ms \biggr]
  \end{align}
  where $j,k$ are the colour indices of the incoming partons in the
  amplitude, and $j',k'$ those in the conjugate amplitude.  The
  tree-level graph for $H$ contributes only to ${}^{1}\!H$, whereas
  one-gluon emission contributes only to ${}^{8}\!H$.  The colour indices
  of $H$ are to be contracted with the colour indices of Wilson lines in
  $S$.  After a corresponding colour decomposition of $S$, the colour
  structure of the cross section involves an additional level of matrix
  multiplication:
  \begin{align}
    \label{Xsect-color-collin}
    B^{c}\, S^{cd,f\bs g}\, A^d\; H_{1}^{f}\, H_{2}^{g} \,,
  \end{align}
  where $f,g$ label the two colour channels of $H_{1,2}$.  This additional
  complication was not discussed in \cite{Diehl:2011yj,Manohar:2012jr},
  where only the tree-level expression of $H$ was considered.

\item \textbf{Soft subtraction in collinear
    factors.}\label{enum:soft-coll-subtr} As a consequence of the
  procedure described in point \ref{enum:subtractions}, the collinear
  factors require subtractions for regions where gluon lines have momenta
  that are not collinear but soft.  As this has not been elaborated on in
  \cite{Diehl:2011yj} we briefly explain it here.

  According to point \ref{enum:leading-regions}, soft gluons inside an
  unsubtracted collinear factor couple to other soft gluons, to a
  collinear line, or to the Wilson lines.  Taking the right-moving factor
  for definiteness, $A_{\text{unsub}}(v_A)$ thus has a soft subgraph $S$
  that connects collinear lines with the Wilson lines.  Collinear gluons
  couple to the Wilson lines as well, and because we have a non-abelian
  theory, the order of gluon attachments is relevant.  Let us show that a
  leading contribution is only obtained if, starting at the physical
  parton line, first all collinear gluons couple to a Wilson line and then
  all soft gluons, as shown in figure~\ref{fig:soft-subtraction}a.
  According to our earlier discussion, $v_A$ may have central or large
  negative rapidity.  Since $|v_A\ms \ell_{c}| \gg |v_A\ms \ell_{s}|$ if
  $\ell_c$ is collinear and $\ell_s$ is soft, this ordering has the
  smallest number of eikonal propagators carrying a collinear rather than
  a soft momentum.  Example graphs are shown in
  figure~\ref{fig:Wilson-line-order}.

  \begin{figure}
    \begin{center}
      \subfigure[]{\includegraphics[width=0.48\textwidth]{%
          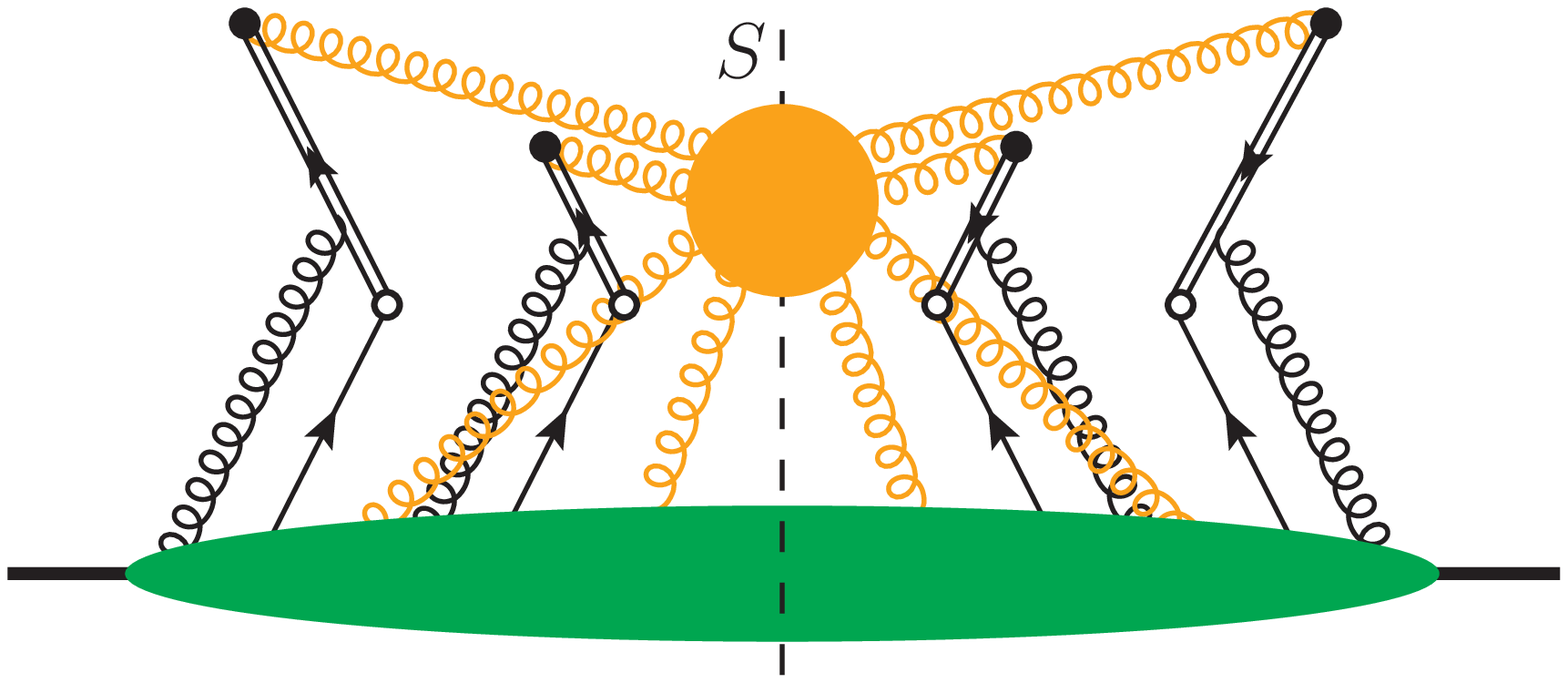}}
      \phantom{x}
      \subfigure[]{\includegraphics[width=0.48\textwidth]{%
          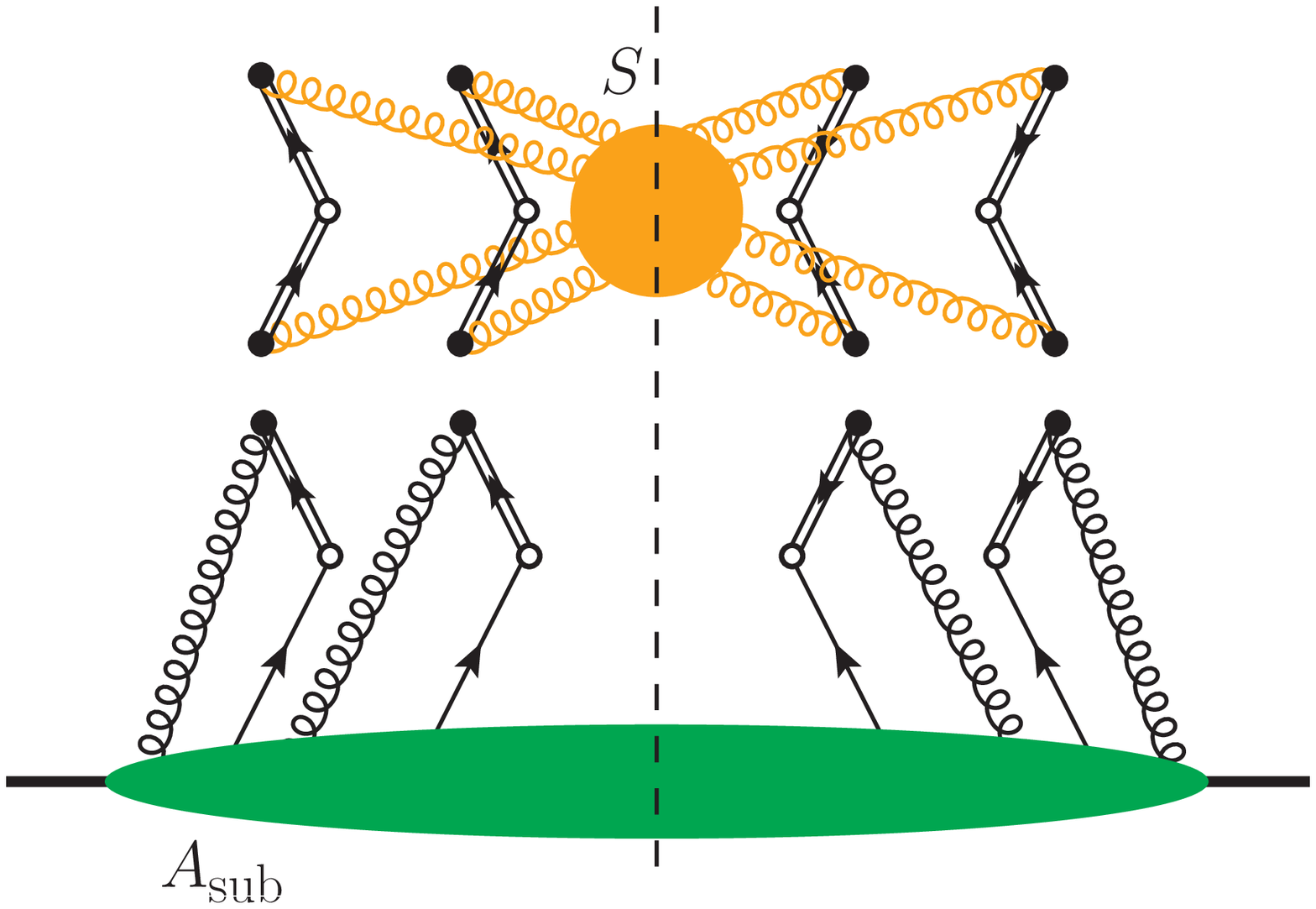}}
      \caption{ \label{fig:soft-subtraction} (a) The unsubtracted
        collinear factor $A_{\text{unsub}}$ with a soft subgraph that
        gives a leading contribution.  (b) Result of the applying
        Grammer-Yennie approximations and Ward identities to graph~a.}
    \end{center}
  \end{figure}

  \begin{figure}
    \begin{center}
      \subfigure[]{\includegraphics[width=0.25\textwidth]{%
          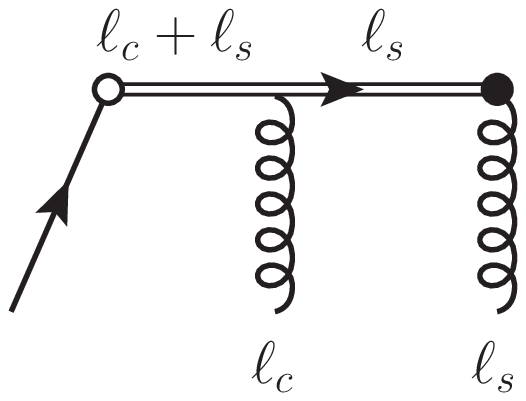}}
      \phantom{xxxx}
      \subfigure[]{\includegraphics[width=0.25\textwidth]{%
          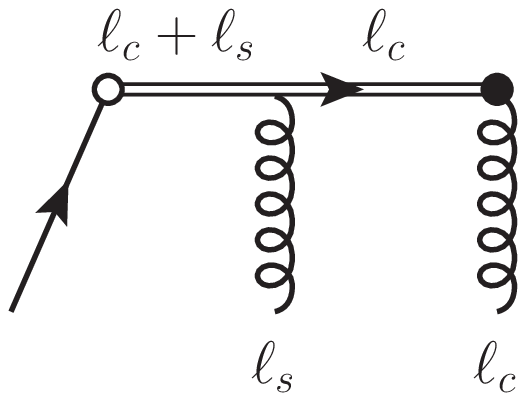}}
      \caption{ \label{fig:Wilson-line-order} The possibilities for one
        soft and one collinear gluon to couple to a Wilson line.  Graph a
        is leading and graph b is subleading.}
    \end{center}
  \end{figure}

  One can now apply the same approximations made for soft gluons that
  couple to the collinear subgraph in the overall process (see
  point~\ref{enum:GY}), provided that one can again show the cancellation
  of Glauber gluon exchange.  The result has the form
  \begin{align}
    \label{subtracted-A}
    A_{\text{unsub}}(v_A) &= S(v_A,v_R) \cdot A_{\text{sub}}  \,,
  \end{align}
  where by construction the subtracted collinear factor $A_{\text{sub}}$
  does not receive a leading contribution from soft momenta.  (It may
  receive leading contributions from hard loop momenta, but as discussed
  in point \ref{enum:subtractions} this will be removed by subtractions in
  the hard factors.)
  From \eqref{subtracted-A} one readily obtains $A_{\text{sub}} =
  S^{-1}(v_A,v_R) \cdot A_{\text{unsub}}^{}(v_A)$.  Note that one can take
  $v_A$ to be light-like in this expression, since potential rapidity
  divergences are removed by the soft subtraction term.  In the cross
  section for TMD factorisation one has a product
  \begin{align}
    \label{Xsect-colour-prod}
    & B_{\text{sub}}^T \cdot S(v_{L}, v_{R}) \cdot A_{\text{sub}}^{}
    \nonumber \\
    &\quad = B^T_{\text{unsub}}(v_B) \cdot S^{-1}(v_{L},v_{B}) \cdot
    S(v_{L},v_{R}) \cdot S^{-1}(v_A,v_R) \cdot A_{\text{unsub}}^{}(v_A)
  \end{align}
  of matrices in colour space.  A corresponding expression is obtained for
  collinear factorisation from \eqref{Xsect-color-collin}.

  For TMD factorisation in single hard scattering, different schemes have
  been proposed in the literature regarding the choice of Wilson lines.  A
  vector $v_A = v_B$ with central rapidity was chosen in
  \cite{Collins:2007ph}, whereas $v_A = v_L$ and $v_B = v_R$ was taken in
  \cite{Ji:2004wu,Ji:2004xq}.  In section 13.7 of \cite{Collins:2011zzd} a
  more complicated scheme involving additional soft factors is presented,
  in which the individual terms in the final factorisation formula do not
  require the explicit removal of Wilson line self-interactions.  $v_A$
  and $v_B$ are taken as light-like in this scheme.  Its analogue for
  double hard scattering has not been studied so far.

\item \textbf{Rapidity evolution.} In general one cannot take all Wilson
  lines lightlike, as mentioned earlier.  The collinear and soft factors
  then depend on the rapidities of the Wilson lines.  This dependence is
  described by differential equations, commonly called Collins-Soper
  equations.  The Collins-Soper equation for the unsubtracted collinear
  factors has been discussed in section 3.4 of \cite{Diehl:2011yj}.  A
  derivation of the rapidity evolution for the soft factor in the same
  framework is still missing.  For the particular rapidity regulator
  introduced within SCET (soft-collinear effective theory) in
  \cite{Chiu:2012ir}, the rapidity evolution of collinear and soft factors
  has been discussed in \cite{Manohar:2012jr}.

  In collinear factorisation, soft and collinear factors are integrated
  over transverse parton momenta (or equivalently, evaluated at
  $\tvec{b}=\tvec{0}$ in transverse configuration space).  In the colour
  singlet channel, one can take all Wilson lines as light-like since
  potential rapidity divergences cancel.  The soft factor then reduces to
  unity, since it involves Wilson line combinations
  $W^\dagger(v,\tvec{b}=\tvec{0}) W(v,\tvec{b}=\tvec{0}) = 1$.  In colour
  nonsinglet channels, one must however keep non-lightlike vectors (or use
  a different rapidity regulator) and retains nontrivial soft factors.

  Factors that depend on a large rapidity difference involve large
  logarithms, called Sudakov logarithms (recall that rapidity is a
  logarithm, see footnote~\ref{rap-def}).  The power of these logarithms
  increases with the order of perturbation theory.  The Collins-Soper
  equations sum these logarithms to all orders and can be used to ensure
  that the final factorisation formula contains no large logarithms in the
  hard factors, which can then be evaluated reliably in fixed-order
  perturbation theory.  The resummed Sudakov logarithms typically lead to
  a suppression of the cross section.  For collinear factorisation this
  means in particular that all channels except for the colour singlet
  channel in the double parton distributions are suppressed, as was
  realised long ago \cite{Mekhfi:1988kj}.  A simple but instructive
  numerical study of the suppression of colour nonsinglet channels is
  given in \cite{Manohar:2012jr}.
    
\item \textbf{UV subtractions.}\label{enum:UV-subtr} Finally, UV
  subtractions are made in the individual factors as necessary, after
  which the regulator of UV divergences can be removed (i.e.\ by taking
  $d$ to $4$ in dimensional regularisation).  It is important to do this
  at the very end, since the rearrangement of soft factors and change of
  auxiliary vectors discussed in point~\ref{enum:soft-coll-subtr} changes
  the nature of the UV singularities as explained in sections 10.8.2 and
  10.11.1 of \cite{Collins:2011zzd}.  Specifically, the limits of taking
  Wilson line rapidities to infinity and $d$ to $4$ do not commute.

  The UV subtractions do not change the overall cross section, where the
  UV divergences in the individual factors cancel, as ensured by the
  subtraction procedure of point \ref{enum:subtractions}.  As mentioned
  earlier, the only UV divergences in the original graphs are those
  removed by the counterterms in the Lagrangian.

  The UV subtracted factors depend on a renormalisation scale $\mu$.  For
  TMD functions, the dependence on that scale is given by simple
  renormalisation group equations with anomalous dimensions multiplying
  the factors themselves.  For collinear parton distributions, one obtains
  DGLAP type equations containing a convolution product of the
  distributions with splitting functions, where the convolution is in
  longitudinal momentum fractions.  For double parton distributions, these
  equations are discussed in
  \cite{Kirschner:1979im,Shelest:1982dg,Snigirev:2003cq,%
    Gaunt:2009re,Ceccopieri:2014ufa}.  Their form is connected with the
  double counting problem noted in point~\ref{enum:SPSvsDPS}, as explained
  in section 5.3.2 of \cite{Diehl:2011yj}.  As usual, the evolution
  equations can be used to ensure that the hard factors do not contain
  large logarithms that would spoil a fixed-order perturbative expansion.

\end{enumerate}


\subsection{Scaling of soft momenta}
\label{sec:soft-scaling}

The typical size of collinear momenta,
$(\ell^+,\ell^-,|\tvec{\ell}|) \sim (Q,\ms \Lambda^2/Q,\ms \Lambda)$ or
$(\Lambda^2/Q,\ms Q,\ms \Lambda)$, is directly connected to the external
kinematics of the process.  Their large components must be of order $Q$ to
enable the hard scattering.  Their small and transverse components
correspond to generic virtualities of order $\Lambda^2$ in collinear
factorisation, where there is no small external kinematic quantity and
$\Lambda$ is a generic nonperturbative scale.  If in TMD factorisation the
measured transverse momenta $\tvec{q}$ are larger than a nonperturbative
scale, then one must take $\Lambda \sim |\tvec{q}|$ as discussed earlier.
The transverse momenta entering the hard scattering must then be of size
$\Lambda$ in order to produce the required $\tvec{q}$ in the final state.

The size of soft momenta is not directly set by external kinematics.  The
Libby-Sterman analysis only specifies that all soft momentum components
vanish in the formal limit $\Lambda\to 0$.  A crucial part of the analysis
is that soft gluons couple to collinear lines.

\begin{figure}
  \begin{center}
      \subfigure[]{\includegraphics[width=0.2\textwidth]{%
          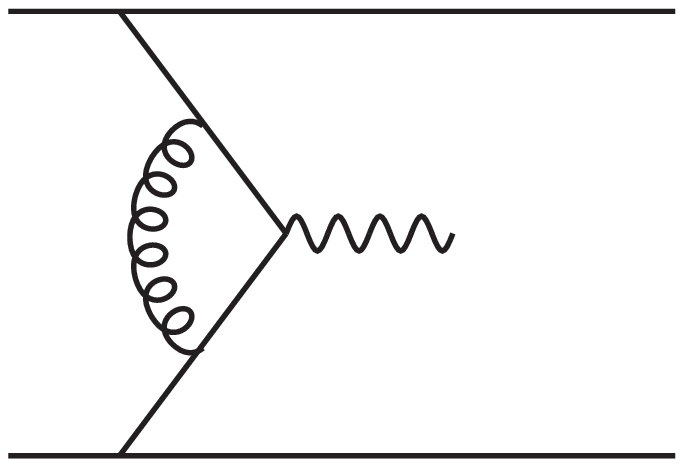}}
      \phantom{xxxx}
      \subfigure[]{\includegraphics[width=0.2\textwidth]{%
          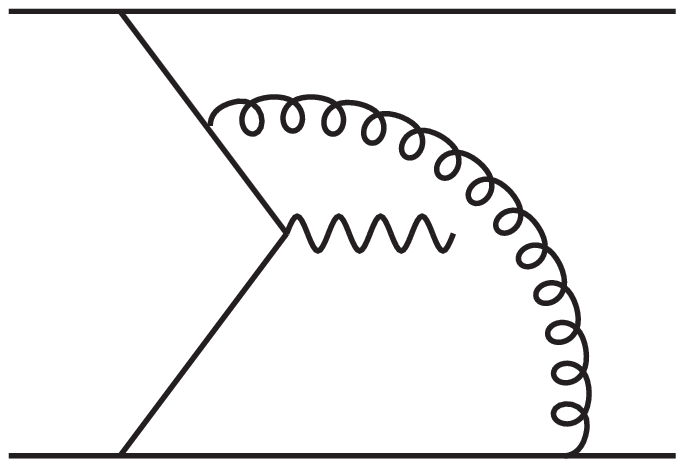}}
    \caption{\label{fig:soft-g-exchange} Examples for soft gluon exchange
      in a simple spectator model.  The graphs are understood to be
      embedded in larger graphs for the cross section of single or double
      Drell-Yan production.}
  \end{center}
\end{figure}

Let us take a closer look at soft momentum scalings that give a leading
contribution to the cross section.  For definiteness we consider the
exchange of a single soft gluon between the collinear subgraphs $A$ and
$B$, with example graphs given in figure~\ref{fig:soft-g-exchange}.  Our
discussion applies likewise to single and double Drell-Yan production.
Leading regions can be conveniently identified by comparing the
expressions for the graph with and without soft gluon exchange:
\begin{align}
  \label{soft-power-counting}
\int \frac{\df^4\ell}{\ell^2}\, A^{(1) ++} \ms B^{(1) --}
~~\text{vs.}~~
A^{(0) +} \ms B^{(0) -} \,.
\end{align}
Here the Lorentz index of the lowest-order expressions $A^{(0)}$ and
$B^{(0)}$ originates from the Fierz decomposition mentioned in
point~\ref{enum:Fierz}, and the additional index in $A^{(1)}$ and
$B^{(1)}$ comes from the exchanged gluon.  The size of each collinear
factor can be determined by boosting to a frame where its momenta scale
like $(\Lambda, \Lambda, \Lambda$).  In such a frame, all components of
$A^{(0)}$ and $B^{(0)}$ scale like $\Lambda^{n}$ (where $n$ is the
appropriate mass dimension), whilst all components of $A^{(1)}$ and
$B^{(1)}$ scale like $\Lambda^{n-1}$.  Boosting back to the overall c.m.\
frame, we find that the largest components of each factor scale like
\begin{align}
  \label{coll-canonical-scaling}
A^{(0) +} ,\, B^{(0) -} &\sim Q \Lambda^{n-1} \,,
&
A^{(1) ++} ,\, B^{(1) --} &\sim Q^2 \Lambda^{n-3} \,.
\end{align}
In \eqref{soft-power-counting} we have already used that other components
of $A$ and $B$ give smaller contributions and neglected these.  A momentum
region in the one-gluon exchange graph thus contributes to the cross
section with the same power as the leading-order graphs if
$\df^4\ell /\ell^2 \sim \Lambda^4 /Q^2$, where $\df^4\ell$ is to be
understood as the size of the integration volume for that region.

The preceding analysis is only valid if $\ell^+$ and $\ell^-$ are at most
of order $\Lambda^2 /Q$, so that they do not change the scaling of momenta
in the collinear factors.  If $\ell^- \sim \Lambda$ then at least one line
in $A^{(1)}$ has momentum scaling like $(Q,\ms \Lambda,\ms \Lambda)$ and
hence a virtuality $Q \Lambda$ instead of $\Lambda^2$.  A corresponding
statement holds for $B^{(1)}$ if $\ell^+ \sim \Lambda$.  The
approximations for collinear momenta discussed in
point~\ref{enum:approximations} of section~\ref{sec:overall-strategy}
remain valid in this case.

A number of different momentum scalings for $\ell$ give a leading
contribution to the cross section.
\begin{enumerate}
\item In the \emph{Glauber} region defined in \eqref{Glauber-scaling} we
  have $\ell \sim (\Lambda^2/Q,\ms \Lambda^2/Q,\ms \Lambda)$, so that
  $\df^4\ell \sim \Lambda^6 /Q^2$ and $\ell^2 \sim \Lambda^2$.  Since
  the collinear factors scale like \eqref{coll-canonical-scaling} in this
  case, we obtain a leading contribution.
\item If all components of $\ell$ are of size $\Lambda^2/Q$ then
  $\df^4\ell \sim \Lambda^8 /Q^4$ and $\ell^2 \sim \Lambda^4 /Q^2$, so
  that we again obtain a leading contribution.  Momenta in this region are
  often called \emph{ultrasoft} in the context of effective theories.

  Since in this region the typical gluon virtuality goes to zero for
  $Q \to \infty$, one may wonder whether nonperturbative effects such as
  confinement invalidate this analysis.  Taking a nonzero gluon mass $m_g$
  to simulate such effects, one finds that ultrasoft scaling only gives a
  leading contribution if $\Lambda^2 /Q \gsim m_g$.
\item If all components of $\ell$ are of size $\Lambda$, which in
  effective theories is often denoted as the
  \emph{soft}\footnote{%
    Throughout this paper we will use the term ``soft'' to denote the full
    region of momenta with components of size $\Lambda$ or smaller.}
  region, we have $\df^4\ell \sim \Lambda^4$ and $\ell^2 \sim \Lambda^2$.
  The scaling in \eqref{coll-canonical-scaling} must be amended for the
  presence of collinear lines with virtuality $Q \Lambda$.  One finds that
  for the graph in figure~\ref{fig:soft-g-exchange}a there is an overall
  penalty factor $(\Lambda/Q)^2$ for one off-shell line in each subgraph
  $A^{(1)}$ and $B^{(1)}$, so that one obtains a leading contribution.  In
  figure~\ref{fig:soft-g-exchange}b the subgraph $A^{(1)}$ has two quark
  lines with virtuality $Q \Lambda$, which results in an overall penalty
  factor $(\Lambda/Q)^3$ and thus in a power suppressed contribution to
  the cross section.
\item The regions where $\ell$ scales like $(\Lambda,\ms \Lambda^2/Q,\ms
  \Lambda)$ or $(\Lambda^2/Q,\ms \Lambda,\ms \Lambda)$ may be regarded as
  hybrids between regions 1 and 3.  With $\df^4\ell \sim \Lambda^5/Q$
  and $\ell^2 \sim \Lambda^2$ and a penalty factor of $\Lambda/Q$ for an
  off-shell line in only one of the two collinear factors, these regions
  contribute to leading power for the appropriate graphs.
\end{enumerate}
This list is not complete, and other scalings such as
$\ell \sim (\Lambda,\ms \Lambda^2/Q,\ms \Lambda \sqrt{\Lambda/Q})$ give a
leading-power contribution as well.  Fortunately, a complete inventory of
all leading momentum scalings is not needed for the factorisation proof.
In the subtraction procedure sketched in point~\ref{enum:subtractions} of
section~\ref{sec:overall-strategy}, a region $R$ is not primarily
characterised by its momentum scaling, but by the approximations $T_R$
appropriate for that region.  A controlled power-law accuracy of these
approximations is an essential ingredient in establishing the method.  For
a detailed discussion we refer to section~10.7 of \cite{Collins:2011zzd}
and to sections~VIII A to D of \cite{Collins:2007ph}.

For the momentum scalings enumerated above, the Grammer-Yennie
approximation \eqref{GY-soft} for a soft gluon coupling to $A$ and its
analogue for a soft gluon coupling to $B$ work in regions 2 and 3, but at
least one of the approximations fails in regions 1 and 4.  To establish
the cancellation of Glauber gluon exchange, we will deform $\ell$ out of
regions 1 and 4 into region 2.


\subsection{Approximations for soft gluons}
\label{sec:soft-approx}

As stated earlier, we approximate a soft gluon momentum $\ell$ entering
the collinear factor $A(\ell)$ by
$\tilde{\ell} = (0, \ell^-, \tvec{\ell})$.  The Grammer-Yennie
approximation
\begin{align}
  \label{GY-soft-again}
  S^{\mu}(\ell)\, A_{\mu}(\tilde{\ell})
  & \approx S^-(\ell)\, \frac{v_R^+}{\ell^-_{\phantom{R}}
    v_R^+ + i\varepsilon}\, \ell^- A^+(\tilde{\ell})
    \approx S_{\mu}(\ell)\,
      \frac{v_R^{\mu}}{\ell v_R^{\phantom{+}} + i\varepsilon}\;
        \tilde{\ell}^{\nu}\bs A_{\nu}(\tilde{\ell})
\end{align}
then fails if $\ell$ is in the Glauber region.

A different kinematic approximation is described in section~10.4.2 of
\cite{Collins:2011zzd}, where
$\tilde{\ell} = (0, \ell^- - \delta^2 \ell^+, \tvec{0})$, with $\delta$
being a parameter of order $\Lambda/Q$.  (The same approximation with
$\delta=0$ was taken in the original work \cite{Collins:1988ig}.)  The
second step of \eqref{GY-soft-again} is then valid even in the
Glauber region.  However, the approximation to neglect $\tvec{\ell}$
inside $A(\ell)$ is not.  The transverse momentum of the soft and
collinear lines is comparable not only in the Glauber region but also in
regions 3 and 4 described in the previous subsection.  In section 3.2 of
\cite{Collins:1988ig} an argument was given why one can nevertheless
neglect $\tvec{\ell}$ in $A$ for single Drell-Yan production, which
required a specific routing of loop momenta.  Let us see why this argument
does not readily generalise to the double Drell-Yan case.

\begin{figure}
  \begin{center}
    \subfigure[]{\includegraphics[width=1\textwidth]{%
        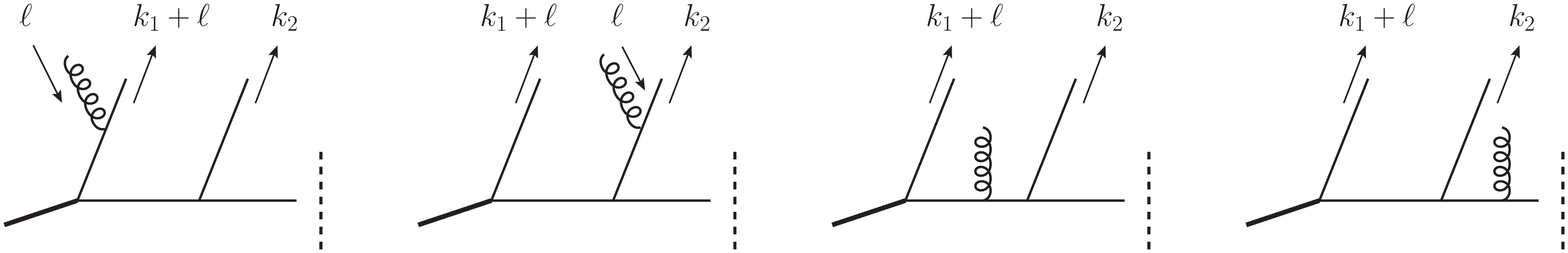}} \\[1em]
    \subfigure[]{\includegraphics[width=0.52\textwidth]{%
        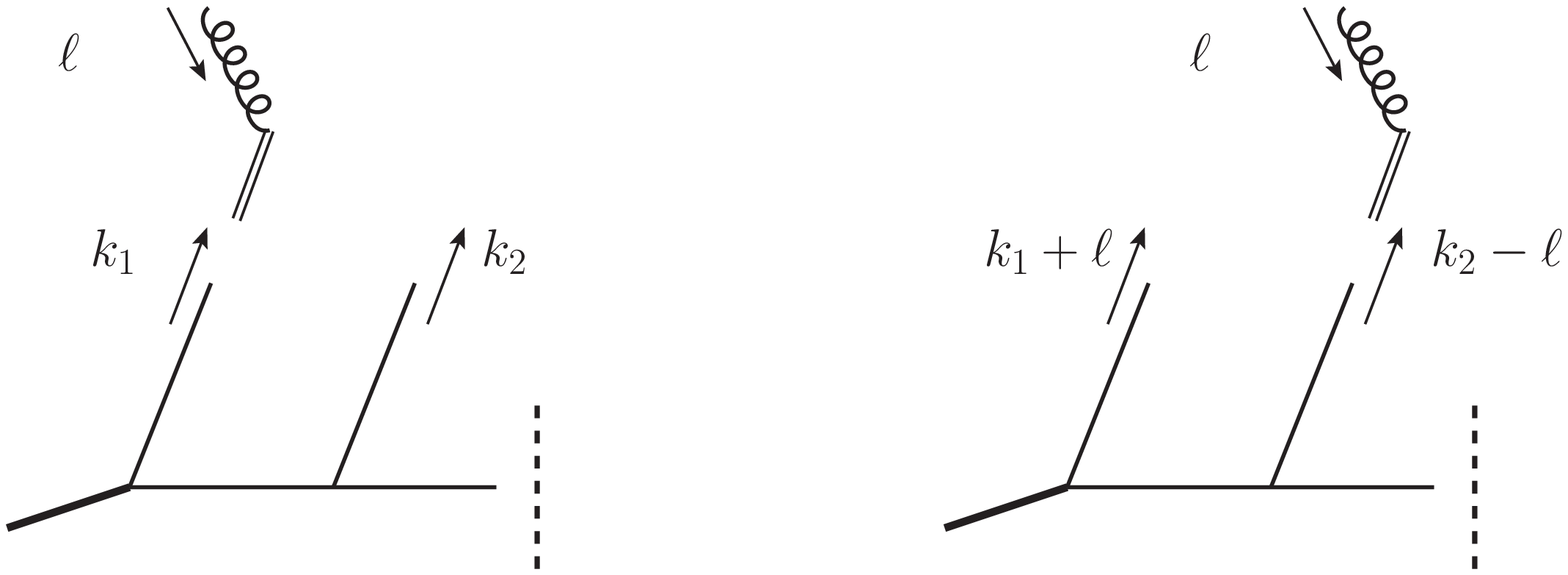}}
    \caption{\label{fig:Ward-id-example} Graphs in a simple spectator
      model for the part of the collinear factor that is left of the
      final-state cut (denoted by the dashed line).  The incoming line in
      each graph represents the hadron, which is colourless and thus does
      not directly couple to gluons. (a) Graphs with one soft gluon
      coupling to the collinear factor.  (b) Result of applying the
      Grammer-Yennie approximation and a Ward identity.}
  \end{center}
\end{figure}

In figure~\ref{fig:Ward-id-example}a we show the graphs for a soft gluon
coupling to $A$ on the left of the final-state cut in a simple spectator
model.  After contour deformation out of the Glauber region and use of the
Grammer-Yennie approximation, a Ward identity converts the sum of these
graphs into collinear factors without a soft gluon, as shown in
figure~\ref{fig:Ward-id-example}b.  As in section 3.2 of
\cite{Collins:1988ig} we have routed the gluon momentum such that it does
not cross the final-state cut.  In order to apply the Ward identity we
must use the same external momentum assignment in each graph, and for
definiteness we have chosen $\ell$ to flow out of the line with $k_1$
rather than the one with $k_2$.  The collinear factor in the first term of
figure~\ref{fig:Ward-id-example}b does not depend on $\ell$, so that we
obtain the correct result whether or not we set $\tvec{\ell}$ to zero in
the original graphs of figure~\ref{fig:Ward-id-example}a (the eikonal
propagators do not depend on $\tvec{\ell}$ and the gluon propagator is to
be truncated in all graphs).  In the case of single hard scattering, this
would be the only term to consider.

The collinear factor in the second term of
figure~\ref{fig:Ward-id-example}b depends however on $\tvec{\ell}$, which
can be of the same size as $\tvec{k}_1$ and $\tvec{k}_2$, as discussed in
the previous subsection.  Neglecting $\tvec{\ell}$ in this factor gives an
incorrect result.  Following the derivation in section 3.2.2 of
\cite{Diehl:2011yj} one finds indeed that in the overall expression of the
cross section, the dependence of this collinear factor on $\tvec{\ell}$ is
crucial to give the correct transverse position of the Wilson lines to
which the soft factor is eventually converted.  This holds true even for
collinear factorisation, where the Wilson lines associated with the
partons of momenta $k_1$ and $k_2$ have a relative transverse distance
$\tvec{y}$ from each other.  Neglecting $\tvec{\ell}$ in the graphs of
figure~\ref{fig:Ward-id-example}a would put all Wilson lines in the soft
factor at the same transverse position, which is incorrect.

\section{Glauber cancellation for one-gluon exchange}
\label{sec:one-gluon}


\subsection{Scope and general method}
\label{sec:scope-and-method}

In this section we demonstrate the cancellation of Glauber gluons in the
double Drell-Yan process at the level of single gluon exchange.  The
momentum $\ell$ of the exchanged gluon can lie in the hard, soft, or
collinear-to-$A$ or $B$ regions, and what we shall show in this section is
that the approximated expressions from these regions reproduce the exact
graph at leading-power accuracy.  There is no need for a separate term for
the Glauber region, where the soft Grammer-Yennie approximation
\eqref{GY-soft} breaks down.  The hard region can be distinguished from
the rest by the fact that it corresponds to generically large transverse
momenta $|\vect{\ell}| \sim Q$ rather than $|\vect{\ell}| \sim \Lambda$.
By considering integrals differential in transverse momenta, and
transverse momenta only of order $\Lambda$, we remove the need to consider
the hard region in the following.

\begin{figure}
\begin{center}
\subfigure[]{\includegraphics[width=0.3\textwidth]{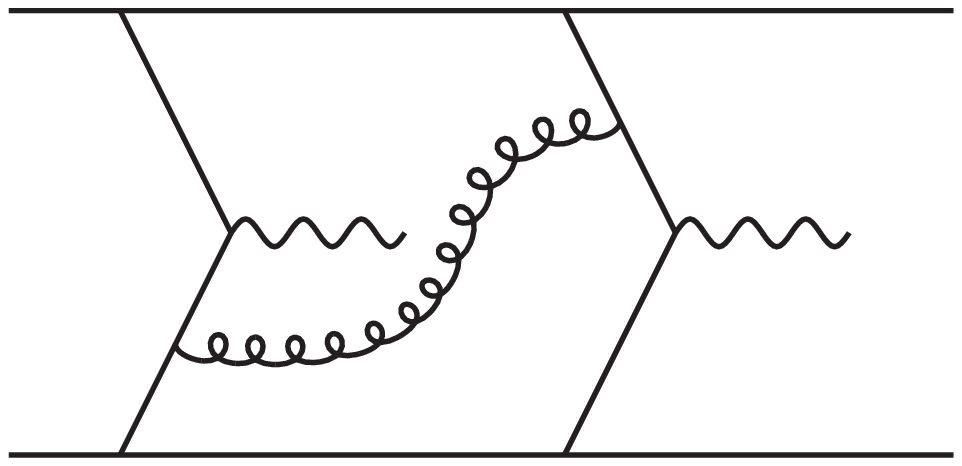}}
\phantom{x}
\subfigure[]{\includegraphics[width=0.3\textwidth]{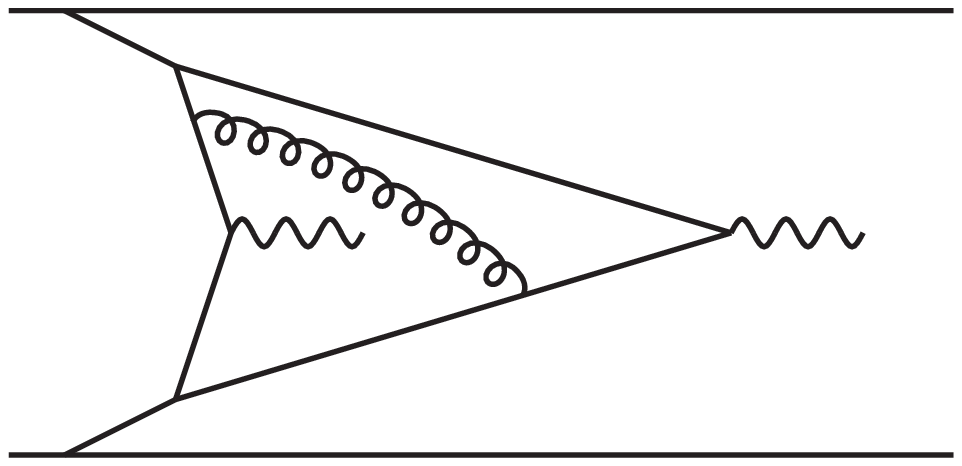}}
\phantom{x}
\subfigure[]{\includegraphics[width=0.3\textwidth]{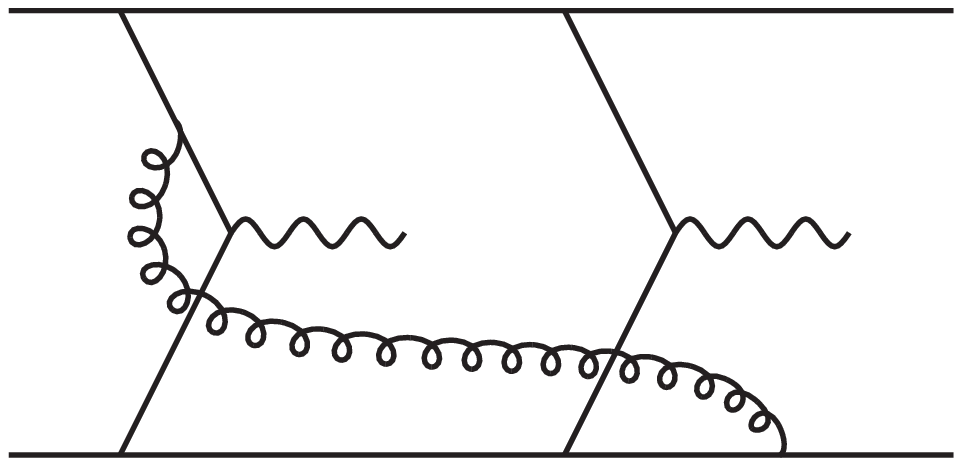}}
\\
\subfigure[]{\includegraphics[width=0.3\textwidth]{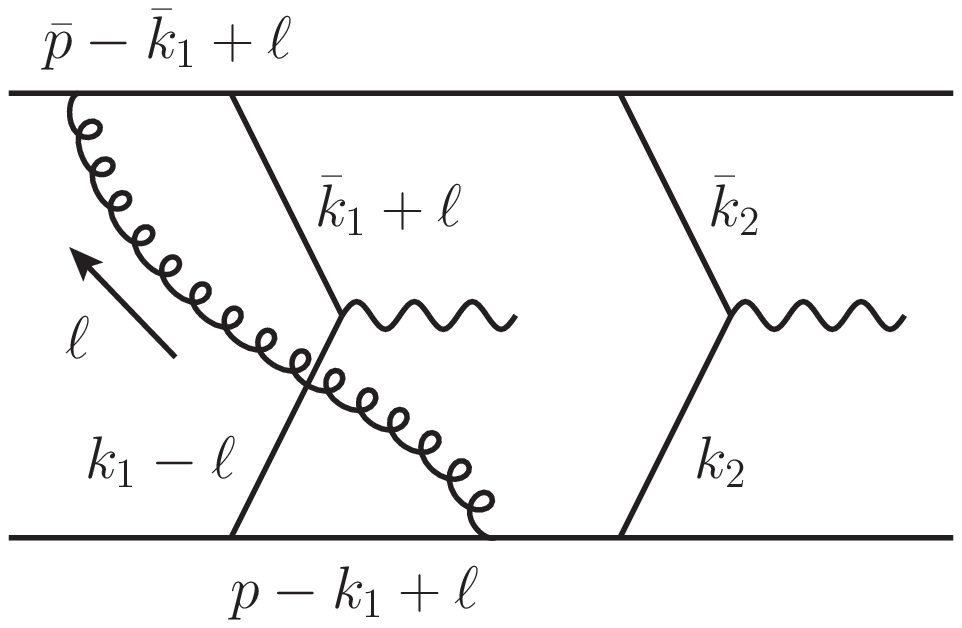}}
\phantom{x}
\subfigure[]{\includegraphics[width=0.3\textwidth]{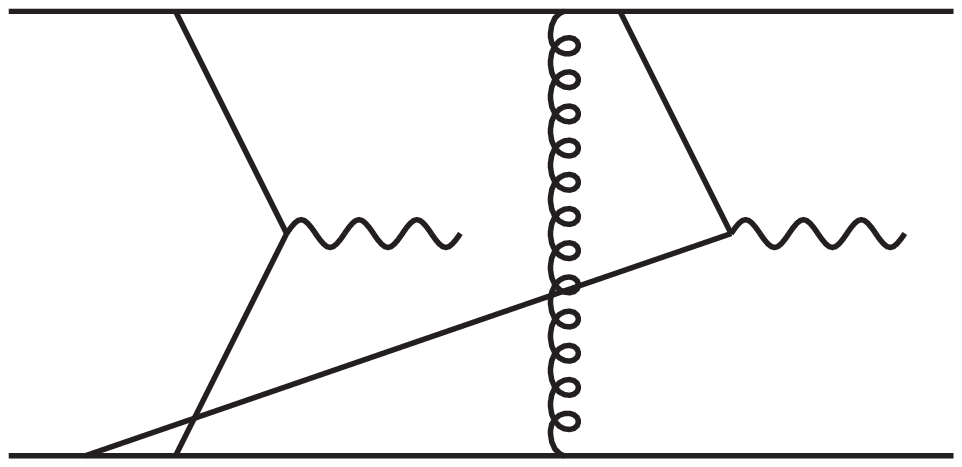}}
\phantom{x}
\subfigure[]{\includegraphics[width=0.3\textwidth]{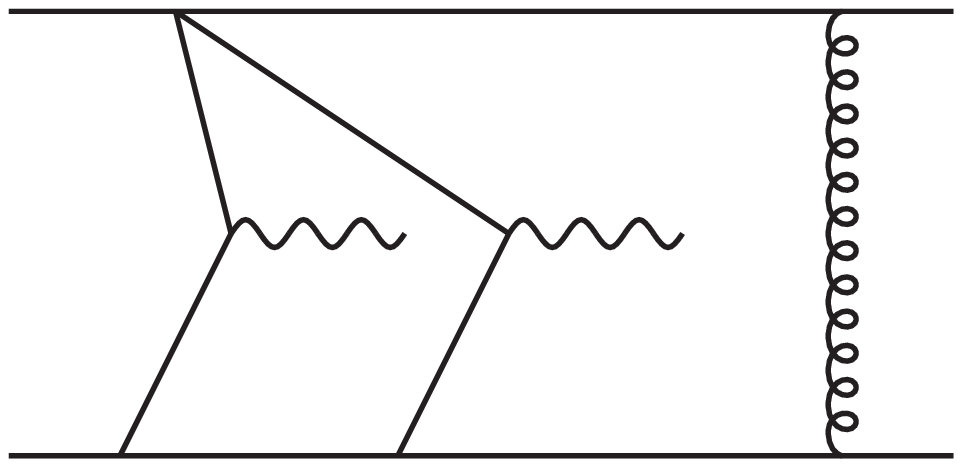}}
\caption{\label{fig:spect-model} Example graphs for the double Drell-Yan
  amplitude within the model described in the text.  In the physical
  process these graphs are embedded in graphs with further spectator lines
  and with a hadron entering each collinear subgraph.}
\end{center}
\end{figure}

For simplicity, we adopt a model with scalar ``partons'' described by a
field $\phi$ and coupling to each other via three- and four-point
vertices.  Gluons couple to these partons as required by gauge invariance,
i.e.\ via the usual scalar QED/QCD-type vertices (see for example
\cite{Srednicki:1019751}).  We thus have a momentum-dependent $g\phi\phi$
vertex and a $gg\phi\phi$ seagull vertex.  Some example graphs for the
double Drell-Yan process in this model are given in
figure~\ref{fig:spect-model}.

We shall not concern ourselves with the detailed colour structure of the
model.  As we shall see, the Glauber cancellation argument depends only on
the analyticity properties of the loop integrands in the gluon momentum,
which in turn is determined only by the denominators of the Feynman and
eikonal propagators.  It does \emph{not} rely on the spin and colour
structure of the model, and hence also applies to graphs in QCD.

The Glauber cancellation argument we shall present here works on a
graph-by-graph basis, with the sum over cuts of the graph taken where
necessary.  We need to consider only graphs in which the gluon extends
between right- and left-moving collinear partons -- graphs in which the
gluon attaches only to right- or left-moving partons are topologically
factorised already.
We classify the partons into active ones and spectators, with active
partons being the ones that attach to a vertex producing an electroweak
gauge boson.  All other partons are spectators.  We thus have
active-active, active-spectator, and spectator-spectator gluon
attachments.

What we shall do in this section corresponds to applying the region
approximants of point~\ref{enum:approximations} in
section~\ref{sec:overall-strategy} in the following order: first we make
contour deformations needed to avoid the Glauber region (with a sum over
cuts where necessary). Then we apply the Grammer-Yennie approximations,
before deforming the contours back to the real axis. Finally the kinematic
approximations and approximations for fermion lines are applied.  With
this order of steps, we do not need to concern ourselves with the
kinematical and fermion line approximations in this section.

To illustrate where problematic pinched poles in $\ell^+$ or $\ell^-$ may
generically arise, it is instructive to consider the example graph in
figure~\ref{fig:spect-model}d.  If $\ell$ is in the Glauber region then
$\ell^-$ is negligible in the right-moving lines above the gauge boson
vertices.  It is also negligible in gluon propagator, which is dominated
by $\tvec{\ell}^2$.  Poles in the small $\ell^-$ region thus result from
the propagator denominators
\begin{align}
[(p-k_1+\ell)^2 + i\epsilon]\, [(k_1-\ell)^2 + i\epsilon]
  &\approx [2 (p-k_1)^+ \ell^- + A + i\epsilon]\,
           [\ms- 2 k_1^+ \ell^- + B + i\epsilon] \,,
\end{align}
where $A = 2 (p-k_1)^+ (p-k_1)^- - (\tvec{p}-\tvec{k}_1+\tvec{\ell})^2$
and $B = 2 k_1^+ k_1^- - (\tvec{k}_1-\tvec{\ell})^2$ are both of order
$\Lambda^2$.  The $\ell^-$ contour is pinched near the origin, between a
pair of poles with real parts of order $\Lambda^2/Q$. We see that the
position of a pole above or below the real axis depends on whether
$\ell^-$ is routed along or against the large and positive plus-momenta
$(p-k_1)^+$ and $k_1^+$ of the collinear lines.

Conversely the poles of $\ell^+$ in the Glauber region are determined by
the following denominator factors:
\begin{align}
[(\bar{p}-\bar{k}_1+\ell)^2 + i\epsilon]\,
   [(\bar{k}_1+\ell)^2 + i\epsilon]
 & \approx [2 (\bar{p}-\bar{k}_1)^- \ell^+ + A + i\epsilon]\,
           [2 \bar{k}_1^- \ell^+ + B + i\epsilon]
\end{align}
with $A = 2 (\bar{p}-\bar{k}_1)^- (\bar{p}-\bar{k}_1)^+ -
(\bar{\tvec{p}}-\bar{\tvec{k}}_1+\tvec{\ell})^2$ and $B = 2 \bar{k}_1^-
\bar{k}_1^+ - (\bar{\tvec{k}}_1+\tvec{\ell})^2$.  In this case $\ell^+$
flows in the same direction as the large positive minus momenta
$(\bar{p}-\bar{k}_1)^-$ and $\bar{k}_1^-$, so that both poles are on same
side of real axis and $\ell^+$ is not pinched at small values.

From this example we see that the pattern of poles and pinches in the
small $\ell^+$, $\ell^-$ region is determined by the flow of large plus or
minus momentum.  The integration over $\ell^+$ is pinched if there is at
least one line where it flows against a collinear momentum with large
minus component, and at least one line where it flows with such a
momentum. An analogous rule holds for $\ell^-$.  This rule will be be
essential in our general proof of the cancellation of the Glauber region
at the one-gluon exchange level.

The remainder of our argument is structured as follows.  In section
\ref{sec:GY-rearrange} we describe the soft and collinear Grammer-Yennie
approximants and give details of a method that allows us to investigate
the validity of the factorisation formula at the one-gluon level.  In
section~\ref{sec:box-graphs} we use this method to show by explicit
calculation that the factorisation formula gives the correct leading-power
expression for a simplified set of double Drell-Yan graphs.  In section
\ref{sec:general-arg} we give an argument that -- after a sum over cuts
where necessary~-- there is no pinched Glauber exchange contribution for
any one-gluon exchange graph within our model.  Finally, in
section~\ref{sec:limitations} we discuss the difficulty to extend this
argument to gluon exchange at higher orders.  As a result, we will pursue
a different approach for our all-order proof of Glauber gluon cancellation
in section~\ref{sec:allorder}.


\subsection{Grammer-Yennie approximation and rearrangement of terms}
\label{sec:GY-rearrange}

For reasons to become obvious shortly, we take the auxiliary vectors for
the soft and collinear Grammer-Yennie approximants to be the same ($v_L =
v_A$, $v_R = v_B$) and call these common vectors $v_L$ and $v_R$ in the
following.  At a later stage of the factorisation proof, this equality can
be lifted using the appropriate Collins-Soper equations.

We group the parton lines in each graph into subgraphs $L$ and $R$, with
$L$ containing the lines above the vertices producing the gauge bosons,
and $R$ containing the lines below.  In the lowest-order graphs with no
gluon $L$ ($R$) contains only left (right) moving collinear lines.  With
left-moving (right-moving) collinear gluon exchange, the line in $R$ ($L$)
between the gluon attachment and the gauge boson vertex is hard. We use
Feynman gauge for the gluon propagator.

Let us consider first a graph in which the gluon couples to two active
partons, as in figure~\ref{fig:spect-model}a and b. The Grammer-Yennie
approximation for the gluon with momentum $\ell$ flowing out of $R$ or
into $L$ respectively reads
\begin{align}
  \label{universal-GY}
R_\mu\ms g^{\mu\alpha} &\to R_\mu\,
   \frac{\ell^\mu v_R^\alpha}{\ell v_R - i\eta'} \,,
& 
g^{\alpha\nu} L_\nu &\to \frac{v_L^\alpha\ms \ell^\nu}{\ell v_L + i\eta}\,
   L_\nu \,.
\end{align}
Since we take the soft and collinear Grammer-Yennie approximants to be the
same, and since the soft auxiliary vectors must have large positive or
negative rapidity, we must take $v_R$ ($v_L$) to have large positive
(negative) rapidity.  Without loss of generality, we can take $v_R^+ =
v_L^- = 1$ (with $v_R^-$, $v_L^+$ much smaller, of order
$\Lambda^2/Q^2$). We allow the infinitesimal parameters $\eta$ and $\eta'$
to be different from the $\varepsilon$ in Feynman propagators, so that in
later calculations we can see that the order in which these parameters
are taken to zero does not matter. The imaginary parts in the eikonal
propagators correspond to past-pointing Wilson lines, as appropriate for
Drell-Yan.  They are chosen such that the pole in $\ell^-$ for the
left-hand term in \eqref{universal-GY}, and the pole in $\ell^+$ for the
right-hand term, lie on the same side of the real axis as the
corresponding active parton propagators.

Whilst the use of light-like auxiliary vectors ($v_R^-, v_L^+ = 0$) in the
Grammer-Yennie approximation may simplify calculations, it can also result
in rapidity divergences as already mentioned in section
\ref{sec:overall-strategy}. We shall in the following investigate the
cancellation of the Glauber region for a single gluon when the auxiliary
vectors are lightlike, spacelike or timelike.  An advantage of spacelike
auxiliary vectors (resulting in spacelike Wilson lines) is that the
additional pole of $1/(\ell^+ v_L^- + \ell^- v_L^+ + i\eta)$ in $\ell^-$
is then on the same side as the one of $1/(\ell^- v_R^+ + \ell^+ v_R^- -
i\eta')$ and hence does not complicate contour deformations, as was
observed in \cite{Collins:2004nx}.  A corresponding statement holds for
the poles in $\ell^+$. The same clearly cannot be said when the Wilson
lines are timelike.

According to the subtraction method discussed in
point~\ref{enum:subtractions} of section~\ref{sec:overall-strategy}, the
following approximant is good for a graph with an active-active gluon
connection in the combined collinear and soft regions of gluon momentum:
\begin{align}
  \label{region-sum}
R_\mu g^{\mu\nu} L_\nu &\to
   R_\mu\, \frac{\ell^\mu v_R^\alpha}{\ell v_R - i\eta'}
   \biggl( g_{\alpha\nu}
     - \frac{v_{L\ms\alpha} \ell_\nu}{\ell v_L + i\eta} \biggr) L^\nu
 + R_\mu \biggl( g^{\mu\alpha}
      - \frac{\ell^\mu v_R^\alpha}{\ell v_R - i\eta'} \Biggr)
   \frac{v_{L\ms\alpha} \ell_\nu}{\ell v_L + i\eta}\, L^\nu
\nonumber \\
 &\quad
  + R_\mu\, \frac{\ell^\mu v_R^\alpha}{\ell v_R - i\eta'}\,
            \frac{v_{L\ms\alpha} \ell_\nu}{\ell v_L + i\eta}\, L^\nu \,.
\end{align}
The first term on the right-hand side is the approximant for left moving
collinear $\ell$, including the soft subtraction. The second term is the
corresponding term for right-moving collinear $\ell$. Finally the third
term is the approximant for the soft region of $\ell$.

In fact, we can make this replacement also for graphs where the gluon
couples to a spectator line on one or both sides.  If it couples for
instance to a spectator in $R$ and an active parton in $L$, as in
figure~\ref{fig:spect-model}c, then the factorisation formula should only
contain the second and third terms in \eqref{region-sum}, which describe
the leading contributions from right moving and soft $\ell$.  The first
term in \eqref{region-sum} can however be added since it is power
suppressed: if $\ell$ is left moving then $R$ is suppressed because it
contains too many off-shell propagators, and if $\ell$ is soft or right
moving, the suppression is ensured by the factor in parentheses contracted
with $L$.  Likewise, in graphs where the gluon couples to spectator lines
on both ends, the first and second term in \eqref{region-sum} are power
suppressed in all relevant regions and may be added without degrading the
accuracy of approximations.  The key advantage of using \eqref{region-sum}
is that it can be rewritten as
\begin{align}
  \label{master-replace}
R_\mu g^{\mu\nu} L_\nu &\to  R_\mu g^{\mu\nu} L_\nu
   - R_\mu \biggl( g^{\mu\alpha}
      - \frac{\ell^\mu v_R^\alpha}{\ell v_R - i\eta'} \Biggr)
   \biggl( g_{\alpha\nu}
      - \frac{v_{L\ms\alpha} \ell_\nu}{\ell v_L + i\eta} \biggr) L^\nu \,.
\end{align}
To show that \eqref{master-replace} is a valid approximation in the
collinear and soft regions of $\ell$ is hence tantamount to showing that
its second term is power suppressed in these regions.  The advantage of
this method, which has long ago been used for factorisation studies in
\cite{Basu:1984ba}, is that it avoids the explicit computation of
complicated loop integrals since one can instead use power counting
arguments.


\subsection{Explicit calculations in a simplified setting}
\label{sec:box-graphs}

Before giving a general proof of Glauber exchange cancellation in the
setting just specified, we find it instructive to perform explicit
calculations for specific graphs.  This will allow us to see how the
general arguments of power counting and complex contour deformations work
in concrete examples.

\begin{figure}
\begin{center}
\subfigure[]{\includegraphics[height=0.35\textwidth]{%
    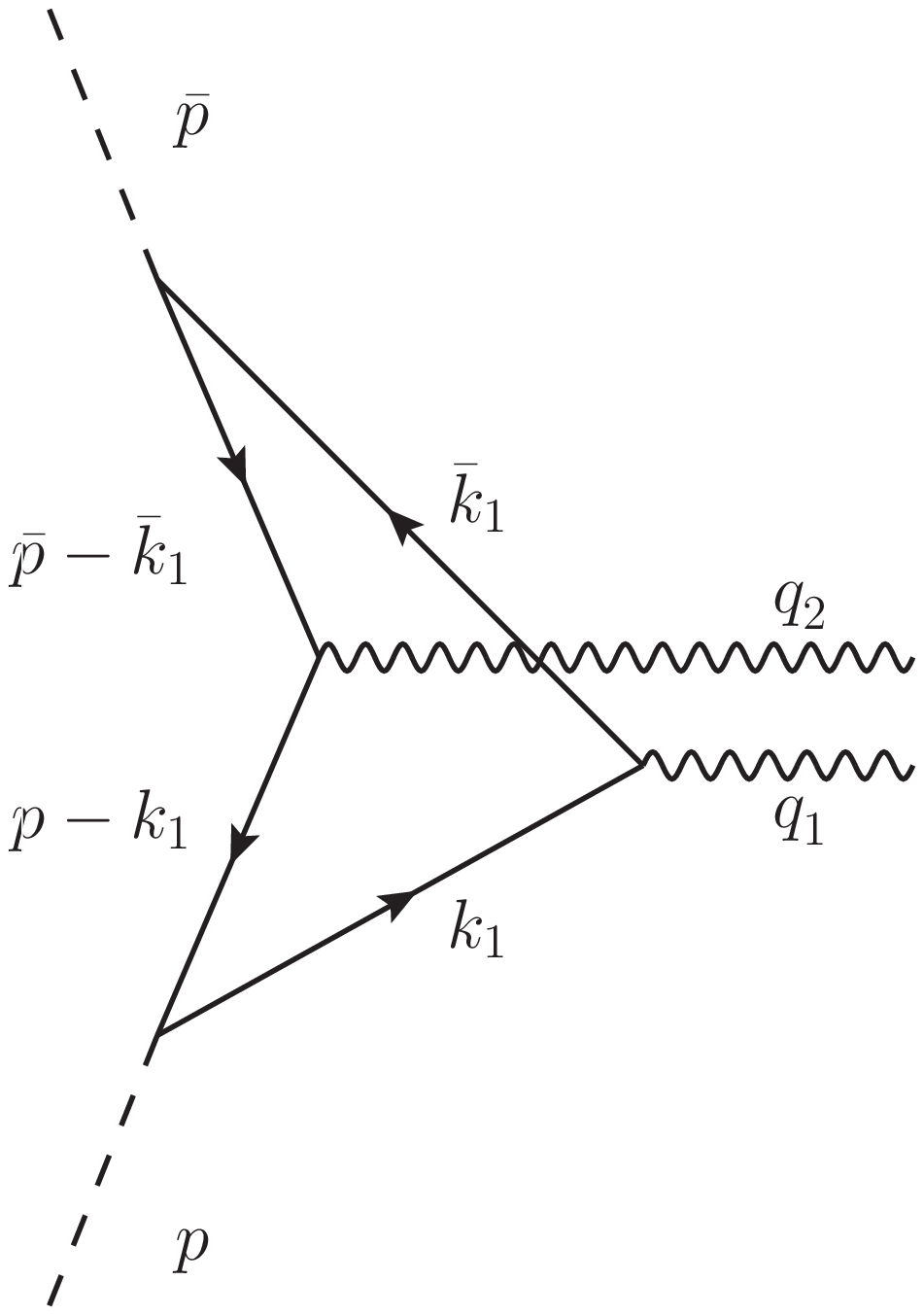}}
\hspace{4em}
\subfigure[]{\includegraphics[height=0.35\textwidth]{%
    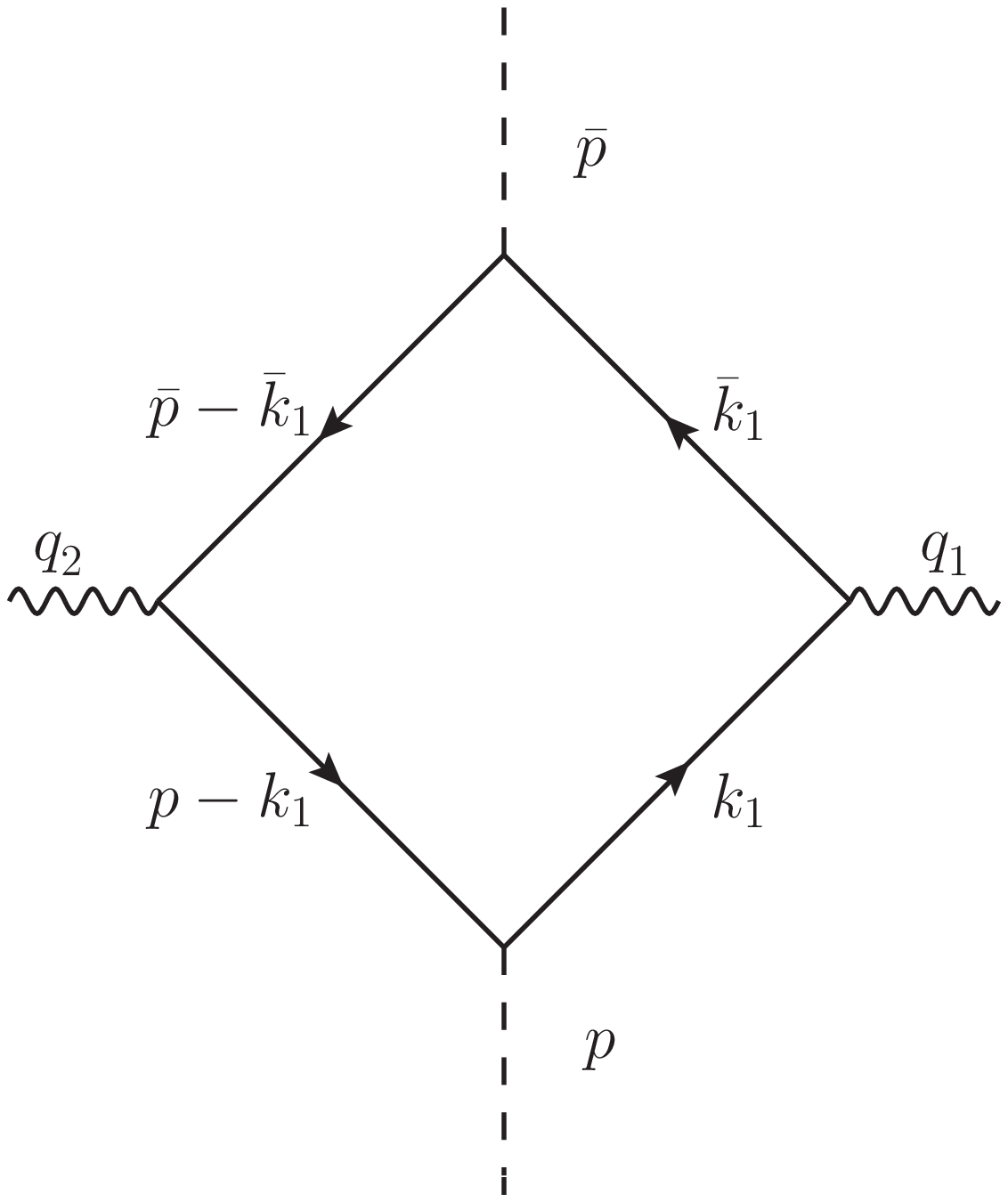}}
\caption{\label{fig:tree-box} (a) Double-Drell Yan graph for the
  simplified model described in section~\protect\ref{sec:box-graphs}. The
  dashed lines are the incoming scalar hadrons, whilst the solid lines are
  scalar partons. (b) Redrawing of (a) used for ease of legibility.  A
  second graph is obtained by reversing the arrows, which denote the flow
  of colour charge.}
\end{center}
\end{figure} 

We restrict our calculation to graphs without parton self-couplings and
complement our model by introducing a colourless scalar ``hadron'' with a
pointlike coupling to a parton pair.  Thus, the basic graph we have to
consider is the box graph in figure~\ref{fig:tree-box} (which has no
spectator partons at all) and the one-gluon corrections to it.  To
regulate collinear divergences we use a small mass $m$ for the scalar
parton, and to regulate infrared divergences at the one-loop level
considered here we give the gluon a mass $\lambda$.

Given the presence of the $gg\phi\phi$ seagull vertex in the model, there
are one-loop corrections to figure~\ref{fig:tree-box} with a gluon tadpole
insertion on any one of the scalar parton lines. The contribution of these
diagrams is proportional to
\begin{align}
\int \frac{\df^{4-2\epsilon} \ell }{\ell^2 - \lambda^2 + i
  \varepsilon} \propto \lambda^{2-2\varepsilon} 
\end{align}
and therefore vanishes in the limit of zero gluon mass, which we take
whenever possible.  

\begin{figure}
  \begin{center}
    \subfigure[]{\includegraphics[width=.22\textwidth]{%
        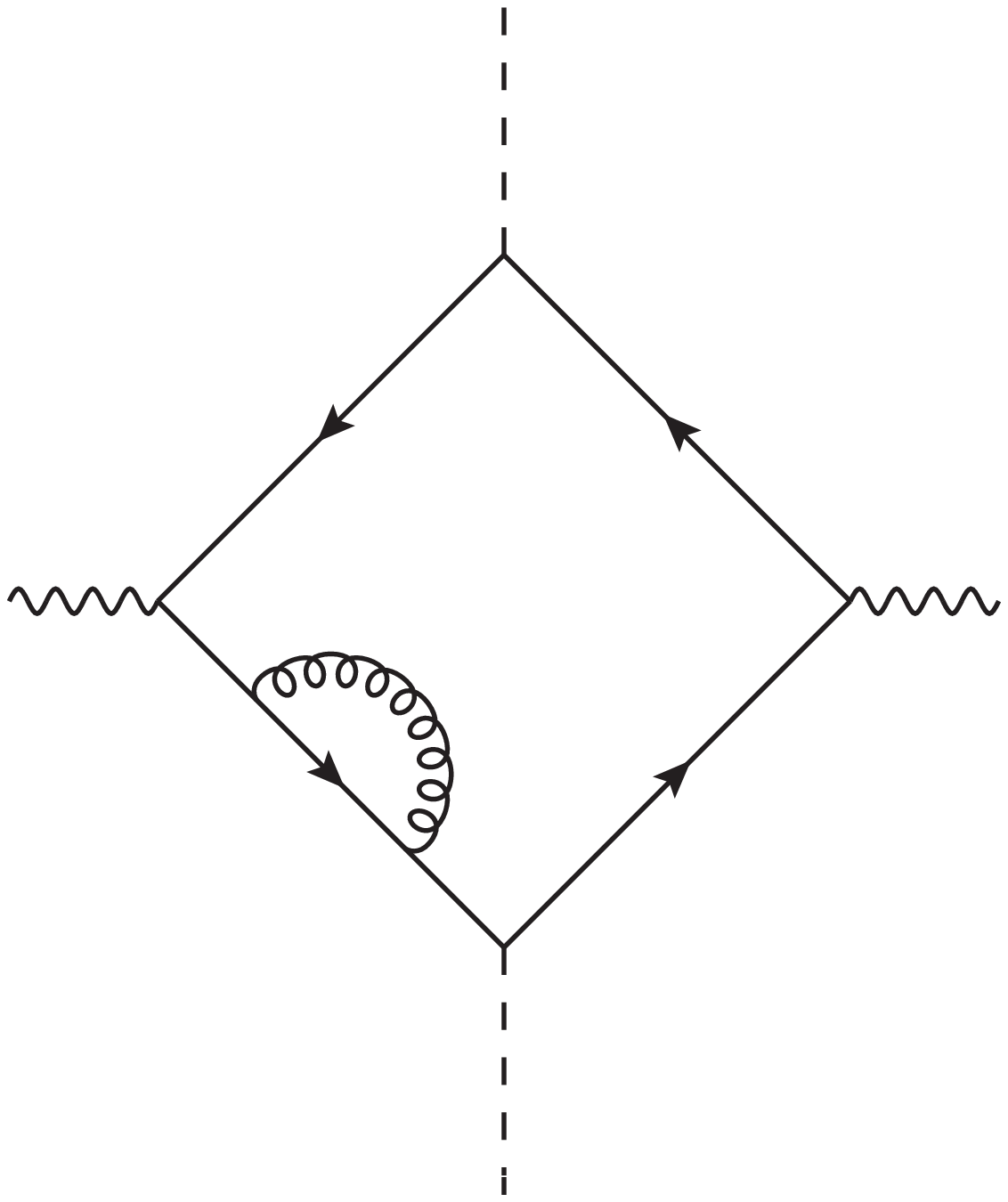}} \phantom{x}
  \subfigure[]{\includegraphics[width=.22\textwidth]{%
      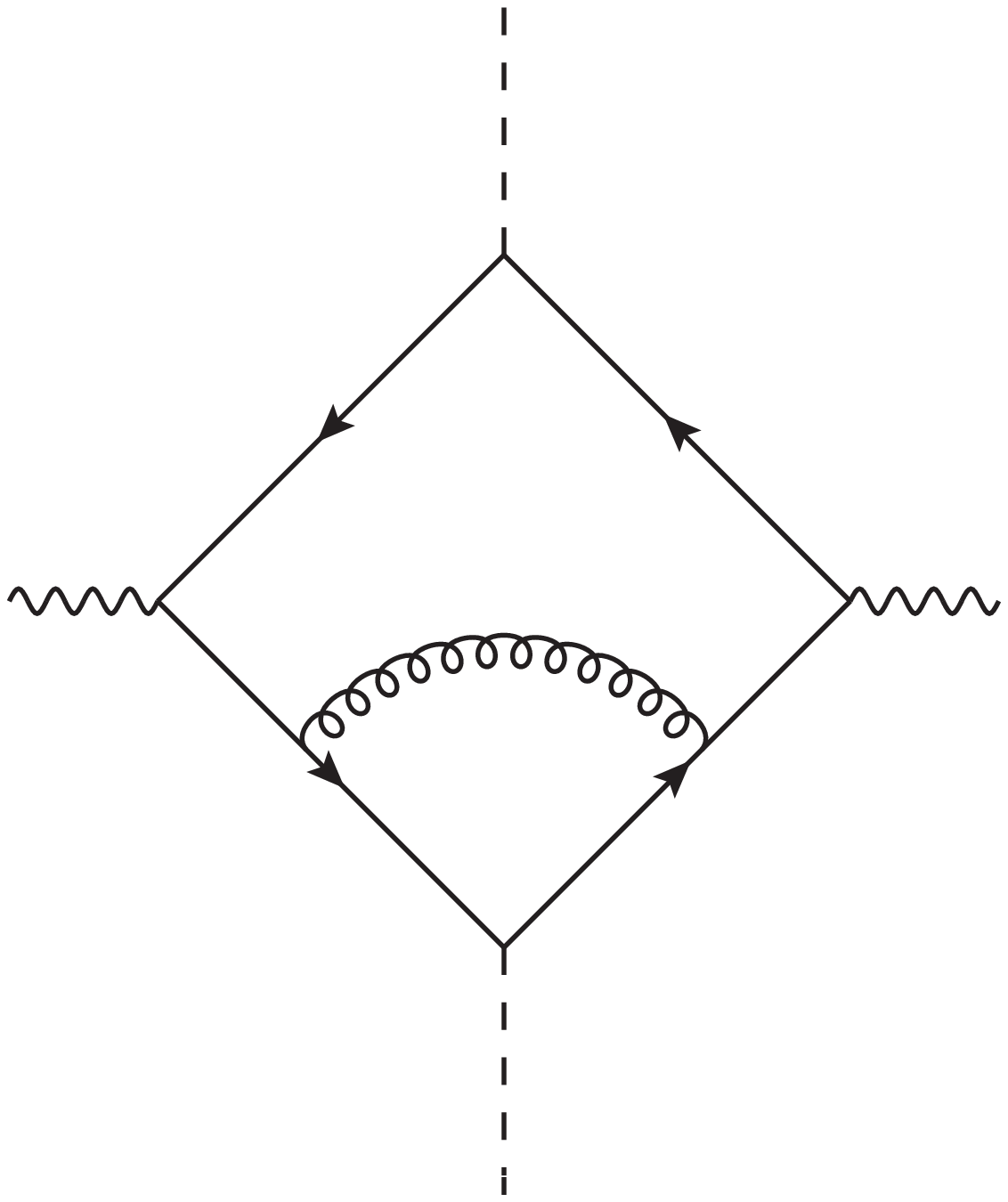}} \phantom{x}
  \subfigure[]{\includegraphics[width=.22\textwidth]{%
      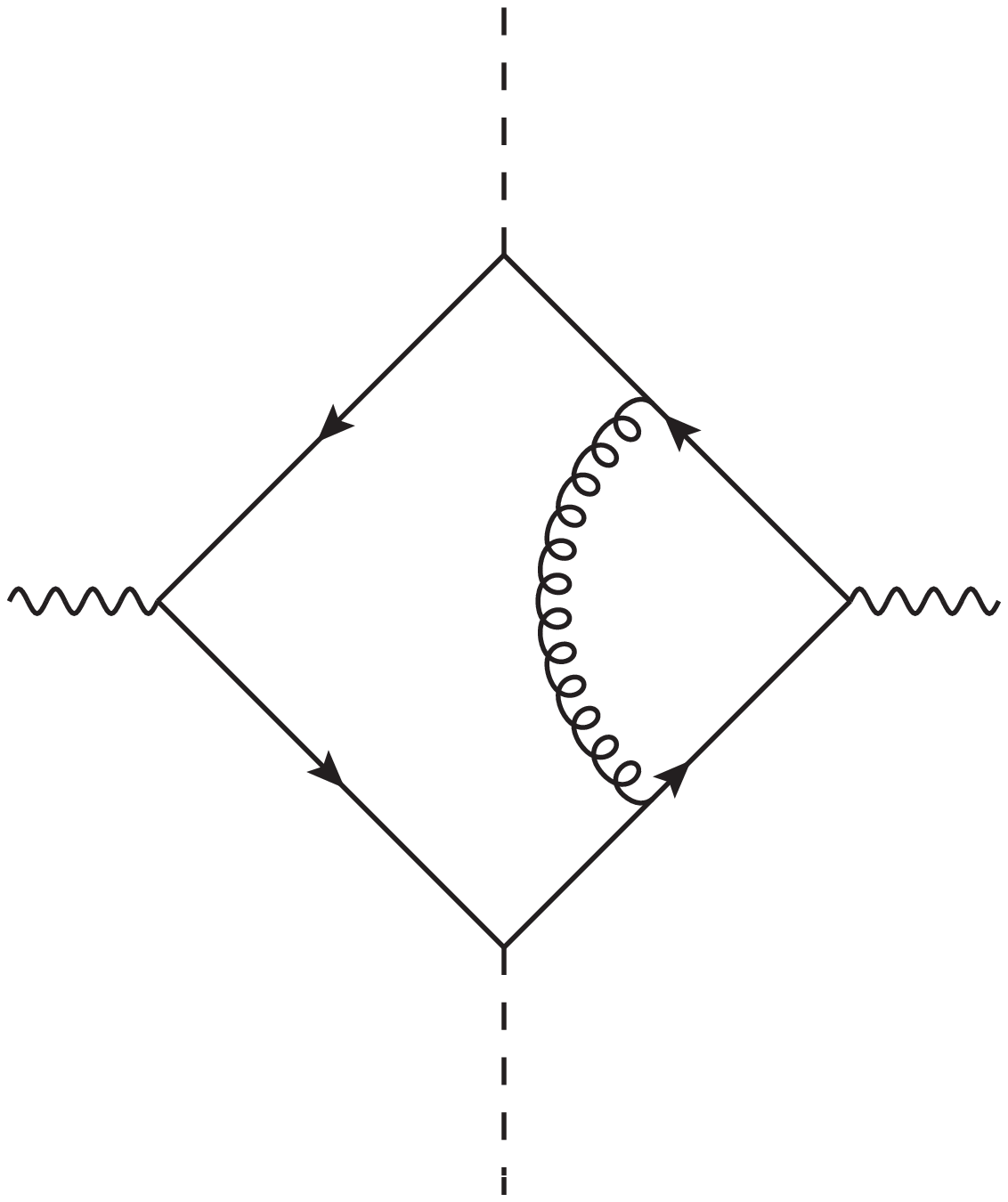}} \phantom{x}
  \subfigure[]{\includegraphics[width=.22\textwidth]{%
      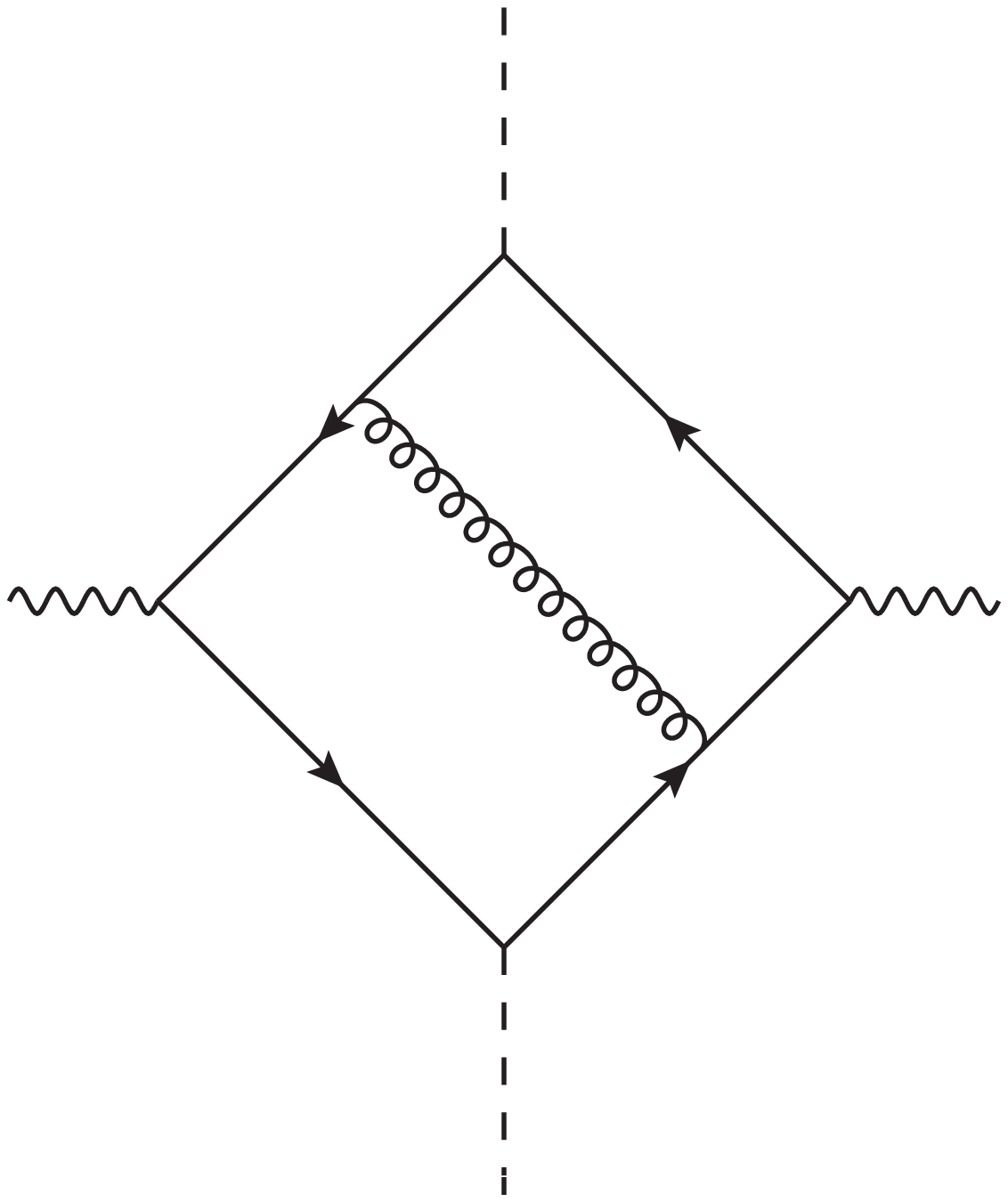}}
  \caption{\label{fig:loop-box} The different topologies of one-gluon
    corrections to the double Drell-Yan amplitude in our model: (a) parton
    self energy, (b) hadron vertex correction, (c) gauge boson vertex
    correction, (d) double box graph, where the gluon connects the two
    hard scatters.}
\end{center}
\end{figure}

The remaining one-gluon corrections to the amplitude are then given by the
graphs in figure~\ref{fig:loop-box} and by graphs with the same topology
obtained by permutation of lines.  The corresponding graphs for the cross
section are given by the interference of these virtual corrections with
the lowest-order graph.  Real gluon emission does not contribute at this
order because one cannot form a colour singlet state out of a single
emitted gluon and two gauge bosons.

We remark in passing that the diagrams in figures~\ref{fig:tree-box} and
\ref{fig:loop-box} appear as subgraphs of more complex graphs possible in
the model of section~\ref{sec:scope-and-method}, although in that context
the incoming particle to the box structure is a coloured parton rather
than a colourless hadron. In this sense, figure~\ref{fig:loop-box}d
appears inside figure~\ref{fig:spect-model}b. Since the calculations of
the following subsections do not depend on the colour factors of the
individual graphs and remain valid if the incoming lines are off-shell
partons, the demonstration of Glauber cancellation in the graphs of
figure~\ref{fig:loop-box} immediately carries over to the more general
case in which these graphs are embedded in more complex structures.  In
that case, graphs with real gluon emission contribute to the cross section
as well, but real gluons cannot be in the Glauber region for kinematic
reasons.

The graphs in which the gluon extends between two collinear partons
travelling in approximately the same direction,
i.e.\ graphs~\ref{fig:loop-box}a and b, factorise topologically as already
mentioned in section \ref{sec:scope-and-method}, and we need not consider
them further. They are part of the one-loop expression of the collinear
factor for right-moving partons.  In the following subsection we will
analyse the most complicated of the remaining graphs, namely the double
box of figure~\ref{fig:loop-box}d. The gauge boson vertex graph c,
discussed in section~\ref{sec:vertex-graph}, will then be comparatively
simple.

As explained in section~\ref{sec:scope-and-method} we will limit our
discussion to the kinematical region where all parton lines in
figure~\ref{fig:tree-box} have left or right moving collinear momenta and
thus transverse momenta of order $\Lambda \ll Q$.  With scalar partons,
this is indeed the only leading momentum region of the graph, provided of
course that the transverse momenta of the gauge bosons are also of order
$\Lambda$.  With spin $1/2$ partons there would be also a leading region
where partons have transverse momenta of order $Q$ and are far off-shell,
so that the graph describes a single hard scattering rather than two.
This problem, mentioned in point~\ref{enum:SPSvsDPS} of
section~\ref{sec:overall-strategy}, does not interfere with the issue of
Glauber gluon cancellation we are focusing on here.

As a further simplification, we replace each electroweak gauge boson by a
scalar particle with a pointlike momentum-independent coupling to two
partons.  This removes the dependence of the scattering amplitude on
polarisation vectors or open Lorentz indices, without affecting the
analytic structure of the graphs.  The results of the following
subsections therefore generalise to the case where spin $1$ bosons are
produced.


\subsubsection{Double box graph}
\label{sec:double-box}

\begin{figure}
\begin{center}
  \includegraphics[width=0.6\textwidth]{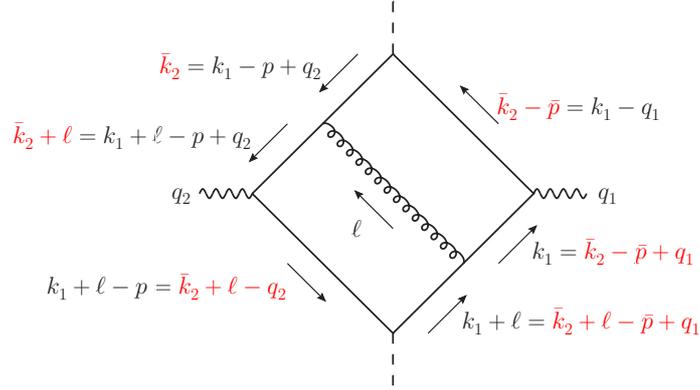}
  \caption{\label{fig:double-box} Double box graph, showing momentum
    labellings and our routing of the extra loop momentum $\ell$.}
\end{center}
\end{figure}

We now turn to the calculation of the graph in
figure~\ref{fig:double-box}, which up to a constant is given by
\begin{align}
 \label{double-box}
\Gamma &= \int \df \bar{k}_2^+ \, \df k_1^-\, \df^{2}\tvec{k}_1^{}
          \int \frac{\df^{4}\ell}{\ell^2 - \lambda^2 + i \varepsilon} \;
           \frac{(2 k_1 + \ell)_{\mu} \ms g^{\mu\nu}
           (2 \bar{k}_2 + \ell)_{\nu}}{%
             [(\bar{k}_2 + \ell)^2 - m^2+ i \varepsilon]\ms
             [(k_1 + \ell - p)^2 - m^2 + i \varepsilon]}
\nonumber \\
 & \quad \times \frac{1}{[(k_1 + \ell)^2 - m^2 + i \varepsilon]\ms
    [k_1^2 - m^2 + i \varepsilon]\ms
    [(\bar{k}_2 - \bar{p})^2 - m^2 + i \varepsilon]\ms
    [\bar{k}_2^2 - m^2 + i \varepsilon]} \,.
\end{align}
Since this graph has no ultraviolet divergence we have set the number of
spacetime dimensions to $4$.  It is understood that
$k_1 + \bar{k}_1 = q_1$, $k_2 + \bar{k}_2 = q_2$, $k_1 + k_2 = p$,
$\bar{k}_1 + \bar{k}_2 = \bar{p}$ and $p+\bar{p} = q_1+q_2$ by momentum
conservation.

We choose a frame where the hadron with momentum $p$ moves fast to the
right and the hadron with momentum $\bar{p}$ fast to the left, with
$\tvec{p} = \bar{\tvec{p}}= \tvec{0}$.  For $k_i$ and $\bar{k}_i$
($i=1,2$) in the collinear region, the momentum components have typical
size
\begin{align}
  \label{coll-reg-powers}
q_i^{\pm} \sim p^+ \sim \bar{p}^- & \sim Q \,,
&
k_i^- \sim \bar{k}_i^+ \sim p^- \sim \bar{p}^+ & \sim \Lambda^2/Q \,,
&
|\tvec{q}_i| \sim |\tvec{k}_i^{}| \sim |\bar{\tvec{k}}_i^{}|
   & \sim \Lambda \,.
\end{align}
The choice of independent integration variables in \eqref{double-box} is
appropriate for power counting since both $\bar{k}_2^+$ and $k_1^-$ can
freely vary over their natural range $\Lambda^2/Q$.

In the collinear momentum region, the lowest order graph in
figure~\ref{fig:tree-box} has all four propagator denominators of order
$1/\Lambda^2$ and an integration volume of order $\Lambda^6/Q^2$, giving
an overall behaviour like $1 /(\Lambda^2 Q^2)$.  It is easy to see that
the Glauber region \eqref{Glauber-scaling} of $\ell$ contributes at the
same order: the numerator factor $(2 k_1 + \ell) (2 \bar{k}_2 + \ell)$ is
of order $Q^2$, the integration volume $\df^4\ell$ of order
$\Lambda^6/Q^2$ and each of the three additional propagators of order
$1/\Lambda^2$.
As discussed in section~\ref{sec:soft-scaling}, both $k_1^-$ and $\ell^-$
can be of order $\Lambda$ as long as $k_1^- + \ell^- \sim \Lambda^2/Q$,
because this only carries the momentum $k_1$ off shell, which is
compensated by a larger integration volume of $\ell^-$ compared with the
Glauber region.  Likewise, $\ell^+$ can also be of order $\Lambda$.
Finally the ultrasoft region with all components of $\ell$ of order
$\Lambda^2/Q$ has a smaller integration volume compared with the Glauber
region, which is compensated by a smaller gluon propagator.  We thus find
that all four soft momentum scalings of $\ell$ enumerated in
section~\ref{sec:soft-scaling} give leading contributions to the graph.
Furthermore, the regions where $\ell$ is right or left collinear give
leading contributions.

As discussed in section~\ref{sec:scope-and-method}, we can establish the
validity of the Grammer-Yennie approximation by showing that the
replacement
\begin{align}
  \label{prop-replace}
g^{\mu\nu} \rightarrow
\biggl( g^{\mu\alpha}
      - \frac{\ell^\mu v_R^\alpha}{\ell v_R - i\eta'} \Biggr)
   \biggl( g_{\alpha}{}^{\nu}
      - \frac{v_{L\ms\alpha}\ms \ell^\nu}{\ell v_L + i\eta} \biggr)
\end{align}
in \eqref{double-box} gives a power suppressed result relative to the
leading behaviour $1 /(\Lambda^2 Q^2)$.


\subsubsection{Lightlike Wilson lines}
\label{sec:lightlike-WL}

For simplicity we first consider the case in which the soft and collinear
Wilson lines are taken lightlike, with $v_R^+ = v_L^- = 1$ and $v_R^- =
v_L^+ = 0$.  The replacement \eqref{prop-replace} then corresponds to the
replacement
\begin{align}
  \label{double-box-full-num}
& (2 k_1 + \ell)_{\mu}\ms g^{\mu\nu} (2 \bar{k}_2 + \ell)_{\nu}
 \,\to\,
 \bigl[ (2 \tvec{\ell}\ms \tvec{k}_1 + \tvec{\ell}^2)
        (2 \tvec{\ell}\ms \bar{\tvec{k}}_2 + \tvec{\ell}^2)
\nonumber \\[0.1em]
 & \qquad
  - 2 \ell^+ k_1^{-} (2 \tvec{\ell}\ms \bar{\tvec{k}}_2^{} + \tvec{\ell}^2)
  - 2 \ell^- \bar{k}_2^{+} (2 \tvec{\ell}\ms \tvec{k}_1^{} + \tvec{\ell}^2)
  - \ell^+ \ell^- \big\{ 4 (\tvec{k}_1 + \tvec{\ell})
                  (\bar{\tvec{k}}_2 + \tvec{\ell}) - \tvec{\ell}^2 \big\}
\nonumber \\
 & \qquad
  + 2 \ell^+ \ell^- (2{k_1}^- + \ell^-) (2 \bar{k}_2^+ + \ell^+) \bigr]
  \,\big/\,  \bigl[ (\ell^- - i \eta') (\ell^+ + i \eta) \bigr]
\end{align}
in \eqref{double-box}.
The original numerator term on the left hand side of this equation is of
order $Q^2$~-- thus when we replace the left hand side by the right hand
side, we only get a leading power contribution when this expression is
also of order $Q^2$. When $\ell$ is soft $\sim (\Lambda, \Lambda,
\Lambda)$ or collinear, the numerator of the right-hand-side is of order
$\Lambda^4$ and the denominator of order $\Lambda^2$, giving a
power-suppressed $\mathcal{O}(\Lambda^2)$ contribution overall.  For the
momentum scalings $\ell \sim (\Lambda^2/Q, \Lambda, \Lambda)$ and $\ell
\sim (\Lambda, \Lambda^2/Q, \Lambda)$, the numerator remains of order
$\Lambda^4$ whilst the denominator grows to order $\Lambda^3/Q$, which is
however not enough to give a leading contribution.  In the ultrasoft
region of $\ell$ the denominator is of order $\Lambda^4/Q^2$ but the
numerator shrinks to order $\Lambda^6/Q^2$, resulting again in a
subleading contribution.  Only when $\ell$ is in the Glauber region, where
the numerator is of order $\Lambda^4$ and the denominator of order
$\Lambda^4/Q^2$, do we get a leading contribution by power counting.

It remains therefore to show that the contribution from the Glauber region
is actually smaller than the naive prediction from power counting.  For
this we can neglect all terms in the numerator of
\eqref{double-box-full-num} that contain either plus- or minus-components,
since in the Glauber region these terms are subleading compared to the
terms containing only transverse components.

Combining the denominators of all Feynman propagators that contain $\ell$
with the help of Feynman parameters we obtain
\begin{align} 
  \label{eq:ll3}
I &= \frac{3!\ms N}{D} \int [\df^4 \alpha]
  \int \frac{\df \ell^-}{\ell^- - i \eta'}
  \int \frac{\df \ell^+}{\ell^+ + i \eta} \;
  \frac{1}{[2 \ell^+ \ell^-
      + 2 a^- \ell^+  + 2 a^+ \ell^-  + A + i \varepsilon]^4}
\end{align}
with
\begin{align}
 \label{abbreviations}
A &= 2 (\alpha_3^{} q_1^+ - \alpha_2^{} q_2^+
      - \alpha_3^{} \bar{p}^+)\ms k_1^-
   + 2 \bar{k}_2^+ (\alpha_1^{} q_2^- - \alpha_1^{} p^- - \alpha_2^{} p^-)
   + 2 (1-\alpha_4^{}) \bar{k}_2^+ k_1^- 
\nonumber \\
 & \quad + 2 \alpha_2^{} q_2^+ p^-
   - \alpha_1^{} (\bar{\tvec{k}}_2^{} + \tvec{\ell})^2
   - (\alpha_2^{} + \alpha_3^{}) (\tvec{k}_1^{} + \tvec{\ell})^2
   - \alpha_4^{} (\tvec{\ell}^2 + \lambda^2) - (1-\alpha_4^{})\ms m^2 \,,
\nonumber \\
  a^+ &= \alpha_3^{} q_1^+ - \alpha_2^{} q_2^+ - \alpha_3^{} \bar{p}^+
       + (1-\alpha_4^{}) \bar{k}_2^+ \,,
\nonumber \\
  a^- & = \alpha_1^{} q_2^- - (\alpha_1^{} + \alpha_2^{}) p^-
       + (1-\alpha_4^{}) k_1^- \,,
\nonumber \\
N &= (2 \tvec{\ell}\ms \tvec{k}_1 + \tvec{\ell}^2)
        (2 \tvec{\ell}\ms \bar{\tvec{k}}_2 + \tvec{\ell}^2) \,,
\nonumber \\
D & = [k_1^2 - m^2 + i \varepsilon]\ms
      [(\bar{k}_2 - \bar{p})^2 - m^2 + i \varepsilon]\ms
      [\bar{k}_2^2 - m^2 + i \varepsilon] \,.
\end{align}
The integration measure for Feynman parameters is
\begin{align}
[\df^n\alpha] &= \df \alpha_1 \cdots \df \alpha_n\;
   \delta(\alpha_1 + \cdots + \alpha_n - 1) &
\text{with $0 \le \alpha_i \le 1$,}
\end{align}
and it is understood that $I$ is still to be integrated over
$\bar{k}_2^+$, $k_1^-$, $\tvec{k}_1^{}$ and $\tvec{\ell}$. Since the
volume of this remaining integration space is $\mathcal{O}(\Lambda^8/Q^2)$
and an overall leading power contribution to $\Gamma$ is
$\mathcal{O}[1/(\Lambda^2Q^2)]$, the integral $I$ has to be of order
$1/\Lambda^{10}$ for a leading power contribution.  Conversely, we must
exclude that $I$ is of order $1/\Lambda^{10}$ in order to establish the
validity of the Grammer-Yennie approximation.

As appropriate for the Glauber region, we only consider the case
$|\tvec{\ell}| \sim \Lambda$ in the following.  Power counting thus gives
$A \sim \Lambda^2$ for any value of the Feynman parameters,\footnote{This
  reflects the fact that the four propagators that have been combined into
  the term raised to the fourth power in \eqref{eq:ll3} are each of order
  $\Lambda^2$ for $\ell^+=\ell^-=0$.}
even if we take masses $\lambda, m \ll \Lambda$.  This implies that
regions where $\alpha_i \ll 1$ for one or several $i$ are power
suppressed: contributions from such regions are suppressed by the phase
space $[\df^n \alpha]$ and cannot be enhanced by having a smaller
denominator.  Only the region where all Feynman parameters are of generic
size can possibly give a contribution of order $1/\Lambda^{10}$ to $I$.
We then have
\begin{align} 
  \label{eq:powercounting}
A & \sim \Lambda^2 \,, &
a^+ , a^-  & \sim Q \,, & 
a^- & > 0 \,,
\end{align}
whereas $a^+$ can have either sign, depending on the relative size of
$\alpha_3^{} q_1^+$ and $\alpha_2^{} q_2^+$.

By power counting, the region where $\ell^+ \sim \ell^- \sim \Lambda^2/Q$
could indeed give a leading contribution to $I$, and we have to show that
this is not the case.  According to the general rule explained in
section~\ref{sec:scope-and-method} the integration over $\ell^-$ has a
pinched pair of poles in that region, whereas the one over $\ell^+$ does
not.

To further analyse $I$ we carry out the integral over $\ell^+$ using
Cauchy's theorem.  For $\ell^- + a^- > 0$ both poles in $\ell^+$ are on
the same side of the real axis and one gets zero.  From the region $\ell^-
+ a^- < 0$ we obtain, picking up the residue at $\ell^+ = - i \eta$ for
simplicity,
\begin{align}  
  \label{eq:ll1}
I &= - 2\pi i\, \frac{3!\ms N}{D} \int [\df^4 \alpha]
      \int^{-a^-}_{-\infty} \frac{\df \ell^-}{\ell^- - i \eta'}\;
      \frac{1}{[2 a^+ \ell^- + A + i \varepsilon]^4} ,
\end{align}
where we have dropped $i\eta$ in the last denominator, since it comes with
the same sign as $i\varepsilon$ and thus does not change the position of
the pole.  We note that the evaluation of Feynman integrals in light-cone
coordinates has to be done with some care, because in some cases the naive
application of Cauchy's theorem can lead to wrong results. In the case at
hand we have cross checked our results with the method described in
appendix \ref{app:feynmanint}.  As explained there, we could not have used
Cauchy's theorem for the $\ell^+$ integration in \eqref{eq:ll3} if we had
kept the full numerator of the graph given in \eqref{double-box-full-num},
because terms with $\ell^+$ or $(\ell^+)^2$ would have cancelled the
denominator $\ell^+ + i\eta$.  Using the method just mentioned, we have
checked that these terms give a power suppressed contribution and can
hence be discarded, as we argued on the basis of power counting before
\eqref{eq:ll3}.

Since $\ell^-$ is at least of size $Q$ in \eqref{eq:ll1}, we have $2 a^+
\ell^- + A \sim Q^2$ or bigger and obtain a power suppressed result for
$I$, as we needed to show.  We observe that in \eqref{eq:ll1} we have two
poles
\begin{align}
\ell^-_1 &= i \eta' \,, &
\ell^-_2 &= - A /(2 a^+) - \text{sgn}(a^+) \ms i \varepsilon \,,
\end{align}
which are pinched for $a^+ > 0$ in agreement with the general rule stated
above.  However, they are far away from the integration region and
therefore do not matter.

We can also explicitly calculate \eqref{eq:ll1} by partial fractioning,
which gives
\begin{align} 
\label{eq:ll2}
I &=  - 2\pi i\, \frac{3!\ms N}{D} \int [\df^4 \alpha]\,
  \biggl\{ \frac{1}{(A + i \varepsilon)^4}  \
      \log{\left| \frac{2 a^+ a^- - A}{2 a^+ a^-} \right|} 
\nonumber \\ 
 &\quad + \frac{i \pi}{(A + i \varepsilon)^4}\,
    \bigl[ \Theta(2 a^+ a^- - A) + \Theta(a^-) - \Theta(a^+) - 1 \bigr]
\nonumber \\ 
  &\quad - \frac{1}{(A + i \varepsilon)^3
     (A - 2 a^+ a^- + i \varepsilon)}
   - \frac{1}{2} \frac{1}{(A + i \varepsilon)^2
     (A - 2 a^+ a^- + i \varepsilon)^2}
\nonumber \\ 
  &\quad - \frac{1}{3} \frac{1}{(A + i \varepsilon)
     (A - 2 a^+ a^- + i \varepsilon)^3} \biggr\}  \,.
\end{align}
Since $a^+a^- \sim Q^2$, the potentially leading terms in \eqref{eq:ll2}
are those proportional to $1/A^4$.  In the generic region of Feynman
parameters, the logarithm gives a suppression by a factor $\Lambda^2/Q^2$
and the $\Theta$ function terms cancel exactly because $a^- > 0$.  This
confirms that $I$ is power suppressed.


\subsubsection{Alternative momentum routings} \label{sec:alteroute}

\begin{figure}[t]
\begin{center}
  \subfigure[]{\includegraphics[width=.32\textwidth]{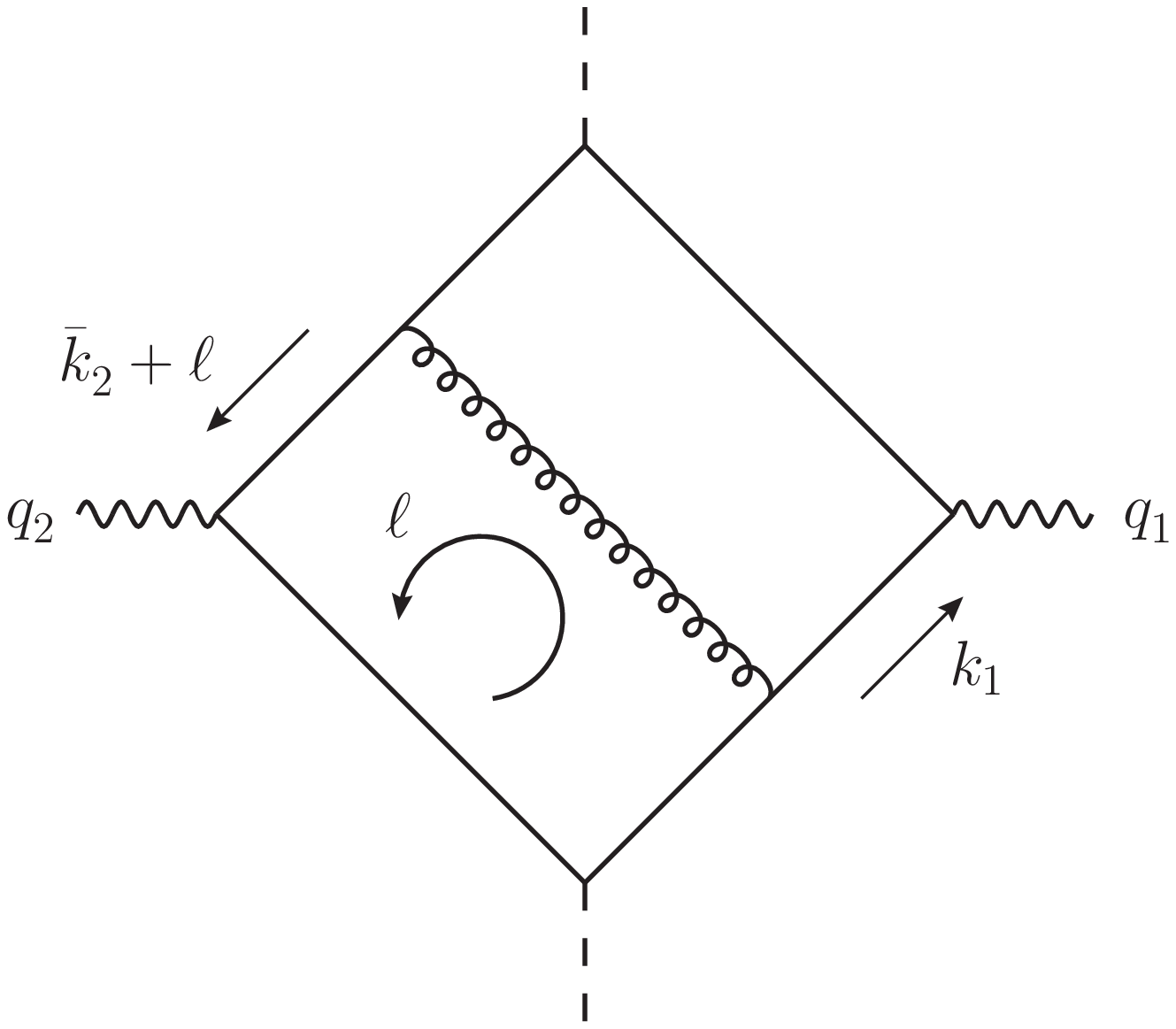}}
  \subfigure[]{\includegraphics[width=.32\textwidth]{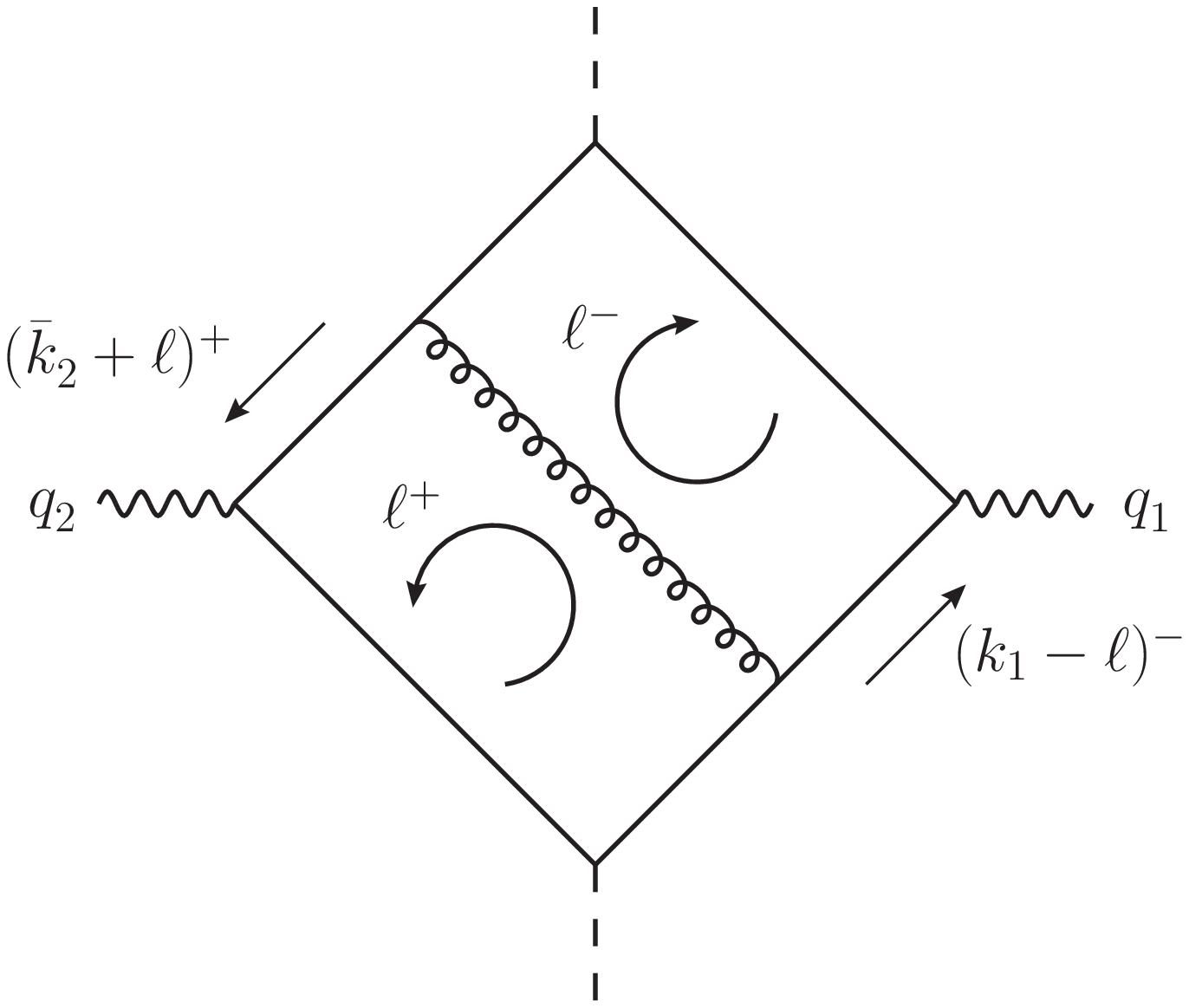}}
  \subfigure[]{\includegraphics[width=.32\textwidth]{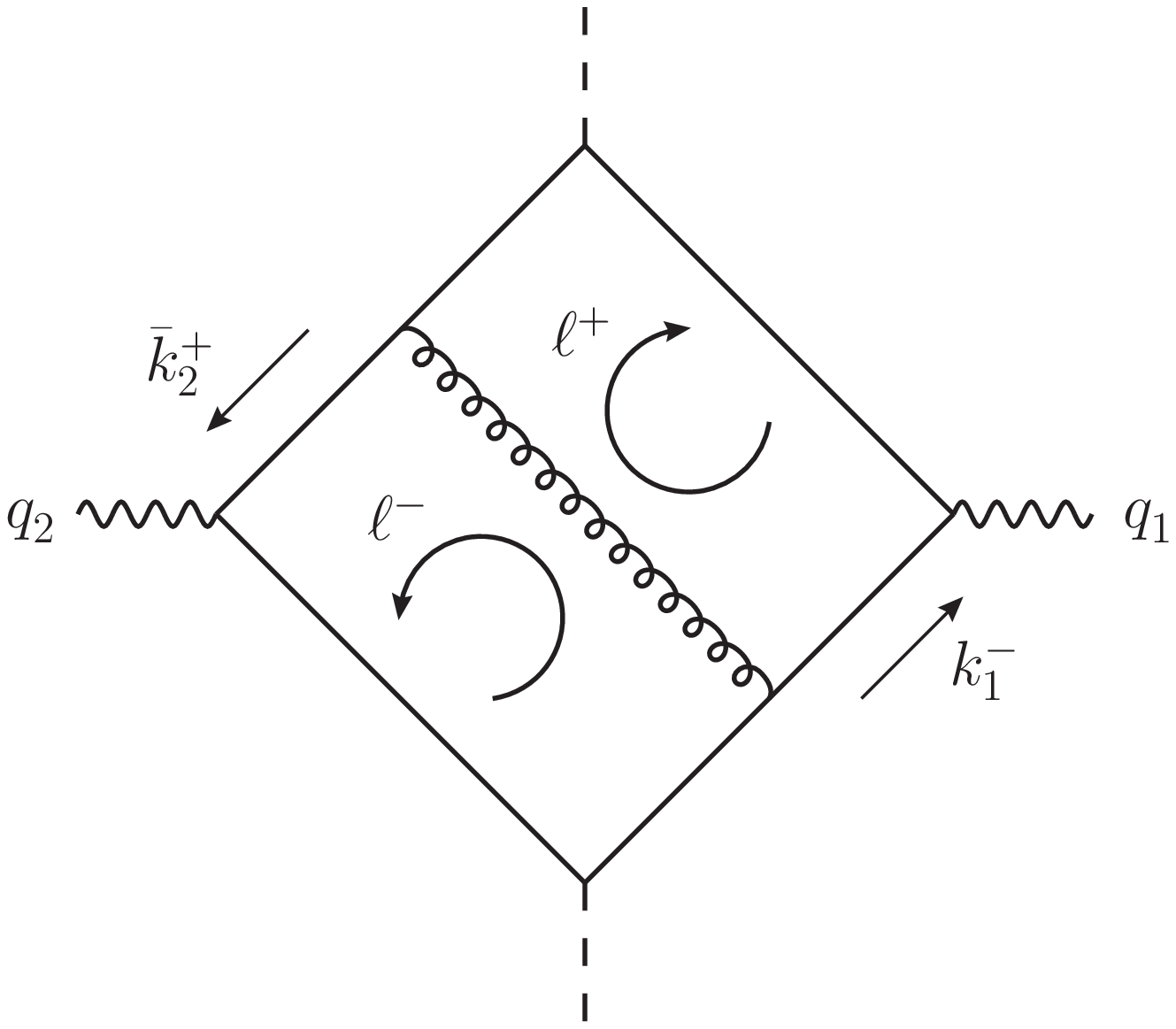}}
  \caption{\label{fig:routings} Different momentum routings for the double
    box graph: (a) routing used in \protect\eqref{double-box}, where
    $\ell^-$ but not $\ell^+$ is pinched in the Glauber region, (b)
    routing for which neither $\ell^+$ nor $\ell^-$ is pinched in the
    Glauber region, (c) routing for which both $\ell^+$ and $\ell^-$ are
    pinched in the Glauber region.}
\end{center}
\end{figure}

We explained in section \ref{sec:scope-and-method} that the pattern of
pinched poles at small $\ell^+$ ($\ell^-$) is determined by the way in
which these loop momentum components flow along the collinear lines with
large minus (plus) momenta. It is therefore interesting to consider
different routings for $\ell^+$ and $\ell^-$ for which their respective
flow along the collinear momenta (and thus the pattern of pinches) is
different.  We will verify by explicit calculation that the final result
for the double box graph, i.e.\ the validity of the factorisation formula,
is independent of the momentum routing. The freedom of choice to re-route
the $\ell^+$ and $\ell^-$ loop momentum components will be essential in
section \ref{sec:general-arg}, where we give a general proof of Glauber
gluon cancellation at the one-gluon level.

The momentum routing in figure~\ref{fig:routings}a has already been
discussed in the previous subsection.  Let us now turn to the routing in
figure~\ref{fig:routings}b, for which neither the $\ell^+$ nor the
$\ell^-$ integration is pinched in the Glauber region.  This momentum
routing can be obtained by the shift $k_1^- \rightarrow k_1^- - \ell^-$ in
\eqref{double-box}.  Using separate sets of Feynman parameters for
combining the propagators containing $\ell^+$ with the gluon propagator on
one hand and for the propagators containing $\ell^-$ on the other hand, we
get
\begin{align}
  \label{alt-routing-1}
 I & = 3!\, 2!\ms N  \int [\df^4 \alpha] [\df^3 \beta]
   \int \frac{\df \ell^-}{\ell^- - i \eta'}
   \int \frac{\df \ell^+}{\ell^+ + i \eta}
\nonumber \\
   & \quad \times
 \frac{1}{[2 \alpha_4^{}\ms \ell^+\ell^- + 2 a^- \ell^+
                +  A + i \varepsilon]^4\,
          [2 b^+ \ell^- + B + i \varepsilon]^3}
\end{align}
with $A$, $a^-$ and $N$ as in \eqref{abbreviations} and
\begin{align}
  \label{abbrev-1}
B &= 2 (\beta_1^{} q_1^+ - \beta_1^{} \bar{p}^+
         - \beta_2^{} \bar{p}^+)\ms k_1^-
     - 2 \bar{k}_2^+ (\beta_2^{} q_1^- - \beta_3^{} q_2^- + \beta_3^{} p^-)
     + 2 \bar{k}_2^+ k_1^-
\nonumber \\
 & \quad + 2 \beta_2^{} \bar{p}^+ q_1^- - \beta_1^{} \tvec{k}_1^2
         - (\beta_2^{}+\beta_3^{}) \bar{\tvec{k}}_2^2 - m^2 \,,
\nonumber \\
b^+ &= - \beta_1^{} q_1^+
       + (\beta_1^{}+\beta_2^{}) \bar{p}^+ - \bar{k}_2^+  \,.
\end{align}
We see that $B \sim \Lambda^2$ for all values of the Feynman parameters
(as is the case for $A$), even if $m$ is neglected.  Hence the regions with
$\beta_i \ll 1$ cannot give a leading contribution to $I$ due to the
suppression from the integration volume $[\df^3\beta]$.  In the generic
region of Feynman parameters, we have
\begin{align}
B & \sim \Lambda^2 \,, &
b^+ & \sim Q \,, &
b^+ & < 0 \,.
\end{align}
Performing again the $\ell^+$ integral using Cauchy's theorem we get
\begin{align}
I &= - 2\pi i\, 3!\, 2!\ms N \int [\df^4 \alpha] [\df^3 \beta]
   \int_{-\infty}^{-a^- /\alpha_4}
   \frac{\df \ell^-}{\ell^- - i \eta'}\;
    \frac{1}{[A + i \varepsilon ]^4\,
             [2 b^+ \ell^- + B + i \varepsilon]^3} \,.
\end{align}
Given that $a^-/\alpha_4$ is of order $Q$ and positive, the $\ell^-$
integration is limited to a region where $2 b^+ \ell^- + B \sim Q^2$ or
bigger, so that we obtain a power suppressed result as before.  With $b^+$
being negative, the two poles in $\ell^-$ are near zero and on the same
side of the real axis, confirming the absence of a pinch in this case.

Finally we turn to the routing in figure~\ref{fig:routings}c. Here, both
the $\ell^+$ and $\ell^-$ integration contours are trapped in the Glauber
region, so that one would expect a leading power behaviour of $I$ in this
case.  Let us see how this can be reconciled with the previous
calculations.  The corresponding momentum routing can be obtained via the
replacement $\bar{k}_2^+ \rightarrow \bar{k}_2^+ - \ell^+$ in
\eqref{double-box}.  Combining the propagators in the same way as above
yields
\begin{align}
  \label{alt-routing-2}
 I & = 3!\, 2!\ms N  \int [\df^4 \alpha] [\df^3 \beta]
   \int \frac{\df \ell^-}{\ell^- - i \eta'}
   \int \frac{\df \ell^+}{\ell^+ + i \eta}
\nonumber \\
   & \quad \times
 \frac{1}{[2 \alpha_4^{}\ms \ell^+\ell^- + 2 a^+ \ell^-
              +  A + i \varepsilon]^4\,
          [2 b^- \ell^+ + B + i \varepsilon]^3}
\end{align}
with $A$, $a^+$ and $N$ as in \eqref{abbreviations}, $B$ as in
\eqref{abbrev-1} and
\begin{align}
  \label{abbrev-2}
b^- & = \beta_2^{} q_1^- - \beta_3^{} q_2^- + \beta_3^{} \bar{p}^-
        - k_1^-  \,.
\end{align}
In the generic region of Feynman parameters, both $a^+$ and $b^-$ are of
order $Q$ but can have either sign.
Now we perform the $\ell^-$ integral using Cauchy's theorem and get
\begin{align}
 I &= - 2\pi i\, 3!\, 2!\ms N \int [\df^4 \alpha] [\df^3 \beta]
   \int_{-a^+ /\alpha_4}^{\infty}
   \frac{\df \ell^+}{\ell^+ + i \eta}\;
   \frac{1}{[A + i \varepsilon]^4\,
            [2 b^- \ell^+ + B + i \varepsilon]^3} \,.
\end{align}
Once again, the poles of the $\ell^+$ integration are near zero. In
contrast to the routing chosen above, however, they can be close to the
integration region.  We have different cases:
\begin{enumerate}
\item $a^+ < 0$ and $b^-$ of either sign: both poles are far away from the
  integration region,
\item $a^+ > 0$ and $b^- > 0$: both poles are below the real axis,
\item $a^+ > 0$ and $b^- < 0$: the integration contour is pinched in the
  Glauber region.
\end{enumerate}
If $\ell^+$ is at least of order $Q$ then $2 b^- \ell^+ + B \sim Q^2$ or
bigger and we obtain a power suppressed integral.  This holds in case 1,
and also in case 2 if we deform the integration contour away from the real
axis on the semi-circle going through the points $-a^+$, $i a^+$, $a^+$.
It holds in case 3 if we deform the contour to the corresponding
semi-circle in the lower half-plane, thus picking up the residue of the
eikonal propagator.  The additional term from the residue is proportional
to $[A + i\varepsilon]^4\, [B + i \varepsilon]^3$, so that $I \sim
1/\Lambda^{10}$ as we expected.  However, this contribution vanishes when
one integrates over either $k_1^-$ or $\bar{k}_2^+$.  To see this, we
rewrite
\begin{align}
  \label{combined-denoms}
[A + i\varepsilon]^{-4}\, [B + i \varepsilon]^{-3}
 &= [2 a^+ k_1^- + 2 (\alpha_1^{} q_2^-
      - \alpha_1^{} p^- - \alpha_2^{} p^-)\ms \bar{k}_2^+
     + \cdots + i\varepsilon]^{-4}
\nonumber \\[0.1em]
 &\quad \times
    [2 (\beta_1^{} q_1^+
        - \beta_1^{} \bar{p}^+ - \beta_2^{} \bar{p}^+)\ms k_1^-
     - 2 b^- \bar{k}_2^+ + \cdots + i \varepsilon]^{-3} \,,
\end{align}
where the ellipses denote terms independent of $k_1^-$ and $\bar{k}_2^+$.
We see that for $a^+ > 0$ the poles of \eqref{combined-denoms} in $k_1^-$
are on the same side of the real axis and thus give a zero integral (note
that $a^+$ depends on $\bar{k}_2^+$ but not on $k_1^-$).  The same result
is obtained if one first integrates over $\bar{k}_2^+$ under the condition
that $b^- < 0$.  We have thus shown again that the Grammer-Yennie
approximation works, but this required integration over several loop
variables.


\subsubsection{Wilson lines with finite rapidity}
\label{sec:non-lightlike-WL}

We now re-analyse the double box graph for Wilson lines along directions
$v_R$ and $v_L$ with a finite rapidity.  According to the general argument
in section~\ref{sec:GY-rearrange}, we expect factorisation to carry over
if we change lightlike Wilson lines into spacelike ones, but for timelike
Wilson lines we have no such expectation.

\paragraph{Spacelike Wilson lines.}
We take spacelike Wilson lines with $v_R^+ = v_L^- = 1$ and $v_L^+ = v_R^-
= - \delta^2$. As discussed in section \ref{sec:GY-rearrange}, we must
take $v_R$ ($v_L$) to have large positive (negative) rapidity, of the same
order as the appropriate collinear particles.  Therefore, we count $\delta
\sim \Lambda/Q$.

In the numerator of the right hand side of \eqref{double-box-full-num},
replacing light-like with space-like Wilson lines only results in
power-suppressed changes in the Glauber region. Some momentum components
$c^+$ are replaced by $v_L\ms c$, some components $c^-$ by $v_R\ms c$, and
the leading power term is still the one with only transverse components
(i.e.\ $N$ given in \eqref{abbreviations}).
In \eqref{eq:ll3} we must make the replacements 
\begin{align}
  \label{spl-WL}
\frac{1}{\ell^- - i \eta'}  \rightarrow
  \frac{1}{\ell^-  - \delta^2 \ell^+ - i \eta'} \,, 
\quad
\frac{1}{\ell^+ + i \eta}   \rightarrow
  \frac{1}{\ell^+  - \delta^2 \ell^- + i \eta} \,.
\end{align} 
The new eikonal propagators have the following poles in the complex
$\ell^+$ plane:
\begin{align}
 \ell^+_{\text{eik},1} &= \delta^2 \ell^- - i  \eta \,,
&
 \ell^+_{\text{eik},2} &= \delta^{-2} \ell^- - i \delta^{-2} \eta' \,.
\end{align}
The first pole remains in the lightlike limit $\delta\to 0$, whereas the
second one migrates to infinity.  Since the two poles lie on the same side
of the real axis, the contour deformations needed to avoid the Glauber
region can be made as for lightlike Wilson lines, and one does expect the
Grammer-Yennie approximation to be valid also in this case.  Let us see
this by explicit calculation of the $\ell^+$ integral.  For $\ell^- + a^-
> 0$, all poles lie on the same side of the real axis and one gets
zero. For $\ell^- + a^- < 0$ we close the contour in the lower half plane.
Approximating $(1- \delta^4) \approx 1$, we then get
\begin{align}
  \label{eq:sl1}
I_{\mathrm{SL}} &= - 2\pi i\, \frac{3!\ms N}{D} \int [\df^3 \alpha]\;
  \biggl\{ \int_{- \infty}^{-a^-}
    \frac{\df \ell^-}{\ell^- - i \eta''}\;
    \frac{1}{[2 \delta^2 (\ell^-)^2 + 2 \ell^- ( a^+ +  \delta^2 a^-)
         +  A + i \varepsilon]^4}
\nonumber \\
 & \qquad\quad
 - \int_{- \infty}^{-a^-}
     \frac{\df \ell^-}{\ell^- - i \eta''}\;
     \frac{1}{[2 \delta^{-2} (\ell^-)^2 + 2 \ell^- (a^+ + \delta^{-2} a^-) 
         + A + i \varepsilon]^4}
   \biggr\}
\end{align}
with $\eta'' = \eta' - \delta^2 \eta$.
Our previous argument for limiting ourselves to the generic region of
Feynman parameters remains valid, so that we can use the properties in
\eqref{eq:powercounting}.  With $a^- > 0$ the ambiguous sign of $\eta''$
does not matter since the pole $\ell^-_{\text{eik}} = i \eta''$ of the
eikonal propagator is outside the integration region.

Since $\ell^- \sim Q$ or larger in the integral \eqref{eq:sl1}, the
integrand is strongly power suppressed, as in the case of lightlike Wilson
lines.  It remains to discuss poles close to the integration path.  With
\eqref{eq:powercounting} we find that that poles of the combined Feynman
propagators are approximately at
\begin{align}
\ell^-_1 &\approx - A /(2 a^+) - \text{sgn}(a^+) \ms i \varepsilon \,,
&
\ell^-_2 &\approx - \delta^{-2} a^+ + \text{sgn}(a^+) \ms i \varepsilon
\intertext{for the first term in \protect\eqref{eq:sl1} and at}
\ell^-_3 &\approx - \delta^2 A /(2 a^-) - i \varepsilon \,,
&
\ell^-_4 &\approx - a^- - \delta^2 a^+ + i \varepsilon
\end{align}
for the second term.  With $A/(2 a^+) \sim \Lambda^2/Q$ and $\delta^2 A/(2
a^-)\sim \Lambda^4/Q^3$, the poles at $\ell^-_1$ and $\ell^-_3$ are close
to zero and thus far away from the integration region.  For $a^+ > 0$ the
pole at $\ell^-_2$ (with a huge real part of order $Q^3/\Lambda^2$) is
infinitesimally close to the integration path but can easily be avoided by
contour deformation.

\begin{figure}[t]
\begin{center}
  \subfigure[]{\includegraphics[height=0.25\textwidth]{%
      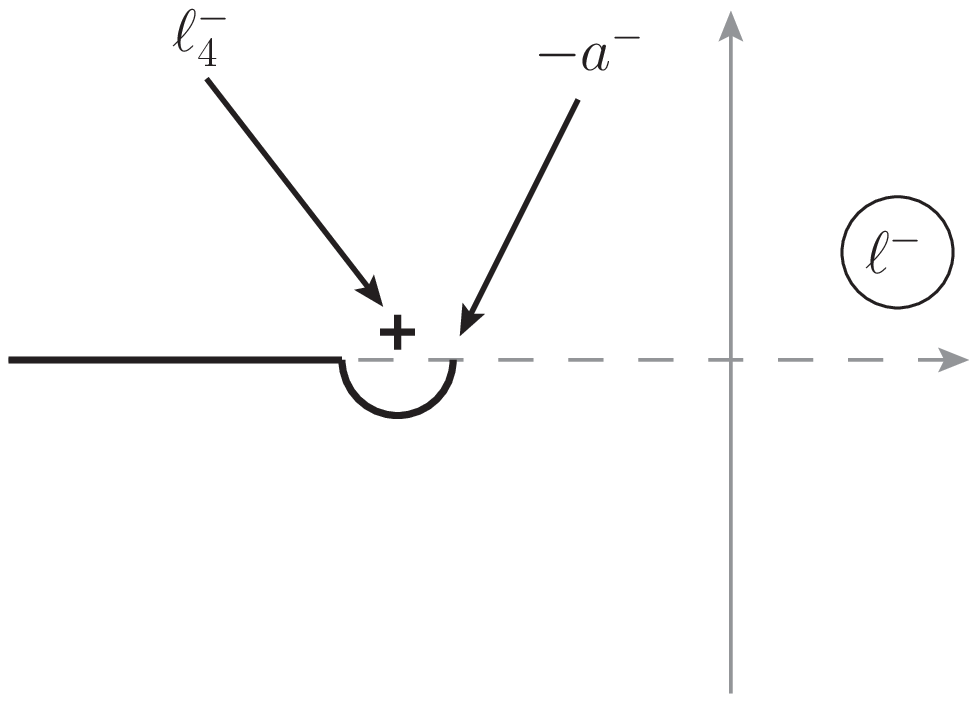}} 
  \phantom{xx}
  \subfigure[]{\includegraphics[height=0.25\textwidth]{%
      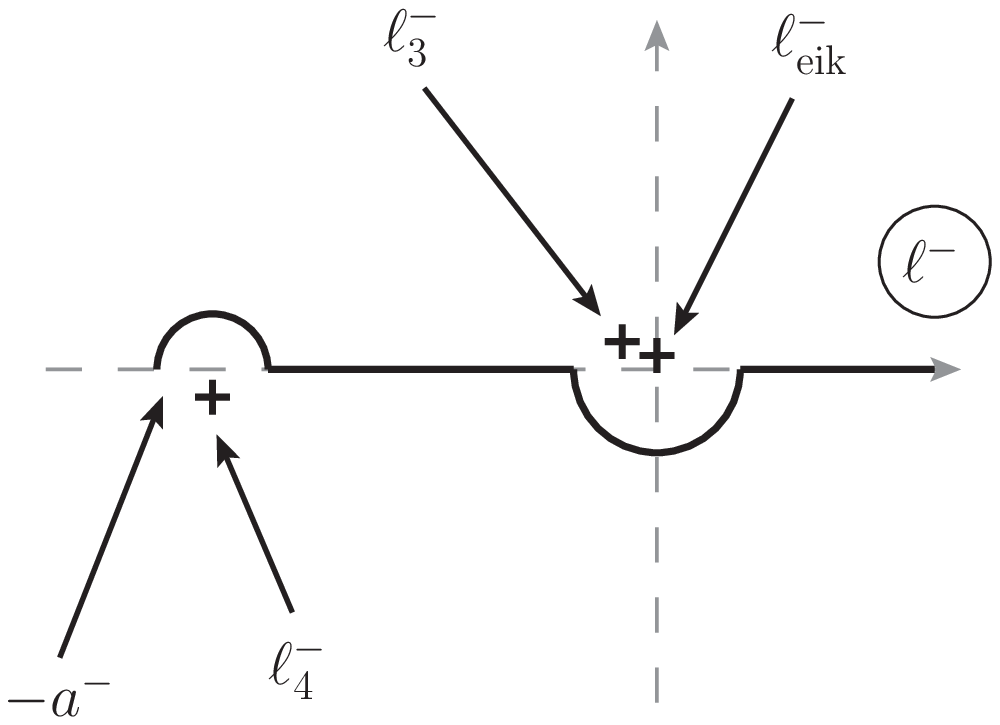}}%
  \caption{\label{fig:poles} Integration paths and nearby poles of the
    second integrals in \protect\eqref{eq:sl1} (panel a) and in
    \protect\eqref{eq:tl1} (panel b).  Both plots are for $a^+ > 0$.}
\end{center}
\end{figure}

The pole at $\ell^-_4$ is located at a short distance of order
$\Lambda^2/Q$ from the endpoint $\ell^- = a^-$ of the integration and
therefore deserves closer examination.  If $a^+ < 0$ then the pole is to
the right of the endpoint, so that the closest point on the integration
path is the endpoint itself.  There we have
\begin{align}
  \label{denom-at-endpoint}
2 \delta^{-2} (\ell^-)^2 + 2 \ell^- (a^+ + \delta^{-2} a^-) + A
 &= - 2 a^+ a^- + A &
\mbox{for $\ell^- = a^-$} \,,
\end{align}
which is of order $Q^2$ and provides a lower limit for the size of the
left hand side on the entire path (i.e.\ an upper limit for the size of
the integrand, where this expression is raised to the fourth power in the
denominator).  This is sufficient to suppress the second integral in
\eqref{eq:sl1}.  If $a^+ > 0$ then the pole is to the left of the
integration endpoint.  It can however be avoided by contour deformation,
and on a semi-circle around $\ell^-_4$ going to the endpoint (see
figure~\ref{fig:poles}a) the upper limit for the denominator based on
\eqref{denom-at-endpoint} remains valid.  Thus the contribution to the
integral is power suppressed, as we anticipated.

\paragraph{Timelike Wilson lines.}
With timelike Wilson lines, the leading numerator factor in the Glauber
region is again given by $N$ in \eqref{abbreviations}.  For the eikonal
propagators, we must make the replacement
\begin{align}
\frac{1}{\ell^- - i \eta'}  \rightarrow 
  \frac{1}{\ell^-  + \delta^2 \ell^+ - i \eta'} \,,
\quad
\frac{1}{\ell^+ + i \eta}   \rightarrow
  \frac{1}{\ell^+  + \delta^2 \ell^- + i \eta} 
\end{align}
in \eqref{eq:ll3}. The $\ell^+$ poles of the  eikonal propagators
\begin{align}
 \ell^+_{\text{eik},1} & = - \delta^2 \ell^- - i  \eta  \,,
&
 \ell^+_{\text{eik},2} & = - \delta^{-2} \ell^- + i \delta^{-2} \eta'
\end{align}
are now on opposite sides of the real axis.

In order to evaluate the $\ell^+$ integral, we close the integration
contour in lower half plane for $\ell^- + a^- < 0$ and take the residue at
$\ell^+ = \ell^+_{\text{eik},1}$, whereas we close it in the upper half
plane for $\ell^- + a^- > 0$ and take the residue at $\ell^+ =
\ell^+_{\text{eik},2}$.  Approximating $(1- \delta^4) \approx 1$ we then
get
\begin{align}
 \label{eq:tl1}
I_{\mathrm{TL}} &= - 2\pi i\, \frac{3!\ms N}{D} \int [\df^3 \alpha]\;
  \biggl\{ \int_{- \infty}^{-a^-}
    \frac{\df \ell^-}{\ell^- - i \eta''}\;
    \frac{1}{[- 2 \delta^2 (\ell^-)^2 + 2 \ell^- ( a^+ -  \delta^2 a^-)
         +  A + i \varepsilon]^4}
\nonumber \\
 & \qquad\quad
 + \int_{-a^-}^{\infty}
   \frac{\df \ell^-}{\ell^- - i \eta''}\;
   \frac{1}{[- 2 \delta^{-2} (\ell^-)^2 + 2 \ell^- (a^+ - \delta^{-2} a^-) 
         + A + i \varepsilon]^4} \,,
   \biggr\}
\end{align}
where now $\eta'' = \eta' + \delta^2 \eta$ is always positive.
In addition to the eikonal pole at
\begin{align}
\ell^-_{\text{eik}} = i \eta''
\end{align}
we now have poles at
\begin{align}
\ell^-_1 &= - A /(2 a^+) - \text{sgn}(a^+)\ms i \varepsilon \,,
&
\ell^-_2 &= \delta^{-2} a^+ + \text{sgn}(a^+)\ms i \varepsilon
\intertext{for the first term in \protect\eqref{eq:tl1} and at}
\ell^-_3 &= \delta^2 A /(2 a^-) + i \varepsilon \,,
&
\ell^-_4 &= - a^- + \delta^2 a^+ - i \varepsilon
\end{align}
for the second term. For the first integral in \eqref{eq:tl1}, the
situation is essentially identical to that for the first integral in
\eqref{eq:sl1} (except that $\ell_2^-$ now has a negative real part for
$a^+ < 0$ rather than $a^+>0$), and we can use the same strategy to avoid
the poles. In the second integral, we have the poles at
$\ell^-_{\text{eik}}$ and $\ell^-_3$ near the origin.  These can be
avoided by contour deformation: a semi-circle around the origin with
radius of order $\Lambda^2/Q$, as in figure~\ref{fig:poles}b, is
sufficient to ensure that the integrand remains power suppressed. Finally,
there is the pole at $\ell_4^-$ near the endpoint of integration.  For
this pole we can use the same logic as in the spacelike case to determine
that it does not result in a leading contribution. In summary then, there
is no leading contribution to $I_{\text{TL}}$, and the factorised
expression for the double box is valid to leading power with timelike
Wilson lines as well.


\subsubsection{Gauge boson vertex correction} 
\label{sec:vertex-graph}

\begin{figure}
\begin{center}
 \includegraphics[width=0.6\textwidth]{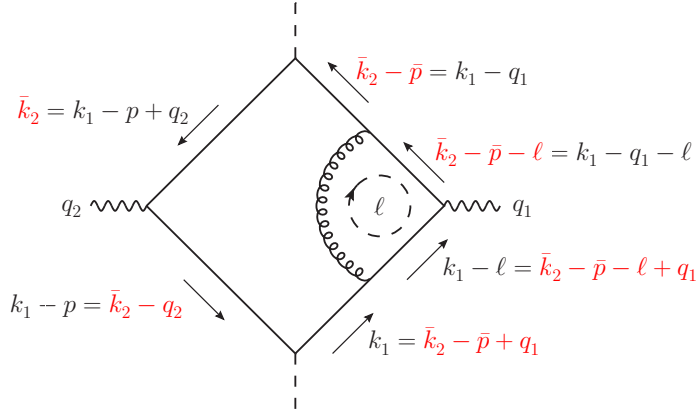}
\caption{\label{fig:photonvertex} Gauge boson vertex correction with
  momentum labellings and our routing of the extra loop momentum $\ell$.}
\end{center}
\end{figure}

We now turn to the vertex correction shown in
figure~\ref{fig:photonvertex}, which is given by
\begin{align}
\Gamma &= \int \df \bar{k}_2^+ \, \df k_1^-\, \df^{2-2\epsilon}\tvec{k}_1^{}
          \int \frac{\df^{4-2\epsilon}\ell}{%
            \ell^2 - \lambda^2 + i \varepsilon}\;
   \frac{(2 k_1 - \ell)_{\mu}\ms g^{\mu\nu} 
        (2 \bar{k}_2 - 2 \bar{p} - \ell)_{\nu}}{%
        [(k_1 - \ell)^2 - m^2 + i \varepsilon]\ms
        [(\bar{k}_2 - \bar{p} - \ell)^2 - m^2 + i \varepsilon]} 
\nonumber \\
& \quad \times \frac{1}{[k_1^2 - m^2 + i \varepsilon]\ms
  [(p - k_1 )^2 - m^2 + i \varepsilon]\ms
  [\bar{k}_2^2 - m^2 + i \varepsilon]\ms
  [(\bar{p} - \bar{k}_2 )^2 - m^2 + i \varepsilon]} \,.
\end{align}
Since the graph has a UV divergence we have kept the loop integrations in
$4-2\epsilon$ dimensions.
The replacement \eqref{prop-replace} now gives
\begin{align} \label{eq:trianglereplacement}
& (2 k_1 - \ell)_{\mu}\ms g^{\mu\nu} (2 \bar{k}_2 - 2 \bar{p} - \ell)_{\nu}
 \,\to\,
 \bigl[ (2 \tvec{\ell}\ms \tvec{k}_1 - \tvec{\ell}^2)  
        (2 \tvec{\ell}\ms \bar{\tvec{k}}_2 - \tvec{\ell}^2)
\nonumber \\[0.1em]
& \qquad 
  - 2 \ell^+ k_1^- (2 \tvec{\ell}\ms \bar{\tvec{k}}_2 - \tvec{\ell}^2)
  - 2 \ell^- (\bar{k}_2^+ - \bar{p}^+)
    (2 \tvec{\ell}\ms \tvec{k}_1 - \tvec{\ell}^2)
  - \ell^+ \ell^- \big\{ 4 (\tvec{k}_1 - \tvec{\ell})
                  (\bar{\tvec{k}}_2 - \tvec{\ell}) - \tvec{\ell}^2 \big\}
\nonumber \\
& \qquad
 + 2 \ell^+ \ell^- (2k_1^- - \ell^-)
     (2\bar{k}_2^+ - 2\bar{p}^+ - \ell^+) \bigr]
  \,\big/\,  \bigl[ (\ell^- - i \eta') (\ell^+ + i \eta) \bigr] \,.
\end{align}
As in the double box case, the leading term on the right hand side of
\eq{trianglereplacement} is the one with only transverse components in the
numerator (when we are in the Glauber region). Thus, we drop all terms
apart from this one.

{
\allowdisplaybreaks[4]
Combining the denominators of the gluon propagator and the scalar parton
propagators containing $\ell$ with the help of Feynman parameters, we get
\begin{align}
I &= \frac{2!\ms N}{D} \int [\df^3 \alpha]
  \int \frac{\df \ell^-}{\ell^- - i \eta'} 
  \int \frac{\df \ell^+}{\ell^+ + i \eta} \;
  \frac{1}{[2 \ell^+ \ell^-
   + 2 a^- \ell^+  + 2 a^+ \ell^-  + A + i \varepsilon]^3}
\end{align}
with
\begin{align}
  \label{abbrev-vertex}
A &=  2 (\alpha_1^{} q_1^+ - \alpha_1^{} \bar{p}^+
        - \alpha_2^{} \bar{p}^+)\ms k_1^-
    - 2 \alpha_2^{} \bar{k}_2^+ q_1^-
    + 2 (\alpha_1^{}+\alpha_2^{}) \bar{k}_2^+ k_1^-
    + 2 \alpha_2^{} \bar{p}^+ q_1^-
\nonumber \\
  &\quad
    - \alpha_1^{} (\tvec{k}_1^{} - \tvec{\ell})^2
    - \alpha_2^{} (\bar{\tvec{k}}_2^{} - \tvec{\ell})^{2}
    - \alpha_3^{} (\tvec{\ell}^2 + \lambda^2)
    - (\alpha_1^{}+\alpha_2^{})\ms m^2 \,,
\nonumber \\
a^+ &= - \alpha_1^{} q_1^+
       + (\alpha_1^{} + \alpha_2^{}) (\bar{p}^+ - \bar{k}_2^+) \,,
\nonumber \\
a^- &= \alpha_2^{} q_1^- - (\alpha_1^{} + \alpha_2^{}) k_1^- \,,
\nonumber \\
N &=  (2 \tvec{\ell}\ms \tvec{k}_1 - \tvec{\ell}^2)  
            (2 \tvec{\ell}\ms \bar{\tvec{k}}_2 - \tvec{\ell}^2) \,,
\nonumber \\
D &= [k_1^2 - m^2 + i \varepsilon] [(p- k_1 )^2 - m^2 + i \varepsilon]\ms
     [\bar{k}_2^2 - m^2 + i \varepsilon]\ms
     [(\bar{p} - \bar{k}_2 )^2 - m^2 + i \varepsilon] \,.
\end{align}
This has the same structure as the expression \eqref{eq:ll3} for the
double box, except that the term with $A$ appears to the third instead of
the fourth power, which was inessential for reaching our conclusion that
\eqref{eq:ll3} was power suppressed.  One can again show that only the
generic region of Feynman parameters can potentially give a leading
contribution, and one finds that the quantities in \eqref{abbrev-vertex}
fulfil the same conditions as stated in \eqref{eq:powercounting} for the
double box.  In addition, $a^+ < 0$ for the vertex correction, whereas it
can have both signs in the double box case.  Hence our arguments regarding
factorisation carry over to the vertex correction graph, for lightlike,
spacelike and timelike Wilson lines.  }


\subsection{General prescription for avoiding the Glauber region}
\label{sec:general-arg}

Let us return to the general case of one gluon exchange in the model
introduced in section \ref{sec:scope-and-method}. As illustrated in
section \ref{sec:alteroute}, we can avoid pinched poles in the Glauber
region in most cases by a judicious separate routing of the plus and minus
components of the gluon momentum $\ell$ through the graph.

To simplify the proof that Glauber gluons do not inhibit factorisation, we
route $\ell^-$ such that it flows in the same direction as the positive
plus momentum of right moving lines whenever possible. In the example of
figure~\ref{fig:spect-model}d we have discussed before, one then routes
$\ell^-$ to flow through the active parton line with momentum $k_2$ rather
than through the line with momentum $k_1$.  Whenever such a routing is
possible for a graph, all poles in $\ell^-$ from the lower part of the
graph are on the same side of the real axis, and one can deform the
integration to the other side, out of the Glauber region to a region where
$|\ell^-| \sim |\tvec{\ell}| \sim \Lambda$.  There are no poles in the
small $\ell^-$ region from the upper part of the graph, where lines are
left moving and have much larger minus momenta. The gluon propagator has
its pole at $2\ell^+\ell^- = \vect{\ell}^2 + \lambda^2$ and thus does not
inhibit the contour deformation just described.  As already discussed, the
spacelike Wilson lines specified in and below \eqref{universal-GY} are
designed to have poles on the same side of the real axis as the
propagators of the right-moving lines and thus do not obstruct deforming
$\ell^-$ back to the real axis after the Grammer-Yennie approximation has
been made.  This is not the case for timelike Wilson lines, whose use
therefore requires additional justification.  We will show in
appendix~\ref{app:timelike} that at the one-gluon exchange level and at
leading-power accuracy they give indeed the same result for the double
Drell-Yan cross section as their spacelike counterparts, generalising what
we found for the specific graphs computed in section~\ref{sec:box-graphs}.

The preceding argument can be repeated to avoid pinched poles in the small
$\ell^+$ region.  Whenever the momentum routing just described is possible
for either $\ell^-$ or $\ell^+$, we can thus perform the Grammer-Yennie
approximation at least on one side of the graph, so that at least one of
the factors in parentheses in \eqref{master-replace} gives a power
suppression.  This is sufficient to ensure that \eqref{region-sum} is a
good approximation throughout the combined collinear and soft momentum
regions, even if one of the $\ell^-$ and $\ell^+$ integrations is trapped
by a pinch in the region $\Lambda^2/Q$, as is the case for
figure~\ref{fig:spect-model}c and~\ref{fig:spect-model}e.\footnote{%
  An analogous argument was used to show factorisation for semi-inclusive
  deep inelastic scattering, where one light-cone component of $\ell$ (but
  not the other) can always be routed to flow only against large collinear
  momenta in the graph, see chapter 12.14.3 of \cite{Collins:2011zzd}.}

It remains to consider the graphs where neither $\ell^-$ nor $\ell^+$ can
be routed in a way to avoid pinched poles.  In the model we are
discussing, this is only the case if both in the upper and in the lower
part of the graph the gluon attaches to a spectator parton that either
goes directly into the final state or only splits into spectator partons
that go into the final state. An example of such a graph is
figure~\ref{fig:spect-model}f.  For these graphs one can use exactly the
same argument as in single Drell-Yan production (reviewed for instance in
\cite{Gaunt:2014ska}).  Deforming the $\ell^+$ and $\ell^-$ contours away
from the poles of the active partons, one crosses poles of spectator lines.
The residues of these poles cancel by a unitarity argument when, for a
given graph of the cross section, one sums over all possible final-state
cuts.


\subsection{Limitations of the present method} 
\label{sec:limitations}

The arguments of the previous section can be rather easily generalised to
the exchange of several gluons in the same model setting, as long as these
gluons do not interact with each other.  However there exist difficulties
in generalising this proof to one that works at all orders in perturbation
theory. These difficulties appear already for single Drell-Yan production.

\begin{figure}
\begin{center}
\includegraphics[height=0.23\textwidth]{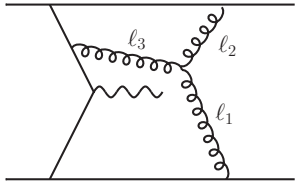}
\caption{\label{fig:limitations} A graph for which the methods of section
  \ref{sec:one-gluon}, in particular the use of
  \protect\eqref{master-replace}, cannot be generalised in an obvious
  manner.}
\end{center}
\end{figure}

As an example consider figure~\ref{fig:limitations}. To construct an
approximation for the combined collinear and soft regions, several
different regions must be considered: in fact there are leading
contributions with three, one or none of the momenta $\ell_1$, $\ell_2$
and $\ell_3$ being soft.  The required approximators and subtraction terms
for these regions do not obviously sum up to give a simple form like
\eqref{master-replace}, which was a key part of our present argument.
For the proof of Glauber gluon cancellation at all orders in the next
section, a different method will therefore be used.

\section{All-order proof of Glauber gluon cancellation}
\label{sec:allorder}

In this section we demonstrate the cancellation of Glauber gluons in
double Drell-Yan production to all orders in QCD perturbation theory.  We
make use of the same techniques that were used to demonstrate the
all-order cancellation of Glauber gluons in the single Drell-Yan process
\cite{Collins:1988ig, Collins:2011zzd} -- in particular, we use the
light-front ordered version of QCD perturbation theory (LCPT).

A subtle point is that the approximations needed to obtain factorisation
are valid in specific regions of the loop momenta, whereas the arguments
that establish the possibility of contour deformations out of the Glauber
region require integration over the \emph{full} range of certain
light-cone components of these momenta.  We therefore need to refine
steps \ref{enum:approximations} and \ref{enum:subtractions} in the general
overview of the proof given in section \ref{sec:overall-strategy}.  This
applies both to single and double Drell-Yan production.
\begin{enumerate}
\item \label{step:1st-approx} The expression of a given graph for the
  double Drell-Yan process is decomposed into a sum over all possible
  regions as in \eqref{subtr-sum}.  For each region we make the
  appropriate kinematic approximations specified in point \ref{enum:kin}
  of section \ref{sec:overall-strategy}, as well as the approximations for
  fermion lines in point \ref{enum:Fierz}.  These approximations are valid
  for all soft gluon momenta, including those in the Glauber region.  In
  accordance with the subtraction formalism, they are made not only in the
  approximant $C_R\ms \Gamma$ for their design region, but also in the
  subtraction terms of that region in approximants for bigger regions.
  The sum $\sum_R C_R\ms \Gamma$, with each term being integrated over the
  full range of all loop momenta, then gives a correct approximation of
  the original graph.

  At this stage we do not yet make the Grammer-Yennie approximations
  necessary for the application of Ward identities.  Before doing so, we
  need to take care of soft gluon momenta in the Glauber region.  We could
  already apply the Grammer-Yennie approximation \eqref{GY-collin} for
  collinear gluons and the associated Ward identity, but we refrain from
  doing so.  This avoids complications of the contour deformation argument
  in the case where the collinear Wilson lines are not strictly lightlike
  (cf.\ our comment in section~\ref{sec:long-pol-coll}).

\item \label{step:sum-cuts} For a given graph $\Gamma$, approximated for a
  given region $R$, we consider the sum over all possible cuts,
  $G_R = \sum_{\text{cuts}} C_R\ms \Gamma$.  We partition the graph into the
  collinear factor $A$ and a remainder factor,
\begin{align}
G_R &= \sum_{\text{cuts}} \int \Bigl[ \prod_j \df^{4-2\epsilon}
  \ell_j \Bigr]\,
  A^{\mu_1 \cdots \mu_n}(\tilde{\ell}_j)\ms R_{\mu_1 \cdots \mu_n}(\ell_j) \,,
\end{align}
  where the $\ell_j$ are the soft momenta entering $A$ from the soft
  subgraph and $\tilde{\ell}_j$ is defined in \eqref{soft-mom-approx}.
  For simplicity, we suppress other momentum arguments in $A$ and $R$, as
  well as the associated integrations.\footnote{To conform with the
    original literature \protect\cite{Collins:1988ig,Collins:2011zzd} we
    call the remainder factor $R(\ell_j)$.  This should of course not be
    confused with the momentum region $R$.}

  The following arguments can all be made at fixed values of transverse
  parton momenta.  Using LCPT we show that, whilst individual cut graphs
  in $G_R$ may have pinched poles at small $\ell_j^-$, the sum over all
  cuts $G_R$ is free from such pinches, and all poles in $\ell_j^-$ that
  originate from $A(\tilde{\ell_j})$ are in the lower half plane.  This
  step is rather involved, and we discuss it in detail in the following
  subsections.

\item \label{step:soft-poles} The amount to which the $\ell_j^-$
  integration can be deformed into the upper half plane is determined by
  the poles of the remainder factor $R(\ell_j)$.  In the unapproximated
  graph, each $\ell_j^-$ is routed to flow through $S$ into $B$, and from
  there back via $H_1$ or $H_2$ into $A$.  After the approximations in
  step \ref{step:1st-approx}, $\ell_j^-$ is set to zero in $B$ and the
  hard subgraphs, so that we only need to worry about the poles of
  $S(\ell_j)$.  For $|\ell_j^{\smash +}| \lsim |\tvec{\ell}_j^{}|$ the
  propagator of the gluon with momentum $\ell_j$ has its pole at
  $|\ell_j^{\smash -}| \gsim |\tvec{\ell}_j^{}|$, and we assume that one
  can route $\ell_j^-$ through $S$ in a way that the same is true for the
  full soft factor.  It would be desirable to have a formal proof that
  this is always possible in momentum regions that give a leading
  contribution to the cross section.  One can then deform the integration
  contour of each soft momentum $\ell_j^-$ out of the Glauber region,
  e.g.\ to a semicircle around the origin in the upper half plane, with
  radius of order $|\tvec{\ell}_j|$.

\item \label{step:subtr} The approximator $C_R$ for a region includes
  subtraction terms for smaller regions $R'$.  According to
  \eqref{subtr-def}, the approximations for both regions $R$ and $R'$ are
  applied to a graph in these terms.  Consider a region $R'$ with a soft
  gluon momentum $\ell_j$ that is collinear or hard in $R$.  Some graphs,
  such as the one in figure~\ref{fig:subt-terms}a are leading in $R$ but
  subleading in $R'$.  It is advantageous to keep the corresponding
  subtraction terms nevertheless, so that one can apply exact Ward
  identities appropriate for the region $R$ at later stages.  In such
  graphs we can make the replacement \eqref{GY-soft} without loss of
  accuracy for the cross section.  For graphs that are leading in $R'$,
  such as the one in figure~\ref{fig:subt-terms}b, we must show that
  $\ell_j$ can be deformed out of the Glauber region, after which
  \eqref{GY-soft} is indeed a good approximation.  One can treat the
  subtraction term for region $R'$ just as the original term for that
  region and follow steps \ref{step:sum-cuts} and \ref{step:soft-poles}
  above.  In the subtraction term, additional approximations are made: in
  figure~\ref{fig:subt-terms}b the transverse gluon momentum
  $\tvec{\ell}_j$ is neglected inside $A'$ and the masses of lines that
  are in $A'$ but not in $A$ are put to zero.  These additional
  approximations do not affect the arguments for contour deformation, and
  we can thus deform the $\ell^-_j$ integration to a semicircle around the
  origin in the upper half plane, with radius of order $|\tvec{\ell}_j|$.

\begin{figure}
\begin{center}
\subfigure[]{\includegraphics[width=0.48\textwidth]{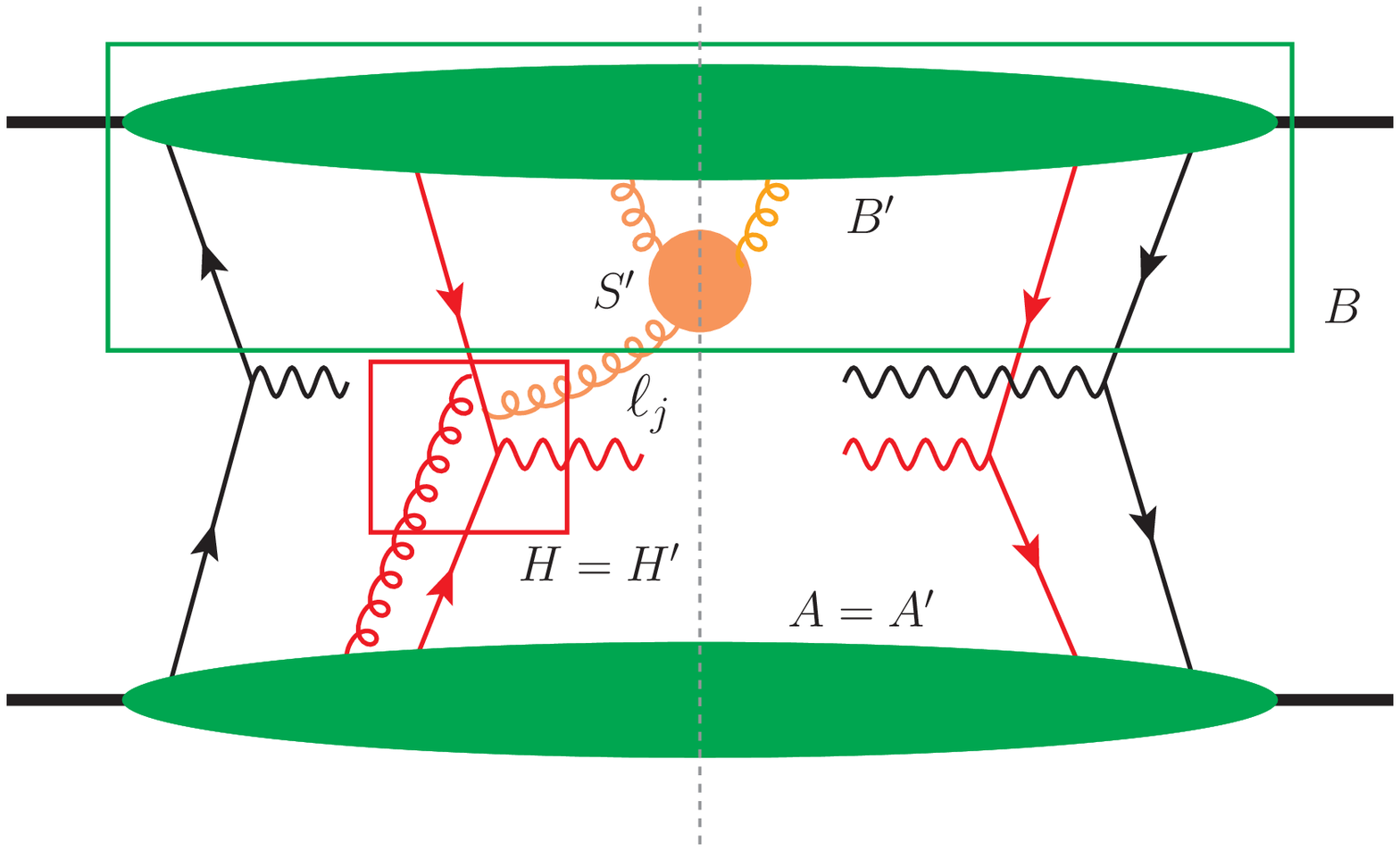}}
\phantom{x}
\subfigure[]{\includegraphics[width=0.48\textwidth]{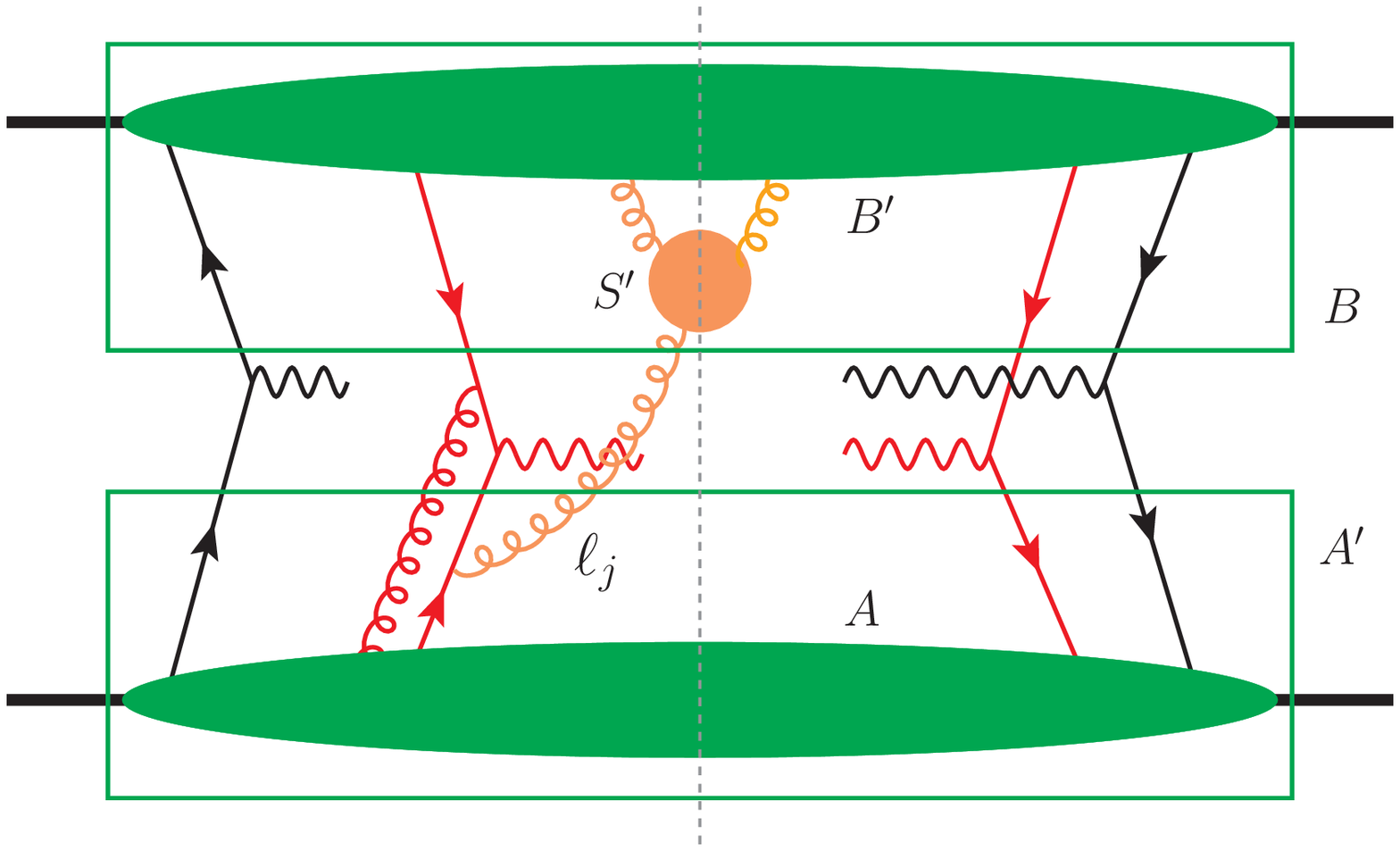}}
\caption{\label{fig:subt-terms} Subtraction terms for a gluon with
  momentum $\ell_j$ that is collinear in region $R$ and soft in region
  $R'$.  The corresponding decomposition into subgraphs $A$, $B$, $H$ or
  $A'$, $B'$, $S'$, $H'$ is indicated (to avoid clutter, $H$ and $H'$ are
  not shown in graph b).  Graph a is power suppressed in region $R'$
  because a soft gluon enters a hard subgraph, whereas graph b is
  leading.}
\end{center}
\end{figure}

\item We repeat steps \ref{step:sum-cuts} to \ref{step:subtr} for soft
  gluons connecting $B$ with $S$, and for subtraction terms for hard or
  collinear gluons that become soft in a smaller region.  The roles of
  plus and minus momenta are now interchanged.  A difference to the
  discussion above is that integrations over the minus components of soft
  gluon momenta in $A$ and $S$ are already deformed, as specified above.
  This does not affect the possibility to deform the integration over the
  plus components of soft gluon momenta in $B$ and $S$.

\item All soft gluon momenta are now deformed out of the Glauber region
  into a region where $\ell_j^+ \sim \ell_j^- \sim |\tvec{\ell}_j^{}|$.
  One can now apply all Grammer-Yennie approximations for soft gluons (see
  points~\ref{enum:GY} and \ref{enum:Glauber} in
  section~\ref{sec:overall-strategy}), as well as all Grammer-Yennie
  approximations for collinear gluons.  This is done in all terms,
  including subtraction terms.  The approximations are valid in their
  design regions; their possible failure in larger regions does not
  degrade the approximation of the cross section because within power law
  accuracy the contribution from these regions is removed by appropriate
  subtraction terms.

  Note that the discussion of subtractions in
  point~\ref{enum:subtractions} of section~\ref{sec:overall-strategy} was
  at the level of individual cut graphs and completely local in momentum
  space.  We now consider the sum over all cuts of a graph, which does not
  pose any problems.  Moreover, the subtractions for soft gluons now work
  only for integrals over $\ell^-_j$ and $\ell^+_j$.  This is because we
  have deformed integration contours in the subtraction term
  $T_R\ms C_{R'} \Gamma$ for a region $R'$ where a gluon is soft, but not
  in the term $T_R\ms \Gamma$ where the corresponding gluon is designated
  as collinear or hard.  The subtraction works correctly for the integral
  because in the integration region where $\ell_j$ is soft, the
  unsubtracted graph $T_R\ms \Gamma$ is correctly approximated by
  $T_R\ms C_{R'} \Gamma$, and for the approximated term we have shown that
  the contour can be deformed.
\end{enumerate}
Having made all Grammer-Yennie approximations, one proceeds with the
factorisation proof as described in section~\ref{sec:overall-strategy},
starting with the deformation of all integration contours back to the real
axis.

In the remainder of this section we give a detailed description of
step~\ref{step:sum-cuts}.  Since LCPT may not be familiar to all readers,
we give a brief summary of the rules of LCPT in section
\ref{sec:LCPTrules}.  We derive these rules from conventional perturbation
theory in section \ref{sec:LCPTderivation}, since an understanding of this
derivation turns out to be important when we consider how to treat the
collinear partons from a proton that enter the hard interactions.  In
section \ref{sec:Step2Details} we present the details of the sum-over-cuts
argument in step~\ref{step:sum-cuts}, starting with a review of the
corresponding procedure for single Drell-Yan production given in
\cite{Collins:1988ig, Collins:2011zzd}.


\subsection{Light-cone perturbation theory (LCPT)}
\label{sec:LCPTrules}

LCPT is a method for computing Green functions or $S$ matrix elements in
which one sums diagrams over all possible orderings of the vertices in the
``light-cone time'' $x^+ = (x^0 + x^3)/\sqrt{2}$.  It is somewhat similar
to old-fashioned time ordered perturbation theory (discussion of which can
e.g.\ be found in \cite{Sterman:1995fz, StermanBook, ThomsonBook}), where
vertices are ordered according to the usual time variable $t = x^0$.

LCPT ultimately gives the same results as conventional covariant
perturbation theory and indeed may be derived from it
\cite{Ligterink:1994tm, Collins:2011zzd}, as we show in the next section.
We here give the rules for LCPT following the conventions of
\cite{Collins:2011zzd}.  The rules may also be found in
\cite{Chang:1968bh, Kogut:1969xa, Yan:1973qg, Mueller:1989hs,
  Zhang:1993dd, KovchegovQCD}, albeit with differing normalisation
conventions.
\begin{enumerate}
\item The diagrams in LCPT are like the diagrams in covariant perturbation
  theory, except that vertices are ordered in the light-cone time
  $x^+$. We sum over all possible orderings in $x^+$ as well as all
  possible graphs. In diagrams we take $x^+$ to increase from left to
  right in the amplitude (i.e.\ to the left of the final state cut) and to
  decrease from left to right in the conjugate amplitude.
\item We assign each line $L$ an on-shell four-momentum $\kappa_L^{}$
  satisfying $\kappa_L^2 = m_L^2$. The plus and transverse components of
  these four-momenta are conserved at vertices, but the minus components
  are not -- these are instead fixed by the on-shell condition, such that
  \begin{equation} \label{on-shell-mom}
    \kappa_L^{} = \left( \kappa_L^+ , \kappa^-_{L}
      = \dfrac{\tvec{\kappa}_L^2+m_L^2}{2\kappa_L^+}, \tvec{\kappa}_L^{}
        \right) \,.
  \end{equation}
\item \label{rule:theta} For each line $L$ we include a factor
  $\frac{1}{2\kappa_L^+} \,\Theta(\kappa_L^+)$, where $\kappa_L^+$ flows
  in the future light-cone time direction.
\item For each loop, we integrate only over the plus and transverse
  components of the loop momentum with measure ${(2\pi)^{1-d}} \int {\df
    \kappa^+ \df^{d-2}\, \tvec{\kappa}}$, where $d=4 - 2\epsilon$.
\item For each intermediate state $\xi$ (that is, set of lines between
  two vertex positions in the light-cone time ordering), we have a factor
  \begin{equation}
    \label{eq:LCdenomdef}
    \dfrac{i}{p_{\xi, \text{inc}}^- - 
      \sum\limits_{L \in \xi}
        \kappa_{L}^- + i\epsilon} \,,
  \end{equation}
  where $p_{\xi, \text{inc}}^-$ is the total external minus momentum
  entering the state from the left (i.e.\ from lower $x^+$) and
  $\kappa_{L}^-$ is the on-shell minus momentum of a line $L$ in $\xi$.
\item Coupling factors at vertices and symmetry factors are the same as
  in Feynman graphs.
\end{enumerate}
Note that strictly speaking these are only the rules for scalar
theories. For theories involving fermion and/or vector boson fields (such
as QCD and QED), the LCPT formulation contains extra vertices associated
with particle exchange(s) that are instantaneous from the point of view of
the ordering in $x^+$ (see e.g. \cite{Kogut:1969xa, Mueller:1989hs,
  Zhang:1993dd}). These vertices are somewhat similar to the new vertices
one encounters connecting collinear particles in SCET
\cite{Bauer:2000yr,Bauer:2001yt}. We will not need the precise structure
of the vertices and hence do not give the corresponding rules here.


\subsection{Derivation of LCPT from covariant perturbation theory}
\label{sec:LCPTderivation}

We now briefly review the derivation of the LCPT rules from covariant
perturbation theory.  We follow section 7.2.3 of \cite{Collins:2011zzd},
although we use a slightly different notation here.  The steps to obtain
the light-front ordered diagrams from a given covariant Feynman graph are
as follows.
\begin{enumerate}
\item Take a covariant Feynman graph calculated according to the usual
  momentum-space rules.
\item Order the vertices $i=1,\ldots,n$. Assign momentum $\kappa_{ij}$ to
  an internal line flowing from vertex $i$ to $j$, and momentum $p_i$ to
  an external line flowing into the graph at vertex $i$.
\item \label{deriv:delta} At each vertex we have a $\delta$ function
  corresponding to the conservation of four-momentum.  Rewrite the minus
  component part of this $\delta$ function as follows:
  \begin{align}
    2\pi \delta\biggl( p_i^- +
    \sum_{j\neq i} \kappa_{ji}^- - \sum_{j\neq i} \kappa_{ij}^- \biggr)
    &= \int \df x_i^+\, \exp\Biggl[ -i x_i^+
       \biggl( p_i^- + \sum_{j\neq i} \kappa_{ji}^- 
               - \sum_{j\neq i} \kappa_{ij}^- \biggr) \Biggr]  \,.
  \end{align}
\item For each vertex there is now an $x_i^+$ which is integrated over the
  full range. Partition the multiple integral over all $x_i^+$ into a sum
  over possible orderings of the $x_i^+$ as
  \begin{align}
    \int \prod_{i=1}^n \df x_i^+
    &= \sum_{\pi} \int \biggl[\, \prod_{i=1}^n \df x_i^+ \biggr] \;
    \Theta\bigl( x_{\pi(n)}^+ - x_{\pi(n-1)}^+ \bigr) \cdots
    \Theta\bigl( x_{\pi(2)}^+ - x_{\pi(1)}^+ \bigr) \,,
  \end{align}
  where we have a sum over all permutations $\pi$ of the $n$ indices,
  i.e.\ over all orderings $x_{\pi(1)}^+ < x_{\pi(2)}^+ < \cdots <
  x_{\pi(n)}^+$ of the light-cone times.
\item For the ordering $i < j$ of vertices use the momentum $\kappa_{ij}$
  rather than $\kappa_{ji}$.
\item For simplicity of notation we now consider only the ordering where
  $x_1^+ < x_2^+ < \cdots < x_n^+$.  For the $\Theta$ functions use the
  momentum-space representation
  \begin{align} \label{eq:thetamta}
    \Theta\bigl( x_{i+1}^+ - x_{i}^+ \bigr)
     &= \int \frac{\df\rho_i^-}{2\pi}\; \frac{i}{\rho_i^- + i\epsilon}\,
          \exp\bigl[ -i (x_{i+1}^+ - x_{i}^+)\ms \rho_i^- \bigr] \,.
  \end{align}
  Performing the integrals over $x_i^+$ gives back a ``momentum
  conservation'' $\delta$ function $\delta\bigl( p_i^- + \sum_{j<i}
  \kappa_{ji}^- - \sum_{j>i} \kappa_{ij}^- - \rho_i^- + \rho^-_{i-1}
  \bigr)$ at the vertex $i$, although this $\delta$ function now includes
  the fictitious momenta $\rho_i^-$ and $\rho_{i-1}^-$.  Using the
  $\delta$ functions to perform the $\rho_i^-$ integrals, we generate the
  ``light-cone energy denominators'' in \eq{LCdenomdef} from the
  denominator factors in \eq{thetamta}:
  \begin{align} \label{eq:denomv1}
    \frac{i}{\rho_i^- + i\epsilon} &=
    \frac{i}{\sum\limits_{j\le i} p_j^-
      - \sum\limits_{j\le i,\, i<k} \kappa_{jk}^- + i\epsilon} \,.
  \end{align}
  This can be shown by induction over $i$, starting at the first
  vertex. The factors in \eq{denomv1} are not yet quite those in
  \eq{LCdenomdef}, due to the fact that the $\kappa^-_{jk}$ are not yet on
  shell.
\item \label{deriv:vertex} Integrate over the minus momenta
  $\kappa_{ij}^-$ of internal lines. There are exactly two places where
  $\kappa_{ij}^-$ occurs: in the light-cone energy denominators, where it
  appears always as $-\kappa_{ij}^- + i\epsilon$, and in the Feynman
  propagator for the appropriate internal line:
  \begin{align} \label{eq:lineprop}
    \frac{i}{2\kappa_{ij}^+\ms \kappa_{ij}^-
        - \tvec{\kappa}_{ij}^2 - m_{ij}^2 + i\epsilon} \,.
  \end{align}

  From this we see that we get a nonzero integral only for
  $\kappa_{ij}^+ > 0$ (this is the origin of the $\Theta$ function in
  rule~\ref{rule:theta} above).  For $\kappa_{ij}^+ > 0$ pick up the pole
  of the propagator denominator, which sets the $\kappa_{ij}^-$ to its
  on-shell value
  \begin{align} \label{eq:OSminus}
    \kappa_{ij,\, \text{os}}^-
    &= \frac{\tvec{\kappa}_{ij}^2 + m_{ij}^2}{2\kappa_{ij}^+}
  \end{align}
  in the energy denominators \eqref{eq:denomv1}.  Note that when
  performing the $\kappa_{ij}^-$ integral we get a factor
  $1/(2\kappa_{ij}^+)$ from the residue of \eq{lineprop}, which is the
  denominator factor in rule~\ref{rule:theta} above.
\end{enumerate}


\subsection{Absence of pinched poles in the Glauber region}
\label{sec:Step2Details}

Having reviewed LCPT, we now show that in the sum over all cuts of a graph
there are no pinched poles in the Glauber region, as announced in
step~\ref{step:sum-cuts} at the beginning of section \ref{sec:allorder}.
We first show how this step works for the single Drell-Yan process, since
the generalisation to double Drell-Yan production turns out to be a small
step on top of this.

\subsubsection{Single Drell-Yan}
\label{sec:single-DY}

\begin{figure}
  \centering
  \includegraphics[scale=0.7]{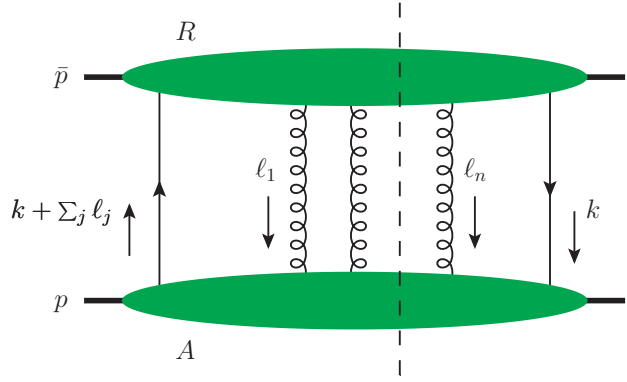}
  \caption{\label{fig:CARpartition} Partitioning of a leading graph and
    region in single Drell-Yan production into a collinear factor $A$ and
    the remainder $R$.  The solid lines are collinear quarks (going from
    $A$ to $H$), whilst the gluons are soft lines (going from $S$ to
    $A$).}
\end{figure}

Let us take a closer look at the sum over cuts $G_R$ for an approximated
graph, which as already discussed can be written as the convolution of the
collinear factor $A$ with a remainder factor $R$, which contains all other
subgraphs.  We start with the case depicted in
figure~\ref{fig:CARpartition}, where we have two physically polarised
partons joining $A$ to $H$ (one in the amplitude, and one in the
conjugate), but no longitudinally polarised gluons.  We postpone the
discussion of the general case with longitudinally polarised gluon
attachments to section~\ref{sec:long-pol-coll}.  The other gluons
exchanged between $R$ and $A$ in figure \ref{fig:CARpartition} are soft
and may be in the Glauber region.

The sum over cuts in $G_R$ is organised as follows.  We consider the
vertices at which the soft gluons enter $A$, and let $V$ denote the
partitioning of these vertices between the amplitude and its conjugate.
For a given $V$, we sum over the set $\mathcal{A}(V)$ of compatible cuts
of $A$, and over the set $\mathcal{R}(V)$ of compatible cuts of $R$.  Then
we sum over all $V$ to get the full set of cuts. Explicitly this gives
\begin{align}
  \label{eq:GLapprox}
G_R &= \int  \dfrac{\df k^+\,
               \df^{d-2}\vect{k}_{}^{\phantom{j}}}{(2\pi)^{d-1}}
   \int \biggl[\, \prod_j \dfrac{\df{\ell}^-_j\,
                      \df^{d-2}\tvec{\ell}_j^{}}{(2\pi)^{d-1}} \biggr] \;\;
     \sum_{V} \sum_{F_A \in \mathcal{A}(V)}
     \int \dfrac{\df k^-}{2\pi}\,
       A_{F_A}^{\mu_1 \cdots \mu_n}(k,\tilde{\ell} _j) 
\nonumber \\
  & \quad \times \sum_{F_R \in \mathcal{R}(V)}
    \int \biggl[\, \prod_j \dfrac{\df\ell_j^+}{2\pi} \biggr]\,
        R^{}_{F_R,\, \mu_1 \cdots \mu_n}(k^+,\vect{k},\ell _j) \,.
\end{align}
Here $A_{F_A}$ ($R_{F_R}$) denotes $A$ ($R$) with the cut $F_A$ ($F_R$).
A crucial point to note is that $A$ is defined to include the propagators
for the external collinear lines with momenta $k + \sum_j \ell_j$ and $k$,
but to exclude the propagators for the external soft lines $\ell_j$, which
are included in $R$.  For simplicity, we shall henceforth omit the Lorentz
indices in $A^{\mu_1 \cdots \mu_n}$ and $R_{\mu_1 \cdots \mu_n}$ and
simply write $A$ and $R$.

In section~\ref{sec:R-no-V} we will show that the factor on the second
line of \eq{GLapprox} is independent of the partitioning $V$.  The sum
over $V$ then only applies to $A$ and gives
\begin{align}
  \label{eq:sumovcuts}
\sum_{V} \sum_{F_A \in \mathcal{A}(V)}
  \int \dfrac{\df k^-}{2\pi}\, A_{F_A}(k,\tilde{\ell} _j) 
  &= \sum_{\text{all }F_A}
     \int \dfrac{\df k^-}{2\pi}\, A_{F_A}(k,\tilde{\ell} _j) \,.
\end{align}
We now consider this expression in LCPT, with light-cone time $x^+$.
Since we do not include the propagators of the soft lines $\ell_j$ in $A$,
we treat these as external lines, with an external momentum $\ell_j$ being
injected at the vertex where the soft line couples to a collinear line
inside $A$.  By contrast, we do include the propagators of the collinear
lines, so we treat these as lines inside $A$ that terminate on a two-point
vertex, whose other line is an external line that carries the collinear
momentum away (and whose associated coupling constant is unity).  These
vertices participate in the light-cone time ordering in the same way as
the other vertices in $A$.  We call them ``hard vertices'' because in the
graph for the physical process they are replaced by vertices at which the
collinear lines enter the hard scattering.
To implement this interpretation we rewrite the integration over $k^-$ as
\begin{equation}
\label{eq:sdy-vertex}
  \textstyle
  \int \df k^-
  \int \df \kappa_i^-\; \delta\bigl( k^- + \sum_j \ell_j^- - \kappa_i^- \bigr) 
  \int \df \kappa_{i'}^-\; \delta\bigl( k^- - \kappa_{i'}^- \bigr) \,,
\end{equation}
treating $\kappa_i$ ($\kappa_{i'}$) as the momentum of the internal
collinear line at the left (right) of the final state cut and the
corresponding $\delta$ functions as associated with the two-point vertices,
as needed in step~\ref{deriv:delta} of the derivation of the LCPT rules in
section \ref{sec:LCPTderivation}.  The integral over the external momentum
component $k^-$ is not used for the derivation of LCPT and will be
performed explicitly.

For a given time ordering $T$ we now partition the states $\xi$ according
to whether they are before the hard vertex $H$ in the amplitude, before
the hard vertex $H'$ in the conjugate amplitude, or in the ``final state''
between $H$ or $H'$ and the cut $F_A$.  States before $H$ go into the
factor $I_T^{}$, those before $H'$ into $I'_T$, and those in the final
state into $F_T^{}$:
\begin{align} \label{eq:CApart}
\int \frac{\df k^-}{2\pi}\, A_{F_A}(k,\tilde{\ell}_j)
  &= \int \dfrac{\df k^-}{2\pi}
      \sum_{T} I_{T}(\tilde{\ell}_j)\,
        F_T(k, \tilde{\ell}_j)\, I'_{T}(\tilde{\ell}_j)
          \times \text{numerator} \,.
\end{align}
The explicit expression of the ``numerator'' in \eq{CApart} is not needed
in the following.  It includes propagator numerators and vertex factors,
as well as a factor $i$ or $-i$ for each light-cone energy denominator
\eq{LCdenomdef}, depending on whether the corresponding intermediate state
is in the amplitude or its conjugate.  This change of sign is compensated
by a change of sign for each interaction vertex in the amplitude or its
conjugate, so that the overall numerator factor does not depend on the
placement of the final state cut $F_A$.  Furthermore, this factor is
independent of $k^-$.  In LCPT, internal vertices and propagator
numerators only depend on the on-shell momenta \eqref{on-shell-mom} of the
relevant lines, so that $k^-$ could only enter the numerator through the
two-point vertex at the end of a collinear line.  The factor for that
vertex is however momentum independent.
The other factors in \eq{CApart} read
\begin{align}
  I_{T}(\tilde{\ell}_j) &= \prod_{\substack{\text{states } \xi \\ \xi < H}}
  \dfrac{1}{p^-
   + \sum\limits_{\substack{\text{vertices} j \\ j < \xi}}\ell_j^-
   - \sum\limits_{\substack{\text{lines } L \\[0.1em] L \in \xi}} \kappa_L^-
   + i \epsilon} \,,
\\
  F_T(k,\tilde{\ell}_j) &=  
  \prod_{\substack{\text{states } \xi \\ H < \xi < F_A}} \dfrac{1}{p^-
   - k^- - \sum\limits_{\substack{\text{vertices } j \\ j > \xi}}\ell_j^-
   - \sum\limits_{\substack{\text{lines } L \\[0.1em] L \in \xi}} \kappa_L^-
   + i \epsilon}
\\ \nonumber
  & \quad \times 2\pi \delta \biggl( p^- - k^-
    - \sum\limits_{\substack{\text{vertices } j \\ j > F_A}} \ell_j^-
    - \sum\limits_{\substack{\text{lines } L \\[0.1em] L \in F_A}} \kappa_L^-
   \biggr)
\\ \nonumber 
  & \quad \times \prod_{\substack{\text{states } \xi \\ F_A < \xi < H'}} 
  \dfrac{1}{p^- -k^- 
    - \sum\limits_{\substack{\text{vertices } j \\ j > \xi}}\ell_j^-
    - \sum\limits_{\substack{\text{lines } L \\[0.1em] L \in \xi}} \kappa_L^-
    - i \epsilon} \,,
\\
I'_{T}(\tilde{\ell}_j)
  &= \prod_{\substack{\text{states } \xi \\ H' < \xi}} \dfrac{1}{p^-
    - \sum\limits_{\substack{\text{vertices } j \\ j > \xi}}\ell_j^-
    - \sum\limits_{\substack{\text{lines } L \\[0.1em] L \in \xi}} \kappa_L^-
    - i \epsilon} \,,
\end{align}
where $p$ is the proton momentum and (in contrast to
section~\ref{sec:LCPTderivation}) we simply number the on-shell minus
momenta of the lines $\kappa_L$ sequentially.
The initial state factors $I_T^{}$ and $I'_T$ only have poles in the lower
half plane for the momenta $\ell^-_j$ entering prior to $H$ or $H'$.
However, the $\ell^-_j$ poles of the factor $F_T$ are pinched and seem to
prevent us from deforming the $\ell^-_j$ integration out of the Glauber
region.  To proceed, we label the $N$ states $\xi_f$ in $F_T$ by an index
$f = 1,\ldots,N$ running from left to right, and abbreviate the sum of
on-shell minus momenta in state $f$ as
\begin{equation}
  D_f = \sum\limits_{\substack{\text{lines }
              L \\[0.1em] L \in \xi_f}} \kappa_L^- \,.
\end{equation}
The sum over all final state cuts $F_A$ for $F_T$ gives
\begin{align}
  \label{eq:FTcutsum}
\sum_{F_A} F_T(k,\tilde{\ell}_j)
  &= \sum_{c=1}^{N} \, \Biggl[ \; \prod_{f=1}^{c-1}
    \dfrac{1}{p^- - k^- - \sum\limits_{j>f} \ell_j^- -D_f + i \epsilon}
\\ \nonumber
& \quad\qquad \times
    2\pi \delta \biggl( p^- - k^- - \sum\limits_{j>c} \ell_j^- - D_c \biggr)
  \prod_{f=c+1}^{N}
    \dfrac{1}{p^- -k^- - \sum\limits_{j>f} \ell_j^- -D_f - i \epsilon}
  \, \Biggr]
\\ \nonumber
& = 
  i \, \Bigg[ \,
    \prod_{f=1}^{N} \dfrac{1}{p^- -k^- - \sum\limits_{j>f} \ell_j^-
     -D_f + i \epsilon}
  - \prod_{f=1}^N \dfrac{1}{p^- - k^- - \sum\limits_{j>f} \ell_j^-
     -D_f - i \epsilon} \, \Biggr] \,,
\end{align}
which is essentially the Cutkosky identity \cite{Cutkosky:1960sp,
  Veltman:1994wz} in the light-cone formalism and can readily be obtained
by using
\begin{align}
  2\pi\delta(x)
   =  i\, \biggl[ \dfrac{1}{x+i\epsilon} - \dfrac{1}{x-i\epsilon} \biggr] \,.
\end{align}
It is now easy to perform perform the integral of \eqref{eq:FTcutsum} over
the external momentum $k^-$ (note that no other factor in \eq{CApart}
depends on this variable):
\begin{align}
  \label{eq:one-or-zero}
\sum_{F_A}  \int \dfrac{\df k^-}{2\pi}\, F_T(k,\tilde{\ell}_j)
  = & \begin{cases} \, 1 \quad \text{if } N =1 \\
                    \, 0 \quad \text{otherwise} \end{cases} \,.
\end{align}
For $N>1$ each product on the r.h.s.\ of \eqref{eq:FTcutsum} decreases
faster than $1/k^-$ at infinity, so that one obtains zero by the theorem
of residues, whereas the result for $N=1$ simply corresponds to the
original $\delta$ function on the l.h.s.
Once we sum over all cuts, $F_T$ thus gives either $0$ or $1$.  The
pinched poles in $\ell_j^-$ are then removed, and we can deform $\ell_j^-$
into the upper half plane as we set out to show.  Notice that if one does
not consider the inclusive Drell-Yan cross section (differential or not in
the transverse momentum of the vector boson), but instead some observable
that is not constant for all the cuts of $F_T$, then this argument fails
and Glauber gluon exchange may not cancel \cite{Gaunt:2014ska,
  Zeng:2015iba}.

As already stated at the beginning of section~\ref{sec:allorder}, the
preceding derivation works not only for the original terms in the region
decomposition of a graph, but also for the subtraction terms.  In the
graph of figure~\ref{fig:subt-terms}b one neglects $\tvec{\ell}_j$ and the
masses of certain lines inside $A'$, which leads to different values of
the on-shell momenta $\kappa_L^-$ in the above expressions, but does not
affect the arguments otherwise.


\subsubsection{Double Drell-Yan}
\label{sec:double-DY}

\begin{figure}
  \centering
  \includegraphics[scale=0.7]{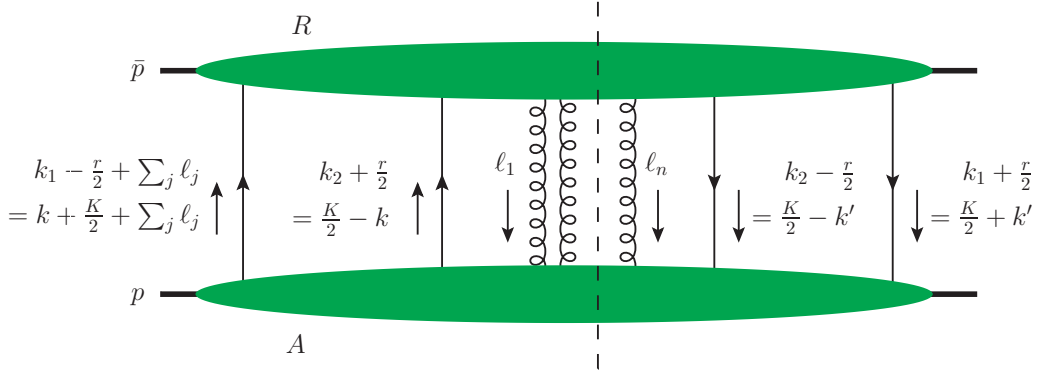}
  \caption{Partitioning of a leading graph and region in double Drell-Yan
    production into a collinear factor $A$ and the remainder $R$.  Note
    that our argument does not depend on which of the four collinear
    partons we chose to route the sum of soft momenta.  We show the
    momentum assignments for double parton scattering found in
    \protect\cite{Diehl:2011yj} (involving $k_1,k_2,r$), as well as the
    alternative assignment (involving $k,k',K$) used later in this
    section. \label{fig:CARpartitionDPS}}
\end{figure}

We now generalise the argument of the previous section to the double
Drell-Yan process.  In this case we have four physically polarised partons
joining $A$ to $H$ (two in the amplitude, and two in its conjugate) --
once again we ignore longitudinally polarised gluon attachments for the
moment.  The analogue of \eqref{eq:GLapprox} now reads
\begin{align} \label{eq:GLDDYapprox}
  G_R &= \int \dfrac{\df k^+_1\, \df^{d-2}\vect{k}_{1}^{}}{(2\pi)^{d-1}} \,
  \dfrac{\df k^+_2\, \df^{d-2}\vect{k}_{2}^{}}{(2\pi)^{d-1}} \,
  \dfrac{\df^{d-2}\vect{r}}{(2\pi)^{d-2}} 
\nonumber \\ 
  & \quad \times \int \biggl[\, \prod_j 
     \dfrac{\df{\ell}^-_j\, \df^{d-2}\tvec{\ell}_j}{(2\pi)^{d-1}} \biggr] \;\;
   \sum_{V} \sum_{F_A \in \mathcal{A}(V)}
     \int \dfrac{\df k_1^-}{2\pi}\, \dfrac{\df k_2^-}{2\pi}\,
          \dfrac{\df r^-}{2\pi}\,
      A_{F_A}(k_1,k_2,r,\tilde{\ell}_j) \big|_{r^+=0} 
\nonumber \\
  & \quad \times \sum_{F_R \in \mathcal{R}(V)}
    \int \biggl[\, \prod_j \dfrac{\df\ell_j^+}{2\pi} \biggr]\,
        R_{F_R}(k^+_1,\vect{k}_1^{},k^+_2,\vect{k}_2^{},-\vect{r},\ell _j) \,.
\end{align}
Detailed justification of this form for $G_R$ can be found in
\cite{Diehl:2011yj}.  A diagram with the momentum assignments is given in
figure \ref{fig:CARpartitionDPS}.  In section~\ref{sec:R-no-V} we will
show that the factor on the last line of \eq{GLDDYapprox} is independent
of the partitioning $V$.  In analogy to the single Drell-Yan process we
can thus perform an unrestricted sum over all cuts $F_A$ of $A$, as in
\eqref{eq:sumovcuts}.

\begin{figure}
  \centering
  \includegraphics[scale=0.7]{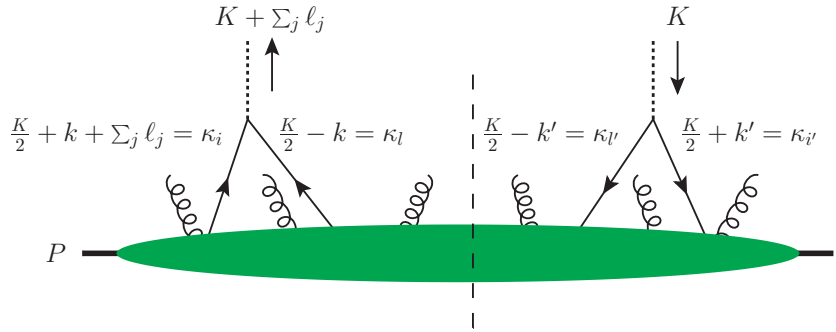}
  \caption{LCPT picture for the collinear factor $A$ in the double
    Drell-Yan process, with collinear lines in the amplitude or its
    conjugate attached together at three-point vertices.  The dotted lines
    are external to the graph and carry away the combined momenta of the
    collinear partons, as explained in the text.  \label{fig:varchange}}
\end{figure}

Now we consider $A$ in LCPT and establish the rules for the vertices to
which the collinear lines attach.  We could straightforwardly copy the
procedure of the previous section, in which case each collinear line would
attach to a two-point vertex leading to an external line.  We would then
consider all light-cone time orderings of the vertices (including allowed
orderings of the two-point vertices), and integrate over the external
minus momenta $k_1^-$, $k_2^-$ and $r^-$.  In Appendix
\ref{app:TimeOrderings} we show how the cancellation of pinched poles
arises in this setting.  Here we take an alternative approach, which
allows us to arrive at the desired result more efficiently.  First, we
make the following change of variables:
\begin{align} \label{eq:k1k2rchg}
  K  &= k_1 + k_2 \,,
&
  k  &= \half\ms (k_1 - k_2 - r) \,,
&
  k' &= \half\ms (k_1 - k_2 + r) \,.
\end{align}
As shown in figure \ref{fig:CARpartitionDPS}, $k^-$ runs up the first
collinear line in the amplitude and down the second, without crossing the
final state cut.  Since the factor $R$ in \eqref{eq:GLDDYapprox} is
independent of $k^-$, we may treat this variable as internal in the
derivation of the LCPT rules (see step~\ref{deriv:vertex} in
section~\ref{sec:LCPTderivation}).  We then have a vertex to which
\emph{both} collinear lines in the amplitude attach, as shown in
figure~\ref{fig:varchange}.  The momentum component $k'^-$ in the
conjugate amplitude can be treated in the same way.  In generalisation of
\eqref{eq:sdy-vertex} we now write the integration over the external minus
momenta as
\begin{align}
\label{eq:ddy-vertex}
\textstyle \int \df k_1^-\, \df k_2^-\, \df r^-
  &= \textstyle \int \df K^-\, \df k^-\, \df k'^-\,
\nonumber \\
&\quad \textstyle \times \int \df \kappa_i^-\, \df \kappa_l^-\;
    \delta\bigl( K^- \!/\ms 2 + k^- + \sum_j \ell_j^- - \kappa_i^- \bigr ) \,
    \delta\bigl( K^- \!/\ms 2 - k^- - \kappa_l^- \bigr ) 
\nonumber \\
&\quad \textstyle \times
  \int \df \kappa_{i'}^-\, \df \kappa_{l'}^-\;
    \delta\bigl( K^- \!/\ms 2 + k'^- - \kappa_{i'}^- \bigr ) \,
    \delta\bigl( K^- \!/\ms 2 - k'^- - \kappa_{l'}^- \bigr ) 
\nonumber \\[0.2em]
&=  \textstyle \int \df K^-
    \int \df \kappa_i^-\, \df \kappa_l^-\;
    \delta\bigl( K^- + \sum_j \ell_j^- - \kappa_i^- - \kappa_l^- \bigr )
\nonumber \\
&\quad \textstyle \times
    \int \df \kappa_{i'}^-\, \df \kappa_{l'}^-\;
    \delta\bigl( K^- - \kappa_{i'}^- - \kappa_{l'}^- \bigr ) \,.
\end{align}
The $\delta$ functions in the last expression are exactly what is required
for introducing three-point vertices according to step~\ref{deriv:delta}
of section \ref{sec:LCPTderivation}.  These vertices are somewhat unusual
in that the plus components (in TMD factorisation also the transverse
ones) of the individual parton lines $\kappa_i$ and $\kappa_l$
($\kappa_{i'}$ and $\kappa_{l'}$) are fixed, but this does not invalidate
the derivation of the LCPT rules.

With this setup we can proceed exactly like in single Drell-Yan
production, having exactly one hard vertex $H$ on the left and one hard
vertex $H'$ on the right of the cut $F_A$.  Summing over all these cuts
and integrating over $K^-$ we obtain the analogue of
\eqref{eq:one-or-zero}, which shows that all pinched poles from final
state interactions cancel in the sum over $F_A$.  We can thus deform the
integration over $\ell_j^-$ into the upper half plane as desired.

Note that it is the integration over $k^-$ that is key in order for us to
be able to attach the two collinear lines in the amplitude to a single
vertex.  In fact, the momentum $k^-$ is Fourier conjugate to the
light-cone time separation $x_{12}^+$ between the fields associated with
these two collinear partons in the operator definition of $A$.  Integrating
over $k^-$ corresponds to setting $x_{12}^+ = 0$.  It is thus not
surprising that the $k^-$ integration allows us to connect the ends of the
two collinear lines together at a single vertex in the LCPT picture.  In
the physical process, this amounts to the two hard scatterings occurring
at the same light-cone time.


\subsubsection{Including longitudinally polarised collinear gluons} 
\label{sec:long-pol-coll}

It is now easy to treat the case in which we have an arbitrary number of
longitudinally polarised collinear gluon attachments between $A$ and $H$,
either in single or in double Drell-Yan production.  For each one of these
gluon lines there is an additional loop momentum $\lambda_k$ in
\eqref{eq:GLapprox} or \eqref{eq:GLDDYapprox}, and the integral over
$\lambda_{\smash{k}}^-$ concerns only the factor $A$ because this
component is set to zero in the hard subgraph.  We can thus use the same
procedure as for the two physically polarised partons in double Drell-Yan
production, attaching the ends of all collinear lines in the amplitude to
one vertex $H$ and the ends of all collinear lines in the conjugate
amplitude to another vertex $H'$.  The argument for the cancellation of
final state poles in soft momentum components $\ell_j^-$ then proceeds as
before.

We remark that, before discussing the contour deformation in $\ell_j^-$,
one could already make the Grammer-Yennie approximation \eqref{GY-collin}
for collinear gluons and then apply the corresponding Ward identities to
remove collinear gluon attachments from the hard factors.  For lightlike
Wilson lines the resulting eikonal propagators
$1 /(\lambda_{k}^+ + i\epsilon)$ can readily be incorporated into the
vertex factors for $H$ and $H'$.  For non-lightlike Wilson lines, however,
the eikonal propagators also depend on $\lambda_{\smash{k}}^-$, which
complicates the derivation of the LCPT rules.  For this reason we postpone
the collinear Grammer-Yennie approximation to a later stage in our present
treatment.


\subsubsection{Independence of the remainder $R$ on partitioning of soft
  vertices} 
\label{sec:R-no-V}

As announced earlier, we now show that the remainder factor $R$ is
independent of the partitioning of soft vertices $V$ in the collinear
factor $A$.  We perform this proof explicitly for the double Drell-Yan
process.  The proof for single Drell-Yan production readily follows from
the derivation given here; a slightly different version of it has been
given long ago in \cite{Collins:1988ig}.

\begin{figure}
  \centering \includegraphics[scale=0.65]{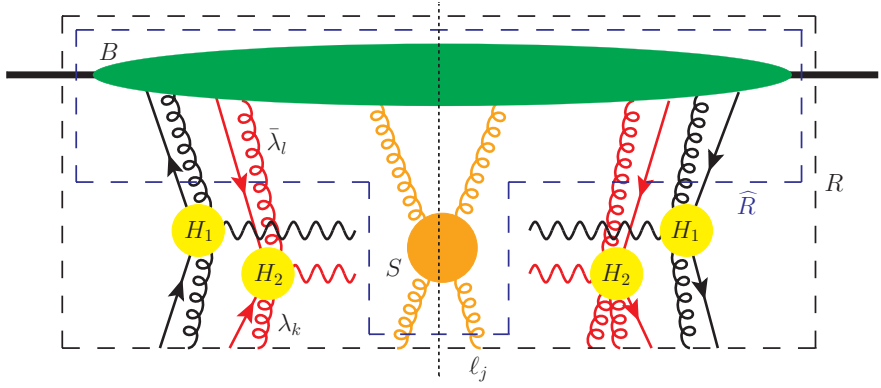}
  \caption{Graphical depiction of the factors $R$ and $\widehat{R}$ in
    \protect\eqref{eq:RandRbar}. \label{fig:dDY_Rbargraph}}
\end{figure}

Consider the factor $R$ for double Drell-Yan production, in the general
case where we have an arbitrary number of longitudinally polarised gluon
attachments to the hard process from both $A$ and $B$. This can be written
as
\begin{align} 
  \label{eq:RandRbar}
& \int \biggl[\, \prod_j \dfrac{\df\ell_j^+}{2\pi} \biggr]\,
  \sum_{F_R \in \mathcal{R}(V)}
  R_{F_R}^{}(k^+_1,\vect{k}_1^{},k^+_2,\vect{k}_2^{}, \bar{\vect{r}},
   \lambda_{\smash{k}}^+,\vect{\lambda}_k^{},\ell _j)
\nonumber \\
& \qquad =
  \int \frac{\df\bar{k}_1^-\, \df^{d-2}\bar{\vect{k}}_1^{}}{(2\pi)^{d-1}}\,
       \frac{\df\bar{k}_2^-\, \df^{d-2}\bar{\vect{k}}_2^{}}{(2\pi)^{d-1}}
  \biggl[\, \prod_l \frac{\df\bar{\lambda}_{\smash{l}}^-\,
             \df^{d-2}\bar{\vect{\lambda}}_l^{}}{(2\pi)^{d-1}} \biggr]\,
     \widehat{R}\bigl( 
       \bar{k}_1, \bar{k}_2, \bar{r}, \bar{\lambda}_l, \ell_j \bigr)
\nonumber \\
& \qquad\qquad \times
    H(k_1^+,\vect{k}_1^{},\bar{k}_1^-,\bar{\vect{k}}_1^{},
      k_2^+,\vect{k}_2^{},\bar{k}_2^-,\bar{\vect{k}}_2^{}, \bar{\vect{r}},
      \lambda_k^+,\vect{\lambda}_k^{},
      \bar{\lambda}_{\smash{l}}^-,\bar{\vect{\lambda}}_l^{}) \,,
\end{align}
where
\begin{align} \label{eq:Rbardef}
& \widehat{R} \bigl( \bar{k}_1, \bar{k}_2, \bar{r},
                     \bar{\lambda}_l, \ell_j \bigr) 
\\ \nonumber
 &\quad = \sum_{F_{BS} \in \mathcal{R}(V)}
    \int \dfrac{\df\bar{k}_1^+}{2\pi} \frac{\df\bar{k}_2^+}{2\pi}
    \frac{\df \bar{r}^+}{2\pi}
    \biggl[\, \prod_l \frac{\df\bar{\lambda}^+_l}{2\pi} \biggr]
    \biggl[\, \prod_j \frac{\df \ell_j^+}{2\pi} \biggr]
    \bigl( BS \big)_{F_{BS}}
       \bigl( \bar{k}_1, \bar{k}_2, \bar{r}, \bar{\lambda}_l, \ell_j
       \bigr) \Big|_{\bar{r}^- = 0} \,.
\end{align}
A diagrammatic depiction of $R$ and $\widehat{R}$ is given in figure
\ref{fig:dDY_Rbargraph}.  The factor $H$ in \eq{RandRbar} contains both
hard scatterings $H_1$ and $H_2$ of figure \ref{fig:dDY_Rbargraph}, plus
all appropriate $\delta$ functions for momentum conservation in the hard
subgraphs.  It is summed over all possible final state cuts of $H_1$ and
$H_2$ (in TMD factorisation there is only one such cut, but there can be
several cuts in collinear factorisation with unobserved jets).  The
quantities $\lambda_k$ are the momenta of the longitudinally polarised
gluons entering $H$ from $A$, and $\bar{\lambda}_l$ are the corresponding
momenta of the longitudinally polarised gluon connections between $H$ and
$B$.

Let us consider $\widehat{R}$ in LCPT, with the light-cone time now being
$x^-$ (rather than $x^+$ as in our analysis of the factor $A$).  Note that
we have the product of the $B$ and $S$ subgraphs in $\widehat{R}$, and
that we make kinematic approximations for soft momenta when they enter
$B$, neglecting their minus components.  In the LCPT formulation these
approximations only adjust the values of on-shell plus momenta
$\kappa_L^+$, the precise values of which do not matter in the arguments
to follow.  We therefore can treat the soft and collinear lines in
$\widehat{R}$ in the same manner as we would without the approximations.

In \eq{Rbardef} we have integrations over the plus momenta of all external
parton lines, i.e.\ all collinear lines (the physically polarised lines
plus the longitudinally polarised gluons) and all soft lines.  We can
hence take all external parton lines in the amplitude and tie their ends
together at a single vertex $H$, using the same argument as in the
previous subsections.  Likewise, we can tie together all external lines in
the conjugate amplitude at a single vertex $H'$. There is a remaining
integral over the total plus momentum $\bar{K}^+$ flowing out of the graph
at $H$ (and, by momentum conservation, flowing back into the graph at
$H'$).  In this representation, the partitioning $V$ determines which soft
lines $\ell_j$ originate in the vertex $H$ and which ones originate in
$H'$.

\begin{figure}
  \centering
\subfigure[]{\includegraphics[scale=0.61]{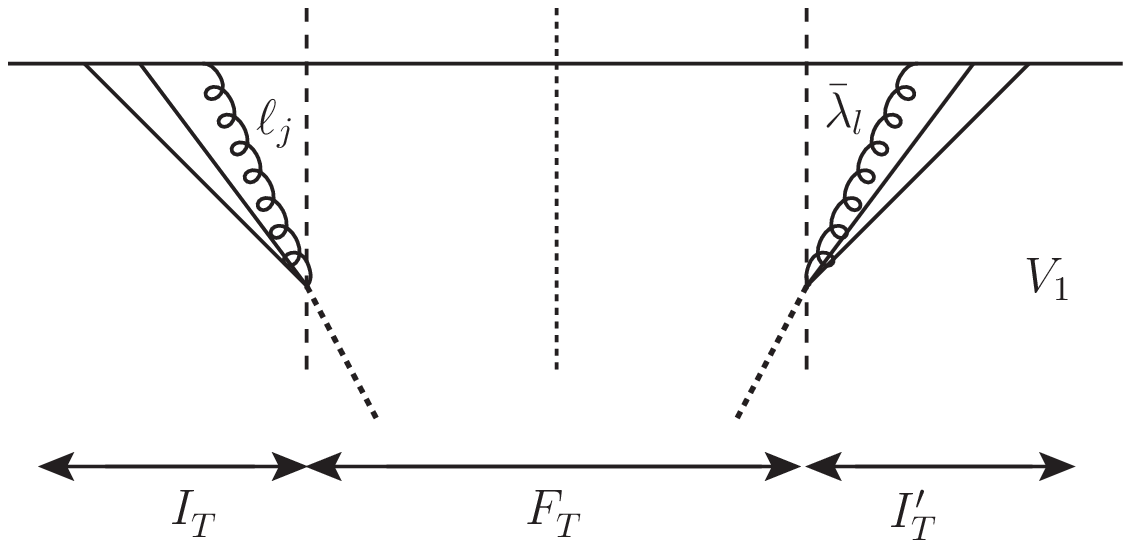}}
\subfigure[]{\includegraphics[scale=0.61]{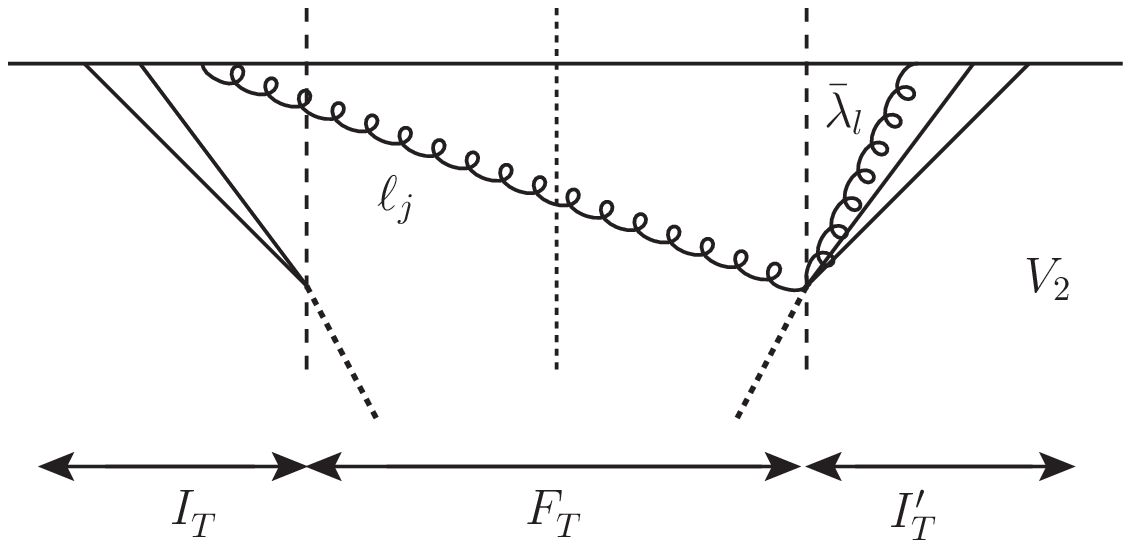}}
\\[1em]
\subfigure[]{\includegraphics[scale=0.61]{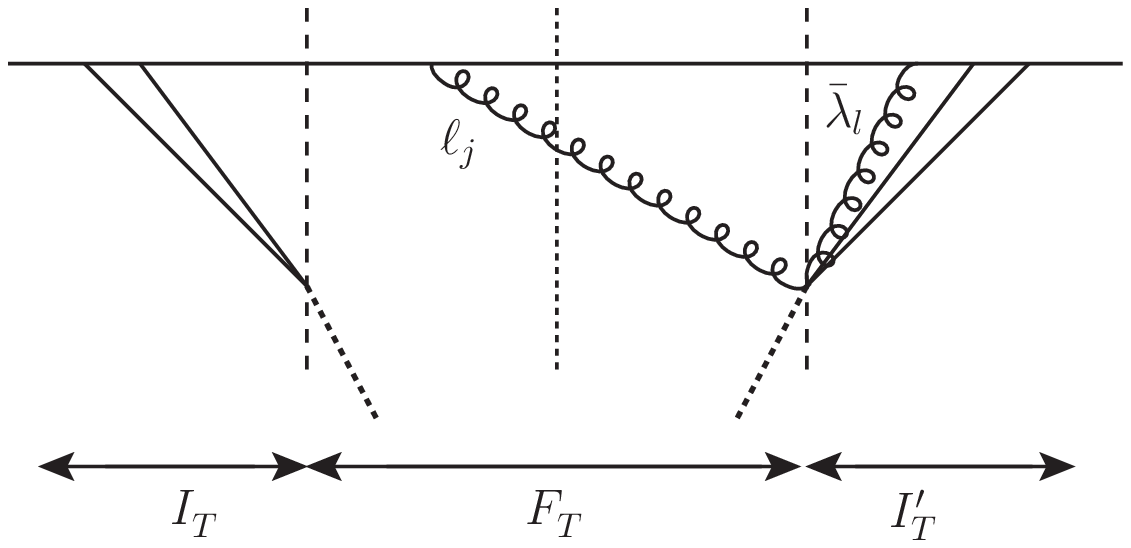}}
  \caption{Examples of LCPT graphs for the factor $\widehat{R}$ in a simple
    setting where the incident hadron is replaced by a single parton line.
    \label{fig:differentVs}}
\end{figure}

As we did for $A$, we partition $\widehat{R}$ for each light-cone time
ordering $T$ into a numerator factor, a factor ${I}_T^{}$ for states that
occur before $H$ in the amplitude, a factor ${I}'_T$ for states that occur
before $H'$ in the conjugate amplitude, and a factor ${F}_T^{}$ for final
states between $H$ or $H'$ and the cut $F_{BS}$.  Only ${F}_T$ depends on
$\bar{K}^+$, so that we can write
\begin{align}
\widehat{R} &=
 \sum_{F_{BS} \in \mathcal{R}(V)} \sum_{T}
    {I}_T \, \Biggl[\, \int \frac{\df  \bar{K}^+}{2\pi}\,
       {F}_T^{}(\bar{K}^+) \,\Biggr] \, {I}'_T \,
    \times \text{numerator} \,.
\end{align} 
Let us consider a given graph and time ordering $T$ of vertices, sum over
the final state cuts $F_{BS}$ that are compatible with $V$, and also
perform the integral over $\bar{K}^+$.  We then get zero unless there is
only a single state in $F_T$, in which case the final state factor is
simply unity.  The argument is completely analogous to the one given in
\eq{FTcutsum}.  In particular, this means that when we have graphs in
which soft gluons arising from an external vertex attach to the final
state, such as in figure \ref{fig:differentVs}c, the sum over cuts always
gives zero.

For the nonzero contributions to $\widehat{R}$ it does not matter whether
a soft gluon $\ell_j$ couples to $H$ or to $H'$, i.e.\ $\widehat{R}$ does
not depend on $V$ as we set out to show.  To see this, consider two
partitionings $V_1$ and $V_2$ that differ only in whether $\ell_j$ enters
the graph in the amplitude or in its conjugate.  For every nonzero graph
in $V_1$ we can find a corresponding nonzero graph in $V_2$ with the same
initial state factors $I_T^{}$ and $I'_T$.  A simple example of
corresponding graphs is given in figure \ref{fig:differentVs}a and b.
Note that even though the final states in graphs a and b are different, we
have $(2\pi)^{-1} \!\int \df\bar{K}^+\, F_T(\bar{K}^+) = 1$ in both cases.

The proof for the independence of $R$ on $V$ in the single Drell-Yan
process is completely analogous, the only difference being that in
$\widehat{R}$ there is only one physically polarised parton (rather than
two) entering the external vertex in the amplitude or its conjugate.

\section{Summary}
\label{sec:sum}

Double parton scattering is an interesting signal and potentially relevant
background process at the LHC.  A factorisation formula
for the double parton scattering cross section was put forward long ago,
based on an analysis of the lowest order Feynman diagrams for the process
and the usual approximations made in collinear factorisation
\cite{Paver:1982yp, Mekhfi:1983az, Diehl:2011yj}. More recently, a similar
analysis was used to obtain a formula for the DPS cross section
differential in final-state transverse momenta \cite{Diehl:2011yj}. For
the double Drell-Yan process, in which one does not have any complications
associated with colour in the final state, several ingredients of a
factorisation proof have been given in \cite{Diehl:2011yj,
  Manohar:2012jr}. Notably missing in these papers is a proof of
cancellation of so-called Glauber gluons, which are soft gluons with much
larger transverse momentum components than longitudinal ones. For these
momentum modes, the approximations required to obtain factorisation break
down.

In the present paper we have filled this gap, demonstrating the
cancellation of Glauber gluons in the double Drell-Yan process (both with
and without measured transverse momenta of the produced gauge bosons).
This was done first at the one-gluon level in a simple model, using
momentum routing arguments and the same unitarity argument that can be
used to show the cancellation of Glauber exchange at the one-gluon level
in single Drell-Yan production (see e.g.~\cite{Gaunt:2014ska}).  Then, an
all-order proof was given, which makes use of light-cone perturbation
theory and generalises the corresponding proof for the single Drell-Yan
process in \cite{Collins:1988ig, Collins:2011zzd}. In the process of
constructing this proof, we also revisited and clarified some issues with
regards to the Glauber cancellation argument and its relation to the rest
of the factorisation proof in the single scattering case.  

It is easy to see that both our proofs --- the one for one-gluon exchange
and the one for all orders --- generalise to other double scattering
processes producing colourless particles, such as the production of a
Higgs boson and a $Z$.  Likewise, both proofs carry over from double
parton scattering to an arbitrary number $n$ of hard scatters (each
producing only colourless particles).  Corresponding cross section
formulae derived from lowest order Feynman diagrams can be found in
\cite{Diehl:2011yj}.  It should not be difficult to adapt the other steps
required for a factorisation proof, laid out in section
~\ref{sec:overall-strategy}, to this case.  What becomes increasingly more
complicated with the number of hard scatters is in particular the colour
structure of multiparton distributions and soft factors, which is not a
problem of conceptual nature but of increased complexity for calculations
and phenomenology.

Even with the results of the present paper, there still remains work to do
to establish the all-order factorisation of the double Drell-Yan process.
Open issues are discussed in section~\ref{sec:overall-strategy}.  Perhaps
the most prominent among them is the question how to treat the double
counting between single and double scattering, and how one should divide
``double splitting graphs'' (e.g.\ the equivalent to
figure~\ref{fig:tree-box} in QCD) between single and double scattering.

Aside from double Drell-Yan production, it would be desirable to
investigate the factorisation properties of other processes, such as the
production of double dijets.  Before analysing such processes with colour
in the final state, one first should investigate the factorisation for the
corresponding single scattering processes, for which to the best of our
knowledge the absence of Glauber gluon effects has not yet been proven.  A
more solid theoretical understanding of double hard scattering with jets
or heavy quarks would be of great practical value, because such processes
have much higher rates than double Drell-Yan production and were already
studied experimentally at the Tevatron \cite{Abe:1993rv, Abe:1997bp,
  Abe:1997xk, Abazov:2009gc, Abazov:2011rd, Abazov:2014qba} and in Run~1
of the LHC \cite{Aad:2013bjm, Chatrchyan:2013xxa, Aaij:2011yc,
  Aad:2014rua, Khachatryan:2014iia, Aaij:2012dz}. Further in-depth studies
can surely be expected from Run~2.

\appendix

\section{Evaluation of Feynman integrals in light-cone coordinates}
\label{app:feynmanint}

As mentioned in section \ref{sec:box-graphs} the evaluation of Feynman
integrals in light-cone coordinates has to be done with some care.
Consider for instance the integral
\begin{align}
  \label{app:example}
  \int \df^4 \ell \;
  \frac{1}{\bigl[ (\ell^0)^2 - \vec{\ell}^{\,2} - M^2
       + i \varepsilon \bigr]^6}
  &= \int \df \ell^+\, \df \ell^-\, \df^2 \tvec{\ell} \;
     \frac{1}{\bigl[ 2 \ell^+ \ell^- - \tvec{\ell}{}^2 - M^2
       + i \varepsilon \bigr]^6} \,. 
\end{align}
The integral on the l.h.s.\ of~\eqref{app:example} is obviously non-zero
while a naive application of Cauchy's theorem for the $\ell^+$ integration
on the r.h.s.\ would yield zero. This apparent paradox is resolved by the
fact that Cauchy's theorem is not applicable in this case, because at the
point $\ell^-=0$ the $\ell^+$ integral is linearly divergent. This issue
has been encountered in the past, and in \cite{Yan:1973qg,Heinzl:2003jy}
it was shown that the proper formula for such integrals reads
\begin{align}
  \label{app:master}
  \int \df \ell^+ \; \frac{1}{[2 \ell^+ \ell^- - M^2 + i \varepsilon]^n}
  &= \frac{(-1)^n}{n-1}\,
     \frac{i \pi \delta(\ell^-)}{[M^2 - i \varepsilon]^{n-1}} \,.
\end{align}
This formula and the methods used to derive it can also be used to
calculate the integral
\begin{align}
  \label{app:todo}
  I & =  \int \frac{\df \ell^-}{\ell^- - i \eta'}
         \int \frac{\df \ell^+}{\ell^+ + i \eta}\;
    \frac{1}{[2 \ell^+ (\ell^- +  a^-)  + 2 a^+ \ell^-
              + A + i \varepsilon]^4}
\end{align}
that we have encountered in section \ref{sec:box-graphs}.  After partial
fractioning with respect to $\ell^+$ we can adapt the methods of
\cite{Yan:1973qg} and obtain the same result as we got using Cauchy's
theorem for the $\ell^+$ integration.  Cauchy's theorem can be used in
this case because at the point $\ell^- = - a^-$ the $\ell^+$ integral
diverges only logarithmically but not linearly.  It can however not be
used if the integrand comes with one or two powers of $\ell^+$ in the
numerator, which cancel the factor $1 /(\ell^+ + i \eta)$.  The method
using \eqref{app:master} combined with partial fractioning continues to
work in this case.


\section{Alternative approach to the double scattering collinear factor}
\label{app:TimeOrderings}

In this appendix we consider the alternative approach, mentioned after
\eqref{eq:GLDDYapprox}, for treating the vertices at the ends of collinear
lines in the double scattering collinear factor $A$.  We take each
collinear line to end on its own two-point vertex, where an external line
carries away the appropriate momentum.  We have to consider all possible
orderings of these two-point vertices in the amplitude and its conjugate,
and a priori we can have any number of interactions and insertions of soft
momenta between the two-point vertices.  This seems to be rather different
from the situation described in section~\ref{sec:double-DY}, where the two
collinear lines are connected together at a vertex, such that nothing can
occur between them.  However, we will see that in the end we obtain the
same result.

\begin{figure}
  \centering
  \subfigure[]{\includegraphics[scale=0.78]{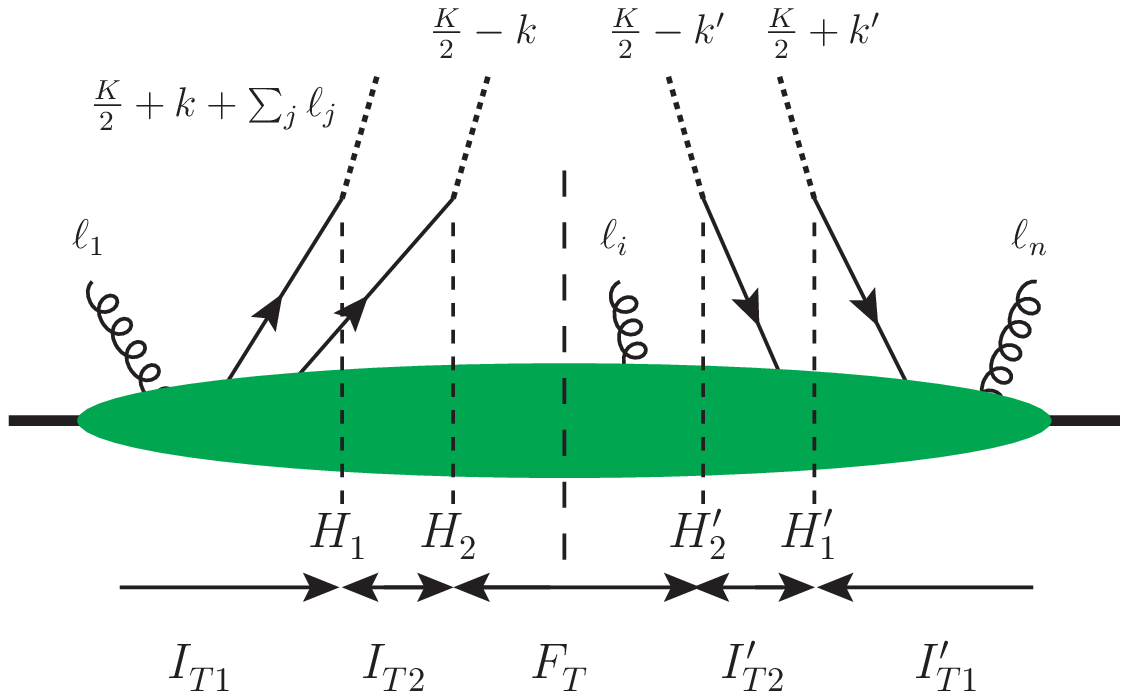}}
  \subfigure[]{\includegraphics[scale=0.78]{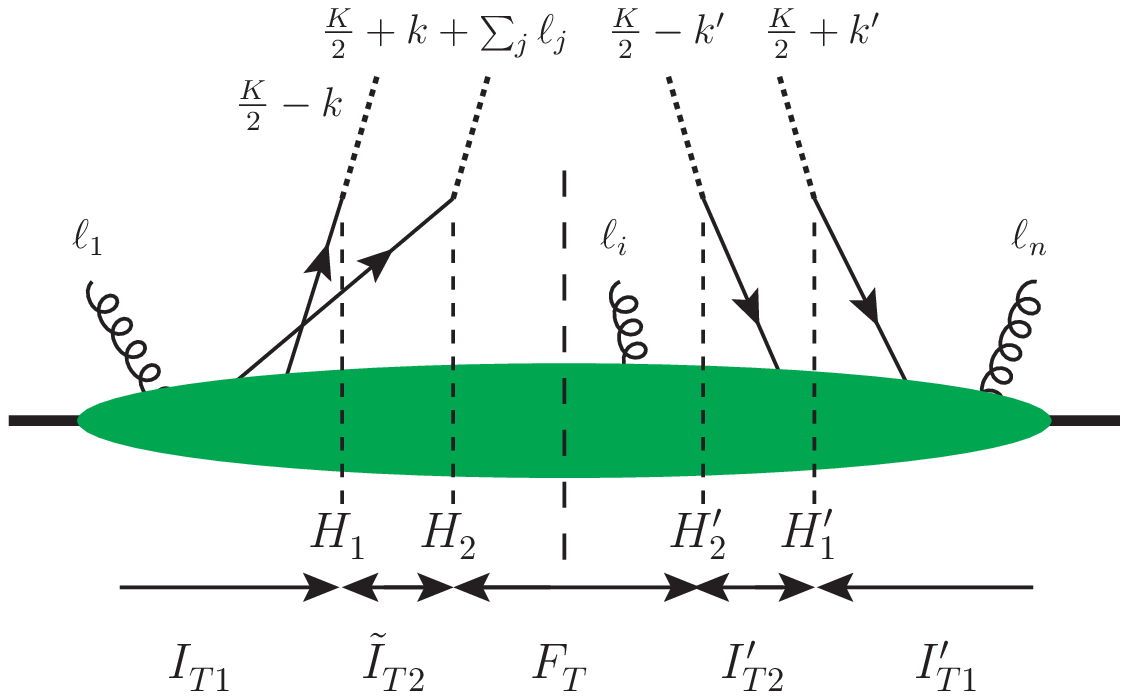}}
  \caption{LCPT graphs for the double scattering collinear factor $A$,
    showing the two possible light-cone time orderings of the hard
    vertices in the amplitude.  The dotted lines attached to the hard
    vertices are external to the graph and carry away the momenta of the
    collinear partons, as explained in the text.  \label{fig:orderings}}
\end{figure}

Let us consider the two LCPT graphs in figure \ref{fig:orderings}, which
are completely identical except for the fact that we have interchanged the
order of the hard vertices in the amplitude.  The internal structure of
the blobs, positions of soft attachments, and light-cone time ordering of
states in the blobs are the same.  To calculate $A$ we must sum over both
graphs.
Now we make the change of variables in \eq{k1k2rchg}.  Then only the
factors $I_{T2}$ and $\tilde{I}_{T2}$ depend on $k^-$, which is to be
integrated over.  For either graph, if there is more than one state
between $H_1$ and $H_2$ (in particular if there is an insertion of a soft
momentum between $H_1$ and $H_2$) then the $k^-$ integration gives zero,
since then there are at least two denominator factors that depend on
$k^-$, which all have their poles on the same side of the real axis.  A
nonzero result is hence only possible if there is exactly one state
between $H_1$ and $H_2$.

We now combine the two graphs.  Since they are identical apart from the
ordering of $H_1$ and $H_2$, the factors $I_{T1}^{}$, $F_T^{}$, $I'_{T1}$
and $I'_{T2}$ are identical between the two and can be factored out.  We
thus need only consider the sum of $I_{T2}$ and $\tilde{I}_{T2}$ for the
one case in which they are nonzero:
\begin{align} 
\label{eq:sumorderings}
\int \dfrac{\df k^-}{2\pi} \, \bigl[ I_{T2} +  \tilde{I}_{T2} \big]
   = \int \dfrac{\df k^-}{2\pi} \biggl[\,
&
   \frac{1}{p^- - k^- - K^-\!/\,2 - \sum_{j>H_1} \ell_j - D_f + i\epsilon}
\nonumber \\
&  + \frac{1}{p^- + k^- - K ^-\!/\,2 + \sum_{j<H_1} \ell_j - \tilde{D}_f
     + i\epsilon} \,\biggr]
 =  -i \,,
\end{align}
where $D_f$ ($\tilde{D}_f$) is the sum of on-shell minus momenta in
$I_{T2}$ ($\tilde{I}_{T2}$).  When we sum over the two orderings, we thus
get a constant factor for the contribution from states between $H_1$ and
$H_2$, which is the same as if we had taken $H_1$ and $H_2$ to be
simultaneous (in light-cone time $x^+$) from the start.  Note that it is
necessary to sum over the two light-cone time orderings in
figure~\ref{fig:orderings}a and b to obtain a convergent and well-defined
integral in~\eq{sumorderings}.

Repeating the procedure for the $k'^-$ integral, we find that we can take
$H'_1$ and $H'_2$ to be simultaneous as well.  In this case the constant
is $+i$ rather than $-i$ as in~\eq{sumorderings}.
Now that the two vertices are simultaneous in the amplitude and its
conjugate, we can apply the argument of sections~\ref{sec:single-DY}
and \ref{sec:double-DY} to show that there are no pinched poles in the
Glauber region.  Note that just as in the approach described there, the
crucial ingredient needed for treating the vertices as simultaneous is the
integration over $k^-$~and~$k'^-$.


\section{Timelike Wilson lines}
\label{app:timelike}

In this appendix, we show the viability of timelike Wilson lines in single
and double Drell-Yan production at the level of one-gluon exchange. Rather
than proving factorisation with timelike Wilson lines directly, we start
from the factorised expression with spacelike Wilson lines, whose validity
we have established in the main body of this paper.  We then show that the
expression with timelike Wilson lines is equivalent to this to
leading-power accuracy.  As in section \ref{sec:one-gluon}, we restrict
our attention to transverse gluon momenta $\vect{\ell}$ of order
$\Lambda$, since this covers the soft and collinear regions where the
Wilson lines are relevant.  We also take the collinear and soft Wilson
lines to have the same directions.  As in section \ref{sec:Step2Details},
we begin with the simpler single Drell-Yan case and later explain how the
argument generalises to double Drell-Yan production.

Let us denote the directions of the left- and right-moving timelike Wilson
lines as $v_{LT} = (\delta^2, 1, \tvec{0})$ and $v_{RT} =
(1, \delta'^2, \tvec{0})$ respectively.  We write their spacelike analogues
as $v_{LS} = (- \delta^2, 1, \tvec{0})$ and $v_{RS} = (1,
- \delta'^2, \tvec{0})$. Here $\delta, \delta' \sim \Lambda/Q$, and it is
understood that right-moving partons have plus components of order $Q$ and
transverse ones of order $\Lambda$.

We first derive an intermediate result that will prove useful later
on. Let $A_{\text{sub}}(v_{Li},v_{Rj})$ with $i,j=T,S$ denote a subtracted
collinear subgraph with a Wilson line along $v_{Li}$ and a Wilson line
$v_{Rj}$ in the soft subtraction term (see~\eqref{subtracted-A}). We now
prove that, for a single gluon attaching to the Wilson line $v_{Lj}$, we
have
\begin{align}
  \label{diff-AT-AS}
  A_{\text{sub}}(v_{LT},v_{RS}) &\approx A_{\text{sub}}(v_{LS},v_{RS})
\end{align}
when the transverse gluon momentum $\tvec{\ell}$ is restricted to be of
size $\Lambda$. Let us first discuss the case where the gluon does not
cross the final-state cut inside $A$.  Before integration over
$\tvec{\ell}$ and up to a global factor irrelevant for our argument, the
subtracted collinear factor reads
\begin{align}
  \label{int-start}
\int \df\ell^-\ms \df\ell^+\,
  \frac{1}{\ell^+ \pm \delta^2 \ell^- - i\varepsilon}\,
  \frac{i}{2\ell^+ \ell^- - \tvec{\ell}{}^2 + i\epsilon}\,
  \biggl[ A^+(\ell) \pm \delta^2 A^-(\ell)
      - \frac{1}{\ell^- - \delta'^2 \ell^+ + i\varepsilon}\,
        \ell A(\ell) \biggr] \,,
\end{align}
where $+\delta^2$ corresponds to $v_{LT}$ and $-\delta^2$ to $v_{LS}$.  We
have taken the gluon momentum to flow from the Wilson line downwards into
$A$, on the left hand side of the final state cut.
Note that as in section~\ref{sec:one-gluon}, we consider here only the
Grammer-Yennie approximations in the soft and collinear factors,
postponing kinematic approximations to a later stage.

To start, we consider the term with $\delta^2 A^-(\ell)$
in \eqref{int-start}.  This is only non-negligible compared to the other
terms if $A^-(\ell) /A^+(\ell)$ is very large, which requires $\ell^- \gg
Q$. The only dominant graphs in $A$ are then those where the gluon couples
to an active parton, since this carries the minimal number of parton lines
off shell (see the discussion in section~\ref{sec:soft-scaling}).
Explicit power counting for such graphs shows that this region is strongly
power suppressed, either due to the gluon propagator or due to the
$\delta^2 \ell^-$ term in the eikonal propagator. Thus, we need not
consider this term further.

We split the remaining term in the square brackets of \eqref{int-start}
into two pieces as follows:
\begin{align} \label{TSsplitto2}
 \biggl[ \left\{ A^+(\ell) - A^+(\tilde{\ell}) \right\} +
         \left\{ A^+(\tilde{\ell}) - \frac{1}{\ell^-
               - \delta'^2 \ell^+ + i\varepsilon}\,
        \ell A(\ell) \right\} \biggr]  \,.
\end{align}
We now show that the sign of the $\delta^2 \ell^-$ term in
\eqref{int-start} is not important for either piece. We begin with the
first piece, which together with the remaining factors and integral is
\begin{align}
  \label{int-coll-region}
\int \df\ell^-\ms \df\ell^+\;
  \frac{1}{\ell^+ \pm \delta^2 \ell^- - i\varepsilon}\,
  \frac{i}{2\ell^+ \ell^- - \tvec{\ell}{}^2 + i\epsilon}\,
  \bigl[ A^+(\ell) - A^+(\tilde{\ell}) \bigr] \,.
\end{align}
For the term $\delta^2 \ell^-$ to be important, the gluon must be left
moving with $\ell^- \sim \delta^{-2} \ell^+$ or larger.  To prevent the
gluon propagator from becoming too large, one must then have $\ell^+ \ll
Q$.  Then, however, the difference in square brackets is negligible, and
so is the integral.  Hence there is no leading contribution from the
region in which the sign of the $\delta^2 \ell^-$ term matters.

We now move on to the second piece in \eqref{TSsplitto2}, which gives
\begin{align}
  \label{int-final}
& \int \df\ell^-\ms \df\ell^+\;
  \frac{1}{\ell^+ \pm \delta^2 \ell^- - i\varepsilon}\,
  \frac{i}{2\ell^+ \ell^- - \tvec{\ell}{}^2 + i\epsilon}\,
  \biggl[ A^+(\tilde{\ell})
      - \frac{1}{\ell^- - \delta'^2 \ell^+ + i\varepsilon}\,
        \ell A(\ell) \biggr]
\\
&\qquad = 2\pi \int_{0}^\infty \df\ell^-\;
    \frac{1}{\tvec{\ell}{}^2 \pm 2 \delta^2 (\ell^-)^2 - i\varepsilon}\,
    \biggl[ A^+(\tilde{\ell})
     - \frac{2\ell^-}{2 (\ell^-)^2 - \delta'^2\ms \tvec{\ell}{}^2
                      + i\varepsilon}\,
       \ell A(\ell) \biggr]_{\ell^+ = \, \tvec{\ell}^2 /(2\ell^-)} \,.
\nonumber
\end{align}
We have performed the integration over $\ell^+$ using Cauchy's theorem.
With all eikonal propagators having their poles in $\ell^+$ above the real
axis (for this it is essential to take $v_R$ spacelike) we have closed the
integration contour in the lower half plane, only picking up the pole of
the gluon propagator if $\ell^- > 0$.
In the second line of \eqref{int-final}, the subtraction term leads to a
suppression for $\ell^- \sim |\tvec{\ell}|$ or larger, whereas for
$\ell^- \ll |\tvec{\ell}|$ the terms with $\pm \delta^2$ in the eikonal
propagators are negligible. Thus the sign of the $\delta^2 \ell^-$ term is
not important here either.

The preceding arguments also work if the gluon couples to a Wilson line
left of the final state cut and then crosses the cut; in that case the
gluon propagator in \eqref{int-start},
\eqref{int-coll-region} and the starting expression of \eqref{int-final}
is to be replaced by $2\pi \delta(\ell^2)\, \Theta(\ell^-)$.  This
completes our proof of \eqref{diff-AT-AS}.

Note that this result cannot be obtained by power counting alone.  For
$|\tvec{\ell}| \sim \Lambda$, the region with $\ell^- \sim \Lambda^2/Q$
and $\ell^+ \sim \Lambda^4/Q^3$ gives a leading-power contribution to
\eqref{int-start} since $\ell^-$ is too small for the Grammer-Yennie
approximation in the subtraction term to work, and the smallness of the
eikonal propagator compensates the smallness of the integration volume of
$\ell^+$ in that region, where one cannot neglect $\pm \delta^2 \ell^-$.
The position of poles in $\ell^+$ was essential to pick up only the pole
of the gluon propagator using Cauchy's theorem, and then the mass-shell
condition for $\ell$ excluded the above dangerous region.

Now let us consider the soft and collinear-to-$A$ or $B$ approximants to
the full set of graphs in which one gluon extends between the left and
right moving collinear sectors. We collect together graph approximants in
which a gluon attaches in all different places to the upper and lower
parts of the graph, with these attachments being either to the left or to
the right of the final-state cut in each case. This allows us to use Ward
identities to convert contractions of $\ell^\mu$ with $A_\mu(\ell)$ or
with $B_\mu(\ell)$ to collinear factors $A$ and $B$ without an external
gluon.

We write $A^{i}$ and $B^{i}$ with $i=T,S$ for the unsubtracted collinear
factor with one gluon coupling to a time- or spacelike Wilson line, and
$S^{ij}$ for a graph contributing to the one-loop expression of the soft
factor with time- or spacelike Wilson lines (with the gluon coupling to
$v_L$ and $v_R$ on a definite side of the final state cut). The difference
between the factorised expressions with timelike and spacelike auxiliary
Grammer-Yennie vectors (or Wilson lines), after the sum over gluon
attachments as specified, is
\begin{align} \label{diff-T-S1}
\Delta &=  (A^T - A\ms S^{TT}) B + A (B^T - S^{TT} B) + A\ms S^{TT} B
\nonumber \\
  & \quad - (A^S - A\ms S^{SS}) B - A (B^S - S^{SS} B) - A\ms S^{SS} B  \,.
\end{align}
We know that the factorised expression with spacelike Wilson lines is a
good approximation to the full expression, so if this difference is
negligible, the expression with timelike Wilson lines is good, too.

The result \eqref{diff-AT-AS} for subtracted collinear factors reads $A^T
- A\ms S^{TS} \approx A^S - A\ms S^{SS}$ in this notation, and its analogue
for left collinear factor reads $B^T - S^{ST} B \approx B^S - S^{SS} B$.
Using this in \eqref{diff-T-S1}, we obtain
\begin{align}
  \label{diff-T-S}
\Delta &\approx - A (S_{TT} + S_{SS} - S_{TS} - S_{ST}) B \,.
\end{align}
It is easy to evaluate the graphs for the one-loop soft factor explicitly
(see e.g.\ section~3.3.1 of~\cite{Diehl:2011yj}).  We find that, up to
power corrections, the result for each graph is a factor independent of
$v_L$ and $v_R$ times $\log(- v_L^-/v_L^+ + \sigma_L^{}\ms i \varepsilon)
+ \log(- v_R^+/v_R^- + \sigma_R^{}\ms i \varepsilon)$, where
$\sigma_{L,R}^{} = -1$ ($+1$) if the corresponding eikonal line is left
(right) of the final state cut.  Plugging this result
into \eqref{diff-T-S}, we get zero.
This result can also be shown in a simple way as follows. Up to global
factor the combination of soft factors can be written either as
\begin{align} \label{TS-softfacts}
\int \df\ell^+\ms \df\ell^- &
\biggl[ \frac{1}{\ell^+ + \delta^2 \ell^- - i\varepsilon}
      - \frac{1}{\ell^+ - \delta^2 \ell^- - i\varepsilon} \biggr]
\frac{i}{2\ell^+ \ell^- - \tvec{\ell}{}^2 + i\epsilon}
\nonumber \\
  \times &
\biggl[ \frac{1}{\ell^- + \delta'^2 \ell^+ + i\varepsilon}
      - \frac{1}{\ell^- - \delta'^2 \ell^+ + i\varepsilon} \biggr]
\end{align}
or as its analogy with the gluon propagator replaced by
$2\pi \delta(\ell^2)\, \Theta(\ell^-)$. The differences of eikonal
propagators yield a power suppression unless
$\ell^+ \lsim \delta^2 \ell^-$ (for the first square bracket) and
$\ell^+ \gsim \ell^- /\delta'^{2}$ (for the second square
bracket). However, both conditions cannot be satisfied at the same time,
so one always has a suppression in \eqref{TS-softfacts}.

Let us now move to the double Drell-Yan process.  In each collinear factor
we now have twice as many Wilson lines to which the single gluon may
attach.  Demonstrating that \eqref{diff-AT-AS} still holds at the
one-gluon level is a straightforward copy of the single Drell-Yan
argument, provided that one routes $\ell$ through the relevant Wilson
line, down into $A$ through the gluon and back up through the parton
associated with the Wilson line.  In the formulae analogous
to \eqref{diff-T-S1} and \eqref{diff-T-S}, both $A^{i}$ and $B^{j}$ are
now summed over the two possible gluon attachments to Wilson lines.
Likewise, the one-loop expression for $S^{ij}$ in \eqref{diff-T-S1}
and \eqref{diff-T-S} now contains a sum over the four possibilities how
the gluon can attach to the different left- and right-moving Wilson lines
(always on a definite side of the final-state cut).  For each of these
possibilities, the longitudinal structure of the loop integral is exactly
the same as for single Drell-Yan production.  The argument
of \eqref{TS-softfacts} works on a diagram-by-diagram basis, and of course
also for the relevant sums over graphs.  Thus our overall argument carries
over to the double Drell-Yan process.

The simple structure in \eqref{diff-T-S} is only obtained with a single
exchanged gluon. We can therefore not decide from the above whether
timelike Wilson lines are generally suitable for factorisation. However,
the proof in the present section shows that a counter-example would
require at least two exchanged gluons.


\acknowledgments

We are indebted to John Collins for numerous helpful discussions.  Thanks
go to Maarten Buffing and Tomas Kasemets for their careful reading of the
manuscript.  Some of the calculations for this work were done with FORM
\cite{Kuipers:2012rf}, and the figures were produced with JaxoDraw
\cite{Binosi:2003yf}.
We acknowledge support by BMBF (grants 05P12WRFTE and 05P15WRCAA).
J.G.\ acknowledges financial support from the European Community under the
“Ideas” program QWORK (contract 320389).


\phantomsection
\addcontentsline{toc}{section}{References}

\bibliographystyle{JHEP}
\bibliography{factdDY}

\end{document}